\documentclass[amsmath, amssymb, aip, jmp, reprint]{revtex4-2}

\usepackage[top=1in, bottom=1.25in, left=1.25in, right=1.25in]{geometry}

\usepackage{bbold}
\usepackage{mathrsfs}
\usepackage{tikz}
\usetikzlibrary{shapes.geometric}
\usetikzlibrary{decorations.markings}

\newtheorem{lemma}{Lemma}
\newtheorem{proposition}{Proposition}

\begin{document}

\title{Knotted 4-regular graphs II: Consistent application of the Pachner moves}
\author{Daniel Cartin}

\begin{abstract} 
	A common choice for the evolution of the knotted graphs in loop quantum gravity is to use the Pachner moves, adapted to graphs from their dual triangulations. Here, we show that the natural way to consistently use these moves is on framed graphs with edge twists, where the Pachner moves can only be performed when the twists, and the vertices the edges are incident on, meet certain criteria. For other twists, one can introduce an algebraic object, which allow any knotted graph with framed edges to be written in terms of a generalized braid group.
\end{abstract}

\maketitle

\section{Introduction}

The quantum states in loop quantum gravity (LQG) are spin networks -- graphs whose edges are labeled with representations of $SU(2)$. The graph underlying each spin network is embedded into a three-dimensional manifold, usually chosen to be $\mathbb{R}^3$ or $S^3$. If we focus purely on the graph aspect of the state, then this embedding leads to more variety than the space of abstract graphs. Specifically, two graphs $G_1$ and $G_2$ may have the same connectivity between vertices (i.e. the same adjacency matrix), but differences in the embedding can lead to non-isomorphic spatial graphs. In other words, there may be no diffeomorphism of the embedding space that takes $G_1$ to $G_2$. Thus, a description of LQG states must include information about the graph itself. When discussing an embedded graph, it is typical to use a graph diagram, where the embedded graph is projected onto a two-dimensional plane; the vertices of the graph can generically be rearranged so that there is a pair of incident edges above, and a pair below, the projection plane. A given embedded graph can have many graph diagrams, so care must be used to ensure the discussion is independent of the choice of diagram. As we will see, the need for such consistency will motivate the results given in this paper.

The classical picture of a configuration space for general relativity is in terms of a manifold. Graphs with particular properties can be pictured as discrete versions of such manifolds by thinking of the dual triangulation of the graph. Three-dimensional manifolds describe spatial slices, so their triangulations consist of tetrahedra (or 3-simplices) whose triangular faces are glued together. To go from a graph to a triangulation, the vertices of the graph are dual to the tetrahedra, while the graph edges are dual to the faces. Since each 3-simplex has four faces, this work focuses on the study of 4-regular graphs, where there are four edges incident to each vertex. Unlike the situation for two-dimensional manifolds, the duality just given is not sufficient to construct a triangulation from a given graph; however, the viewpoint of this paper focuses on the graphs as fundamental. For now, we note that the duality between 4-regular graphs and 3-dimensional triangulations is sufficient to define at least a pseudo-manifold. In addition, we consider graphs which have both multiple edges between two vertices, or self-loops for a single vertex. Properly, these are usually called pseudographs, but for simplicity, they will be referred to as graphs from now on. This implies the need for generalized triangulations (as defined in Burton~\cite{Bur11}), with the possibility of identifications between faces of the same tetrahedron, or two tetrahedra may have more than one face glued together.


\begin{figure}[hbt]
\centering
\begin{tikzpicture}[> = latex]


\matrix[column sep = 0.25 cm]{

	\coordinate (v1) at (0, -1/3, {sqrt(8/9)});
	\coordinate (v2) at ({sqrt(2/3)}, -1/3, {-sqrt(2/9)});
	\coordinate (v3) at ({-sqrt(2/3)}, -1/3, {-sqrt(2/9)});
	\coordinate (v4) at (0, 1, 0);

	\draw (v1) node [below left] {$b$} -- (v2) node [right] {$d$};
	\draw (v1) -- (v3) node [left] {$c$};
	\draw (v1) -- (v4) node [above] {$a$};
	\draw [dotted] (v2) -- (v3);
	\draw (v2) -- (v4);
	\draw (v3) -- (v4);

&

	\begin{scope}[->, font = \footnotesize]

		\draw (0, 0.25) -- node [above] {1-4 move} (1, 0.25);
		\draw (1, -0.25) -- node [below] {4-1 move} (0, -0.25);

	\end{scope}

&

	\coordinate (v0) at (0, 0, 0);
	\coordinate (v1) at (0, -1/3, {sqrt(8/9)});
	\coordinate (v2) at ({sqrt(2/3)}, -1/3, {-sqrt(2/9)});
	\coordinate (v3) at ({-sqrt(2/3)}, -1/3, {-sqrt(2/9)});
	\coordinate (v4) at (0, 1, 0);

	\draw (v1) node [below left] {$b$} -- (v2) node [right] {$d$};
	\draw (v1) -- (v3) node [left] {$c$};
	\draw (v1) -- (v4) node [above] {$a$};
	\draw [dotted] (v2) -- (v3);
	\draw (v2) -- (v4);
	\draw (v3) -- (v4);

	\begin{scope}[thick, dashed]

		\draw (v0) node [above right] {$e$} -- (v1);
		\draw (v0) -- (v2);
		\draw (v0) -- (v3);
		\draw (v0) -- (v4);

	\end{scope}

&

	\draw [dashed] (0, -1) -- (0, 1);

&

	\coordinate (v1) at (0, -1/3, {sqrt(8/9)});
	\coordinate (v2) at ({sqrt(2/3)}, -1/3, {-sqrt(2/9)});
	\coordinate (v3) at ({-sqrt(2/3)}, -1/3, {-sqrt(2/9)});
	\coordinate (v4) at (0, 1, 0);
	\coordinate (v4p) at (0, -1, 0);

	\draw [fill = gray!45, draw = none] (v1) -- (v2) -- (v3) -- (v1);

	\draw (v1) node [below left] {$b$} -- (v2) node [right] {$d$};
	\draw (v1) -- (v3) node [left] {$c$};
	\draw (v1) -- (v4) node [above] {$a$};
	\draw [dotted] (v2) -- (v3);
	\draw (v2) -- (v4);
	\draw (v3) -- (v4);

	\draw (v1) -- (v4p) node [below] {$a'$};
	\draw (v2) -- (v4p);
	\draw [dotted] (v3) -- (v4p);

&

	\begin{scope}[->, font = \footnotesize]

		\draw (0, 0.25) -- node [above] {2-3 move} (1, 0.25);
		\draw (1, -0.25) -- node [below] {3-2 move} (0, -0.25);

	\end{scope}

&

	\coordinate (v1) at (0, -1/3, {sqrt(8/9)});
	\coordinate (v2) at ({sqrt(2/3)}, -1/3, {-sqrt(2/9)});
	\coordinate (v3) at ({-sqrt(2/3)}, -1/3, {-sqrt(2/9)});
	\coordinate (v4) at (0, 1, 0);
	\coordinate (v4p) at (0, -1, 0);

	\draw [fill = gray!30, draw = none] (v4) -- (v1) -- (v4p) -- (v4);
	\draw [fill = gray!60, draw = none] (v4) -- (v2) -- (v4p) -- (v4);
	\draw [thick, dashed] (v4) -- (v4p);

	\draw (v1) node [below left] {$b$} -- (v2) node [right] {$d$};
	\draw (v1) -- (v3) node [left] {$c$};
	\draw (v1) -- (v4) node [above] {$a$};
	\draw [dotted] (v2) -- (v3);
	\draw (v2) -- (v4);
	\draw (v3) -- (v4);

	\draw (v1) -- (v4p) node [below] {$a'$};
	\draw (v2) -- (v4p);
	\draw [dotted] (v3) -- (v4p);

\\
};

\end{tikzpicture}
\caption{\label{Pach-tet}Pachner moves acting on tetrahedra for the triangulation of a three-dimensional manifold. For the 1-4 move, the vertex $e$ is added, along with edges $ae, be, ce, de$. For the 2-3 move, the face $bcd$ is removed, and the edge $aa'$ is added, along with the faces $aba', aca',$ and $ada'$.}
\end{figure}
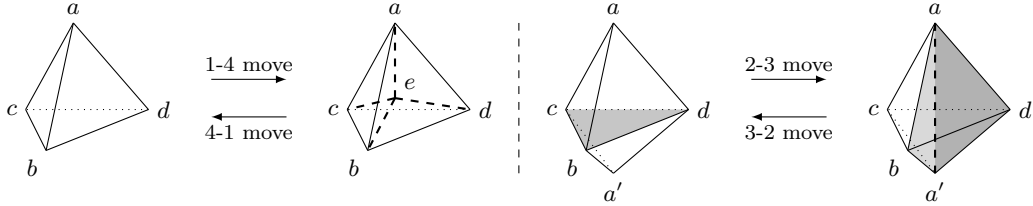

Since the spin networks represent kinematic states of quantum gravity, there must be dynamical rules underpinning any particular model. A frequent choice is the use of graph moves based on the Pachner moves on triangulations, which ensures independence from the choice of the discretization used. For a given manifold, the Pachner moves are sufficient to pass between any two triangulations of the manifold~\cite{Pac91}. How the Pachner moves act on 3-simplices is shown in Figure \ref{Pach-tet}. Because of the duality between tetrahedra and 4-valent graphs, one can define equivalent moves that change one graph to another. These will also be referred to as Pachner moves; when the context is unclear, it will be spelled out whether the moves are those of graphs or their dual triangulations. However, when one tries to use these graph Pachner moves, it can lead to inconsistencies. In particular, the same move acting on a graph can give different final graphs, depending on the graph diagram chosen. This issue will lead to the choice of framed graphs as the natural setting for defining the Pachner moves on knotted 4-regular graphs. In particular, the twists on framed edges provide exactly the necessary information for unambiguous results after any Pachner move is applied. The main goal of this work is to spell out how to define these moves consistently, as well as some consequences of this method.

The plan of this paper is now summarized. In Section \ref{unframe}, some important notions for graphs are defined. In particular, the notation used for graph diagrams is given, along with the generalized Reidemeister moves -- a finite sequence of such moves starting with one graph diagram will obtain another graph diagram for the same embedded graph. This section also shows an example of how the Pachner moves, applied to two graph diagrams of the same graph, can give diagrams for inequivalent graphs. This situation is rectified in Section \ref{frame}, where we turn to framed edges for the graph. These framed edges allow for twists along the edges, which is sufficient information to consistently apply the Pachner moves. A related notion to framed graphs is the 3-line graph, which demonstrates explicitly information from the dual triangulation. In addition, the definition of a 3-line graph associated to a knotted framed 4-regular graph allows a topological invariant to be defined for the graph. However, some framed edges do not permit the definition of a 3-line graph; this occurs when the twist is not naturally dual to the gluing of 3-simplex faces. How to handle this issue is demonstrated in Section \ref{gen-braid}. There, the definition of an added algebraic object allows one to define 3-line graphs for all possible twists. From this, just as any knot or link can be written as a closed braid, so with this extra object, any 3-line graph can be written as the closure of an expanded braid group. This work is summed up in Section \ref{conclude}, and questions for further research are discussed.

\section{Pachner moves on knotted graphs}
\label{unframe}

\subsection{Graph definitions}

Since many of the terms in graph theory have differing definitions, depending on the reference, here we spell out specifically what is meant by particular terms. Suppose that $G = G(E, V)$ is an abstract graph with a set of vertices $V$, and a set of edges $E$ between the vertices. The graphs considered here may have multiple edges between the same two vertices, or have self-loops, where the edge connects twice to the same vertex. It is convenient to think of each edge as made of two half-edges, each of which is incident to a vertex and another half-edge. The number of half-edges incident to a particular vertex is known as its {\it valence} or {\it degree}. Then, graphs where all vertices have the same valence are called {\it regular}. All graphs in this work will be 4-regular, meaning that every vertex has four half-edges incident to it; self-loops thus count as two half-edges incident to the vertex. Let $f: G \to M$ be an embedding of the abstract graph $G$ into a three-dimensional manifold $M$; we will assume that $M$ is $\mathbb{R}^3$ or $S^3$, and not worry about the details of the choice. The image $f(G)$ is referred to as a {\it knotted graph} or {\it spatial graph}. By choosing a plane in $M$, the graph embedding can be projected onto the plane to give a {\it graph diagram} $D$. There are two reasons that edges in the graph diagram may intersect. One of these is the location of a vertex, the other a {\it crossing}, where one edge passes over the other. The term {\it node} is used to refer to a member of the set of vertices and crossings of the diagram. These definitions recapitulate those given in previous work~\cite{Car22}, which will be referred to as Paper I from now on.

\begin{figure}[hbt]
\centering
\begin{tikzpicture}[align = center]
\matrix[column sep = 1 cm]{

&

	\node {Physical\\representation};

&

	\node {Schematic\\representation};
	
&

	\node {Dual\\tetrahedron};

\\

	\node {$\oplus$ vertex state};

&


	\node [cylinder, draw, minimum height = 0.5 cm, minimum width = 0.05 cm, aspect = 0.5,
		left color = gray!70, right color = gray, middle color = gray!50, label = {left : $A$}] at ({-sqrt(8) / 6}, 0) {};

	\node [cylinder, draw, minimum height = 0.5 cm, minimum width = 0.05 cm, aspect = 0.5, rotate = 180,
		left color = gray!70, right color = gray, middle color = gray!50, label = {left : $C$}] at ({sqrt(8) / 6}, 0) {};

	\node [cylinder, draw, minimum height = 0.5 cm, minimum width = 0.05 cm, aspect = 0.5, rotate = -90,
		left color = gray!70, right color = gray, middle color = gray!50] at (0, {-sqrt(8) / 6}) {};

	\draw [ball color = gray!50] (0, 0, 0) circle (0.5);

	\node [cylinder, draw, minimum height = 0.5 cm, minimum width = 0.05 cm, aspect = 0.5, rotate = 90,
		left color = gray!70, right color = gray, middle color = gray!50, label = {right : $B$}] at (0, {sqrt(8) / 6}) {};

	\node [cylinder, draw, minimum height = 0.5 cm, minimum width = 0.05 cm, aspect = 0.5, rotate = -90,
		left color = gray!70, right color = gray, middle color = gray!50, label = {right : $D$}] at (0, {-sqrt(8) / 6}) {};

&


	\draw (-0.5, 0) node [left] {$A$} -- (0.5, 0) node [right] {$C$};
	\draw [fill = white] (0, 0) circle (0.15);
	\draw (0, -0.5) node [below] {$D$} -- (0, 0.5) node [above] {$B$};

&

	\draw [fill = gray!30] (-1, 0) node [left] {$c$} -- (1, 0) node [right] {$a$} -- (0, 1) node [above] {$d$} -- cycle;
	\draw [fill = gray!30] (-1, 0) -- (1, 0) -- (0, -1) node [below] {$b$} -- cycle;
	\draw [dashed] (0, 1) -- (0, -1);

\\

	\node {$\ominus$ vertex state};

&


	\node [cylinder, draw, minimum height = 0.5 cm, minimum width = 0.05 cm, aspect = 0.5, rotate = -90,
		left color = gray!70, right color = gray, middle color = gray!50, label = {left : $B$}] at (0, {sqrt(8) / 6}) {};

	\node [cylinder, draw, minimum height = 0.5 cm, minimum width = 0.05 cm, aspect = 0.5, rotate = 90,
		left color = gray!70, right color = gray, middle color = gray!50, label = {left : $D$}] at (0, {-sqrt(8) / 6}) {};

	\draw [ball color = gray!50] (0, 0, 0) circle (0.5);

	\node [cylinder, draw, minimum height = 0.5 cm, minimum width = 0.05 cm, aspect = 0.5, rotate = 180,
		left color = gray!70, right color = gray, middle color = gray!50, label = {right : $A$}] at ({-sqrt(8) / 6}, 0) {};

	\node [cylinder, draw, minimum height = 0.5 cm, minimum width = 0.05 cm, aspect = 0.5,
		left color = gray!70, right color = gray, middle color = gray!50, label = {right : $C$}] at ({sqrt(8) / 6}, 0) {};

&


	\draw (0, -0.5) node [below] {$D$} -- (0, 0.5) node [above] {$B$};
	\draw [fill = white] (0, 0) circle (0.15);
	\draw (-0.5, 0) node [left] {$A$} -- (0.5, 0) node [right] {$C$};

&

	\draw [fill = gray!30] (-1, 0) node [left] {$c$} -- (0, 1) node [above] {$d$} -- (0, -1) node [below] {$b$} -- cycle;
	\draw [fill = gray!30] (1, 0) node [right] {$a$} -- (0, 1) -- (0, -1) -- cycle;
	\draw [dashed] (-1, 0) -- (1, 0);

\\
};
\end{tikzpicture}
\caption{\label{vert-states}The $\oplus$ and $\ominus$ vertex states, along with their physical representations, the schematic representation used in graph diarams, and the corresponding tetrahedra in the dual simplicial complex.}
\end{figure}
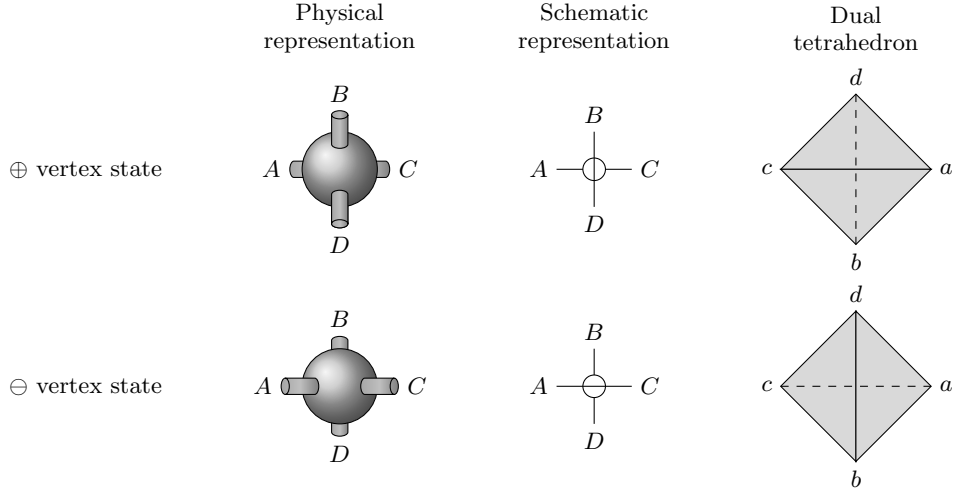

To show a particular graph diagram visually, the notation developed by Wan~\cite{Wan07} is used. Without loss of generality, one can chose the projection plane of the graph diagram so that it passes through vertices in such a way that two of the incident edges are above the plane, and the other two are below. By rotating the vertex, it projects into one of the two possibilities shown in Figure \ref{vert-states}. There, the set $\{A, B, C, D\}$ gives the four edges incident to the vertex. The choice of how the vertex is intersected by the projection plane gives the two {\it vertex states} $\oplus$ and $\ominus$. Figure \ref{vert-states} shows three ways of thinking about these states. First, if we imagine the vertex as a 2-sphere	 of finite radius, and the edges incident to the vertex with finite thickness as well, then the vertices corresponding to the two vertex states are shown in the left-hand column of Figure \ref{vert-states}. This physical picture will align with the framed graphs defined in Section \ref{3-line}. When showing a graph diagram, the notation used in the middle column is used. For the $\oplus$ state, edges $B$ and $D$ are those above the projection plane, while for the $\ominus$ state, edges $A$ and $C$ lie above the plane. The dual tetrahedra for each vertex state are shown in the right-hand column. Here, the four vertices are labeled from the set $\{a, b, c, d\}$. Each dual face is labeled by the set of three lower-case letters not corresponding to the upper-case edge letter. So, for example, the face $abc$ in the triangulation is dual to the edge $D$ in the graph.

In order to describe a given graph diagram, we define a {\it Dowker-Thistletwaite} (or DT) sequence in a manner similar to that used for knots. To do this, the graph diagram must be one where an Eulerian circuit can be found through the diagram, such that the circuit always passes through each vertex from an incident edge to the edge opposite it. This kind of graph diagram is known as a proper graph diagram. As shown in Paper I, any choice of Eulerian circuit through a graph diagram can be changed into proper form by combinations of reversing a portion of the circuit or using the RV move, as needed. After this is done, and the graph diagram is in proper form, the DT sequence is defined as follows.

\begin{enumerate}

	\item For a graph with $N$ nodes, the sequence is made of pairs of node labels drawn from the set $\{0, \cdots, 2N - 1\}$.
	
	\item The abbreviated form of the DT sequence will be a listing of odd numbers, given in order of their even counterpart labels in each pair.
	
	\item A crossing is a negative crossing type if the edge with the even label passes under the edge with the odd label, and a positive crossing has the reverse; a number representing a crossing with have either a $+$ or $-$ superscript to indicate the crossing type.

	\item Likewise, the vertices are denoted by $\ell$ or $u$, based on the crossing in the knot diagram corresponding to the vertex. A portion of the Eulerian circuit through a vertex is considered the top edge if it goes between the two incident edges above the projection plane. If the even edge passes under the odd label in the knot diagram, the vertex is labeled as $\ell$; otherwise, the vertex is labeled $u$.

\end{enumerate}
For any given knotted graph, there are many equivalent DT sequences, since the graph diagram is invariant under modifying the node labels $x$ by $x \mapsto x + b$ or $x \mapsto b - x$ (both modulo $2N$) for any number $1 \le b \le 2N - 1$. The standard form used will be the sequence in lowest lexicographic order, where the superscripts are ordered as $\ell, - +, u$.

In Section \ref{3-line}, it will be convenient to generalize the edges of the graph. In the previous paragraph, each edge $e \in E$ was a one-dimensional line between two vertices (or the same vertex, if a self-loop). These edges can be extended into a {\it framed} edge, where the cross-section of the edge is a triangle. Note that this is a non-standard definition of a framed edge (see, e.g. Elhamdadi, Hajij, and Istvan~\cite{Elh-Haj-Ist19}), but is the one used by Markopoulou and Pr\'emont-Schwarz~\cite{Mar-Pre-Sch08}. A two-dimensional basis can be defined on the edge cross-section, where one of the vectors is chosen to be parallel to one of the triangle sides, with the second vector perpendicular to the first, pointing in the direction of the triangle vertex not on the original triangle side. These framed edges can have twists, where the vector basis rotates in the embedding space as one travels from one of tbe incident vertices to the other. A graph with framed edges will be referred to as a framed graph. Building on the notion of a framed edge, for each knotted 4-regular graph one can define a {\it 3-line graph}, where each single edge is replaced by three {\it lines} between vertices; the group of three lines is called a {\it leg}. These lines represent the sides of the triangular cross-section of a framed edge, or equivalently, the edges of the faces in the triangulation dual to the graph. Each leg is dual to a triangular face. As with graphs, we will consider graph diagrams for the 3-line graph, where the lines are projected onto a plane. Thus, the lines will generically have crossings, as will the vertices (to represent the vertex state). Examples of these 3-line graphs are given at various points later in the paper.

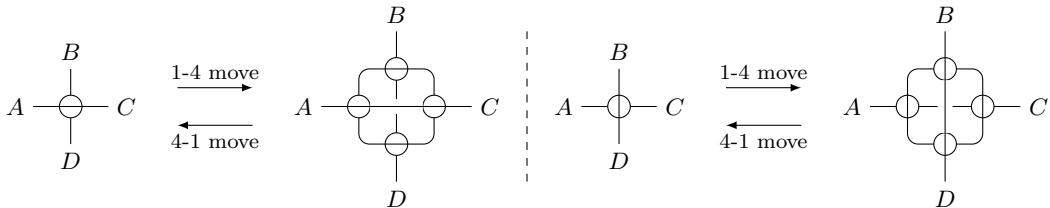
\begin{figure}[hbt]
\centering
\begin{tikzpicture}[> = latex]
\matrix[column sep = 0.25 cm, row sep = 0.5 cm]{


	\draw (0, -0.5) node [below] {$D$} -- (0, 0.5) node [above] {$B$};
	\draw [fill = white] (0, 0) circle (0.15);
	\draw (-0.5, 0) node [left] {$A$} -- (0.5, 0) node [right] {$C$};s

&

	\begin{scope}[->, font = \footnotesize]

		\draw (0, 0.25) -- node [above] {1-4 move} (1, 0.25);
		\draw (1, -0.25) -- node [below] {4-1 move} (0, -0.25);

	\end{scope}

&


	\draw (0, 1) node [above] {$B$} -- (0, -1) node [below] {$D$};

	\draw [fill = white] (-0.5, 0) circle (0.15);
	\draw [fill = white] (0, 0.5) circle (0.15);
	\draw [fill = white] (0.5, 0) circle (0.15);
	\draw [fill = white] (0, -0.5) circle (0.15);

	\draw [rounded corners] (-0.5, -0.15) -- (-0.5, -0.5) -- (0.5, -0.5) -- (0.5, -0.15)
		(-0.5, 0.15) -- (-0.5, 0.5) -- (0.5, 0.5) -- (0.5, 0.15);

	\draw [draw = white, double = black, double distance between line centers = 3 pt, line width = 2.6 pt] (-0.35, 0) -- (0.35, 0);

	\draw (-1, 0) node [left] {$A$} -- (-0.35, 0);
	\draw (0.35, 0) -- (1, 0) node [right] {$C$};
	
&
	\draw [dashed] (0, -1) -- (0, 1);
&


	\draw (-0.5, 0) node [left] {$A$} -- (0.5, 0) node [right] {$C$};
	\draw [fill = white] (0, 0) circle (0.15);
	\draw (0, -0.5) node [below] {$D$} -- (0, 0.5) node [above] {$B$};

&

	\begin{scope}[->, font = \footnotesize]

		\draw (0, 0.25) -- node [above] {1-4 move} (1, 0.25);
		\draw (1, -0.25) -- node [below] {4-1 move} (0, -0.25);

	\end{scope}

&


	\draw (-1, 0) node [left] {$A$} -- (1, 0) node [right] {$C$};

	\draw [fill = white] (-0.5, 0) circle (0.15);
	\draw [fill = white] (0, 0.5) circle (0.15);
	\draw [fill = white] (0.5, 0) circle (0.15);
	\draw [fill = white] (0, -0.5) circle (0.15);

	\draw [rounded corners] (-0.15, -0.5) -- (-0.5, -0.5) -- (-0.5, 0.5) -- (-0.15, 0.5)
		(0.15, -0.5) -- (0.5, -0.5) -- (0.5, 0.5) -- (0.15, 0.5);

	\draw [draw = white, double = black, double distance between line centers = 3 pt, line width = 2.6 pt] (0, -0.35) -- (0, 0.35);

	\draw (0, 1) node [above] {$B$} -- (0, 0.35);
	\draw (0, -0.35) -- (0, -1) node [below] {$D$};

\\
};
\end{tikzpicture}
\caption{\label{1-4-graph}1-4 and 4-1 Pachner graph moves.}
\end{figure}

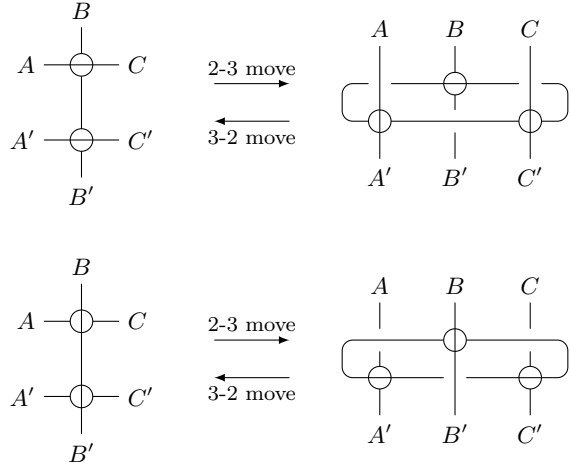
\begin{figure}[hbt]
\centering
\begin{tikzpicture}[> = latex]
\matrix[column sep = 0.5 cm, row sep = 0.5 cm]{


	\draw (0, -1) node [below] {$B'$} -- (0, 1) node [above] {$B$};

	\draw [fill = white] (0, -0.5) circle (0.15);
	\draw [fill = white] (0, 0.5) circle (0.15);

	\draw (0.5, -0.5) node [right] {$C'$} -- (-0.5, -0.5) node [left] {$A'$};
	\draw (0.5, 0.5) node [right] {$C$} -- (-0.5, 0.5) node [left] {$A$};

&

	\begin{scope}[->, font = \footnotesize]

		\draw (0, 0.25) -- node [above] {2-3 move} (1, 0.25);
		\draw (1, -0.25) -- node [below] {3-2 move} (0, -0.25);

	\end{scope}

&

	
	\draw (0, -0.75) node [below] {$B'$} -- (0, -0.4) (0, -0.1) -- (0, 0.75) node [above] {$B$};
	\draw (0, 0.25) [fill = white] circle (0.15);
	\draw [rounded corners] (0, 0.25) -- (0.85, 0.25) (1.15, 0.25) -- (1.5, 0.25) -- (1.5, -0.25) -- (1.15, -0.25)
		(0.85, -0.25) -- (-0.85, -0.25) (-1.15, -0.25) -- (-1.5, -0.25) -- (-1.5, 0.25) -- (-1.15, 0.25) (-0.85, 0.25) -- (0, 0.25);
	
	\draw (-1, -0.25) [fill = white] circle (0.15);
	\draw (1, -0.25) [fill = white] circle (0.15);
	
	\draw (-1, 0.75) node [above] {$A$} -- (-1, -0.75) node [below] {$A'$};
	\draw (1, 0.75) node [above] {$C$} -- (1, -0.75) node [below] {$C'$};

\\


	\draw (0.5, -0.5) node [right] {$C'$} -- (-0.5, -0.5) node [left] {$A'$};
	\draw (0.5, 0.5) node [right] {$C$} -- (-0.5, 0.5) node [left] {$A$};

	\draw [fill = white] (0, -0.5) circle (0.15);
	\draw [fill = white] (0, 0.5) circle (0.15);

	\draw (0, -1) node [below] {$B'$} -- (0, 1) node [above] {$B$};

&

	\begin{scope}[->, font = \footnotesize]

		\draw (0, 0.25) -- node [above] {2-3 move} (1, 0.25);
		\draw (1, -0.25) -- node [below] {3-2 move} (0, -0.25);

	\end{scope}

&

	
	\draw (-1, 0.75) node [above] {$A$} -- (-1, 0.4) (-1, 0.1) -- (-1, -0.75) node [below] {$A'$};
	\draw (1, 0.75) node [above] {$C$} -- (1, 0.4) (1, 0.1) -- (1, -0.75) node [below] {$C'$};
	
	\draw (-1, -0.25) [fill = white] circle (0.15);
	\draw (1, -0.25) [fill = white] circle (0.15);
	
	\draw [rounded corners] (0.15, -0.25) -- (1.5, -0.25) -- (1.5, 0.25) -- (0.15, 0.25) (-0.15, 0.25) -- (-1.5, 0.25) --
		(-1.5, -0.25) -- (-0.15, -0.25);
	
	\draw (0, 0.25) [fill = white] circle (0.15);
	\draw (0, -0.75) node [below] {$B'$} -- (0, 0.75) node [above] {$B$};

\\
};
\end{tikzpicture}
\caption{\label{2-3-graph}The 2-3 and 3-2 Pachner graph moves.}
\end{figure}

Previously, we saw in Figure \ref{Pach-tet} how the Pachner moves act on the tetrahedra of a triangulation. Using the mapping between the triangulation and the graph given in Figure \ref{vert-states}, we can now draw the corresponding Pachner moves on (unframed) graphs. The 1-4 move, and its inverse 4-1 move, for the graph is shown in Figure \ref{1-4-graph}; the same for the 2-3 and 3-2 moves are drawn in Figure \ref{2-3-graph}. To aid in presenting results later in Section \ref{3-line}, the 2-3 moves are drawn differently than the corresponding diagrams in Paper I. How the edges connecting the vertices are drawn is based on whether the face common to the dual tetrahedra is above or below the projection plane. In particular, first consider the 1-4 and 4-1 moves on the left-hand side of Figure \ref{Pach-tet}. The single tetrahedron is drawn with the edge $ab$ above the projection plane, and the edge $cd$ beneath. Thus, it is dual to a vertex in the $\ominus$ state. Now, look at the tetrahedron $abce$ resulting from the 1-4 move. In the projection plane used in the left side of Figure \ref{Pach-tet}, the tetrahedra $abce$ and $abde$ are above the tetrahedra $acde$ and $bcde$. The new tetrahedra obtained after the 1-4 move can all be oriented so that they are also in the $\ominus$ state. For example, the tetrahedron $abce$, the edge $ab$ is above the edge $ce$, while for the tetrahedron $bcde$, the edge $be$ is above the edge $cd$. The location of these dual tetrahedra relative to the projection plane informs how the graph edges cross above or below any other edges.

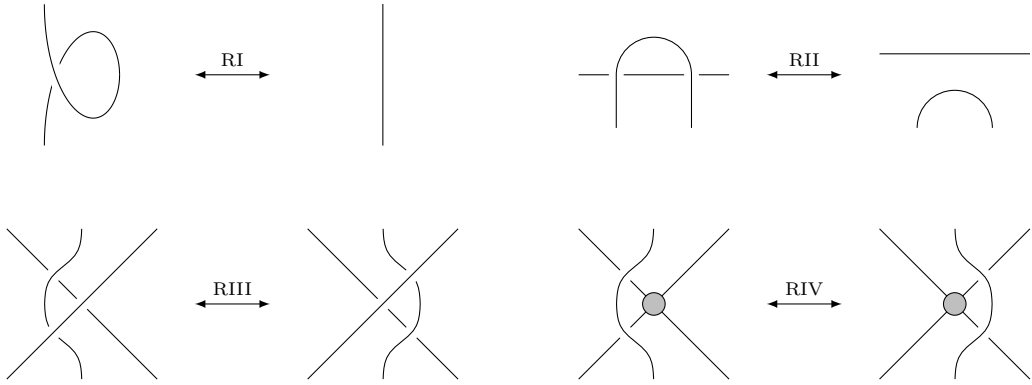
\begin{figure}[hbt]
\begin{tikzpicture}[> = latex, font = \scriptsize]
\matrix[column sep = 1.5 cm, row sep = 1 cm]{



	\begin{scope}[xshift = -1.5 cm]

		\draw [variable = \t, samples = 100, domain = -90 : 90] plot ({0.5 * sin(\t) - 0.5}, {cos(\t) + 0.3 * (3.14 / 180) * \t});
		\draw [variable = \t, samples = 100, domain = 90 : 270, draw = white, double = black, double distance between line centers = 3 pt, line width = 2.6 pt] plot ({0.5 * sin(\t) - 0.5}, {cos(\t) + 0.3 * (3.14 / 180) * \t});
	
		\draw [<->] (1, {0.3 * 3.14 / 2}) -- node [above] {RI} (2, {0.3 * 3.14 / 2});
	
		\draw (3.5, {-0.3 * 3.14 / 2}) -- (3.5, {0.3 * 3 * 3.14 / 2});

	\end{scope}

&


	\begin{scope}[xshift = -3 cm]

		\draw (0, {0.3 * 3.14 / 2}) -- (2, {0.3 * 3.14 / 2});
		\draw [draw = white, double = black, double distance between line centers = 3 pt, line width = 2.6 pt] (0.5, {-0.15 * 3.14 / 2}) -- (0.5, {0.3 * 3.14 / 2}) arc (180 : 0 : 0.5) -- (1.5, {-0.15 * 3.14 / 2});
	
		\draw [<->] (2.5, {0.3 * 3.14 / 2}) -- node [above] {RII} (3.5, {0.3 * 3.14 / 2});
	
		\draw (4, 0.75) -- (6, 0.75);
		\draw (4.5, {-0.15 * 3.14 / 2}) arc (180 : 0 : 0.5);

	\end{scope}

\\
	
	
	\draw (-3, 1) -- (-1, -1);
	\draw [draw = white, double = black, double distance between line centers = 3 pt, line width = 2.6 pt]
		plot [variable = \y, samples = 15, domain = -1: 1] ({-0.5 / (1 + exp(-5 * \y)) - 2}, 0.5 * \y - 0.5);
	\draw [draw = white, double = black, double distance between line centers = 3 pt, line width = 2.6 pt]
		plot [variable = \y, samples = 15, domain = -1: 1] ({0.5 / (1 + exp(-5 * \y)) - 2.5}, 0.5 * \y + 0.5);
	\draw [draw = white, double = black, double distance between line centers = 3 pt, line width = 2.6 pt] (-1, 1) -- (-3, -1);
		
	\draw [<->] (-0.5, 0) -- node [midway, above] {RIII} (0.5, 0);
		
	\draw (3, -1) -- (1, 1);
	\draw [draw = white, double = black, double distance between line centers = 3 pt, line width = 2.6 pt]
		plot [variable = \y, samples = 15, domain = -1: 1] ({0.5 / (1 + exp(-5 * \y)) + 2}, 0.5 * \y - 0.5);
	\draw [draw = white, double = black, double distance between line centers = 3 pt, line width = 2.6 pt]
		plot [variable = \y, samples = 15, domain = -1: 1] ({-0.5 / (1 + exp(-5 * \y)) + 2.5}, 0.5 * \y + 0.5);
	\draw [draw = white, double = black, double distance between line centers = 3 pt, line width = 2.6 pt] (1, -1) -- (3, 1);
	
&


	\draw (-3, 1) -- (-1, -1) (-1, 1) -- (-3, -1);
	\draw [fill = gray!50] (-2, 0) circle (0.15);

	\draw [draw = white, double = black, double distance between line centers = 3 pt, line width = 2.6 pt]
		plot [variable = \y, samples = 15, domain = -1: 1] ({-0.5 / (1 + exp(-5 * \y)) - 2}, 0.5 * \y - 0.5);
	\draw [draw = white, double = black, double distance between line centers = 3 pt, line width = 2.6 pt]
		plot [variable = \y, samples = 15, domain = -1: 1] ({0.5 / (1 + exp(-5 * \y)) - 2.5}, 0.5 * \y + 0.5);
		
	\draw [<->] (-0.5, 0) -- node [midway, above] {RIV} (0.5, 0);
		
	\draw (3, -1) -- (1, 1) (1, -1) -- (3, 1);
	\draw [fill = gray!50] (2, 0) circle (0.15);

	\draw [draw = white, double = black, double distance between line centers = 3 pt, line width = 2.6 pt]
		plot [variable = \y, samples = 15, domain = -1: 1] ({0.5 / (1 + exp(-5 * \y)) + 2}, 0.5 * \y - 0.5);
	\draw [draw = white, double = black, double distance between line centers = 3 pt, line width = 2.6 pt]
		plot [variable = \y, samples = 15, domain = -1: 1] ({-0.5 / (1 + exp(-5 * \y)) + 2.5}, 0.5 * \y + 0.5);

\\
};
\end{tikzpicture}
\caption{\label{Reid-move}The first four generalized Reidemeister moves acting on graphs; not all variations of the moves are shown. For the RIV move, the vertex state is irrelevant, so a generic vertex is shown.}
\end{figure}

\begin{figure}[hbt]
\centering
\begin{tikzpicture}[> = latex, font = \scriptsize]
\matrix[column sep = 1 cm, row sep = 0.5 cm]{


	\draw (0, -0.6) node [below] {$C$} -- (0, 0.6) node [above] {$A$};
	\draw [fill = white] (0, 0) circle (0.15);
	\draw (-0.6, 0) -- (0.6, 0) node [right] {$B$};

	\draw [->, very thick, gray] (-0.3, -0.3) arc (225 : 125 : {0.3 * sqrt(2)});

	\draw [->] (1.4, 0) -- node [above] {RV} (2.4, 0);


	\draw [rounded corners] (2.9, 0) -- (3.9, 0) -- (3.9, 0.2) (3.9, 0.4) -- (3.9, 0.6) node [above] {$A$};
	\draw [fill = white] (3.5, 0) circle (0.15);
	\draw [rounded corners] (3.5, -0.6) node [below] {$C$} -- (3.5, 0.3) -- (4.1, 0.3) node [right] {$B$};

&


	\draw (0, -0.6) node [below] {$C$} -- (0, 0.6) node [above] {$A$};
	\draw [fill = white] (0, 0) circle (0.15);
	\draw (-0.6, 0) -- (0.6, 0) node [right] {$B$};

	\draw [->, very thick, gray] (-0.3, 0.3) arc (135 : 235 : {0.3 * sqrt(2)});

	\draw [->] (1.4, 0) -- node [above] {RV} (2.4, 0);


	\draw [rounded corners] (2.9, 0) -- (3.9, 0) -- (3.9, -0.2) (3.9, -0.4) -- (3.9, -0.6) node [below] {$C$};
	\draw [fill = white] (3.5, 0) circle (0.15);
	\draw [rounded corners] (3.5, 0.6) node [above] {$A$} -- (3.5, -0.3) -- (4.1, -0.3) node [right] {$B$};

\\


	\draw (-0.6, 0) -- (0.6, 0) node [right] {$B$};
	\draw [fill = white] (0, 0) circle (0.15);
	\draw (0, -0.6) node [below] {$C$} -- (0, 0.6) node [above] {$A$};

	\draw [->, very thick, gray] (-0.3, -0.3) arc (225 : 125 : {0.3 * sqrt(2)});

	\draw [->] (1.4, 0) -- node [above] {RV} (2.4, 0);


	\draw [rounded corners] (3.5, 0.6) node [above] {$A$} -- (3.5, -0.3) -- (3.8, -0.3) (4, -0.3) -- (4.1, -0.3) node [right] {$B$};
	\draw [fill = white] (3.5, 0) circle (0.15);
	\draw [rounded corners] (2.9, 0) -- (3.9, 0) -- (3.9, -0.6) node [below] {$C$};

&


	\draw (-0.6, 0) -- (0.6, 0) node [right] {$B$};
	\draw [fill = white] (0, 0) circle (0.15);
	\draw (0, -0.6) node [below] {$C$} -- (0, 0.6) node [above] {$A$};

	\draw [->, very thick, gray] (-0.3, 0.3) arc (135 : 235 : {0.3 * sqrt(2)});

	\draw [->] (1.4, 0) -- node [above] {RV} (2.4, 0);


	\draw [rounded corners] (3.5, -0.6) node [below] {$C$} -- (3.5, 0.3) -- (3.8, 0.3) (4, 0.3) -- (4.1, 0.3) node [right] {$B$};
	\draw [fill = white] (3.5, 0) circle (0.15);
	\draw [rounded corners] (2.9, 0) -- (3.9, 0) -- (3.9, 0.6) node [above] {$A$};

\\
};

\end{tikzpicture}
\caption{\label{RV-move}The RV move for rotating a vertex around one of its incident edges, based on its current vertex state. The edges considered here are unframed; the RV move for framed graphs will be defined later, in Section \ref{frame-Reid}.}
\end{figure}
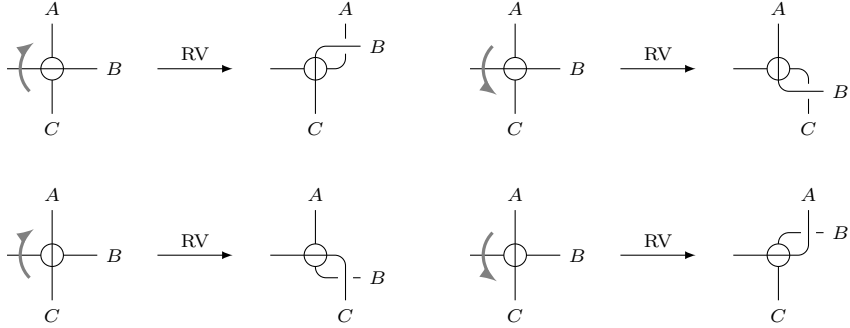

Finally, there are other graph moves between isomorphic graphs. Three of these result from the same moves arising in knot theory, where two knot diagrams of the same knot can be related by a finite sequence of such moves. The last two show how to deal with the presence of vertices; we will refer to the set of all five graph moves as (generalized) Reidemeister moves. Some possibilities for the first four such moves are shown in Figure \ref{Reid-move}, while all variations of the RV move are shown in Figure \ref{RV-move}. Note that these are given for unframed edges. When the edges are framed, the RI and RV moves will be different. In particular, the RV move will create twists on the incident edges, which depend on which edge the vertex is rotated around, and the direction of the rotation. This will be discussed in more detail in Section \ref{frame-Reid}.

\subsection{Inconsistency of the Pachner moves}

\begin{figure}[hbt]
\begin{tikzpicture}[> = latex]
\matrix[column sep = 0.5 cm, row sep = 1 cm]{


	\draw (-1, 0) circle (0.15);
	\draw [fill = white] (0, 0) circle (0.15);


	\draw [rounded corners] (-0.85, 0) -- (0.35, 0) (0.65, 0) -- (1, 0) -- (1, -0.75) -- (-1, -0.75) -- (-1, 0.5) -- (0, 0.5) -- (0, 0.15)
		(0, -0.15) -- (0, -0.5) -- (0.5, -0.5) -- (0.5, 0.75) -- (-1.5, 0.75) -- (-1.5, 0) -- (-1.15, 0);
	

	\draw [->, very thick, gray] (-0.6, -0.3) arc (225 : 125 : {0.3 * sqrt(2)});

&

	\draw [->] (-0.5, 0) -- node [above] {RV move} (0.5, 0);
	
&

	
	\draw [rounded corners] (-1.65, -0.5) -- (-1.75, -0.5) -- (-1.75, 0.75) -- (0.5, 0.75) -- (0.5, -0.5) -- (0, -0.5) -- (0, 0.5) --
		(-1, 0.5) -- (-1, -0.5) -- (-1.35, -0.5);
		
	
	\draw [fill = white] (-1, 0) circle (0.15);
	\draw [fill = white] (0, 0) circle (0.15);
	
	
	\draw [rounded corners] (0.35, 0) -- (-1.5, 0) -- (-1.5, -0.75) -- (1, -0.75) -- (1, 0) -- (0.65, 0);

	
	\draw [dashed] (-1.25, -0.25) rectangle (0.25, 0.25);
		
&

	\draw [->] (-0.5, 0) -- node [above] {2-3 move} (0.5, 0);
	
&

	\draw (0, 0.5) circle (0.15);
	\draw (0.75, 0) circle (0.15);
	\draw (0, -0.5) circle (0.15);
	
	\draw [rounded corners] (-0.15, 0.5) -- (-0.5, 0.5) -- (-0.5, 1.25) -- (1.25, 1.25) -- (1.25, 0.5) -- (0.15, 0.5)
		(0, 0.5) -- (0, 1) -- (0.75, 1) -- (0.75, 0.65) (0.75, 0.35) -- (0.75, 0.15) (0.75, -0.15) -- (0.75, -0.35) (0.75, -0.65) -- (0.75, -1) -- (0, -1)
			-- (0, 0.5)
		(-0.35, -0.5) -- (-0.15, -0.5) (0.15, -0.5) -- (1.5, -0.5) -- (1.5, 1.5) -- (-1, 1.5) -- (-1, -0.5) -- (-0.65, -0.5) (-0.35, -0.5) -- (-0.15, -0.5)
		(0.15, 0) -- (1.35, 0) (1.65, 0) -- (1.75, 0) -- (1.75, -1.25) -- (-0.5, -1.25) -- (-0.5, 0) -- (-0.15, 0);

\\


	\draw (-1, 0) circle (0.15);
	\draw [fill = white] (0, 0) circle (0.15);


	\draw [rounded corners] (-0.85, 0) -- (0.35, 0) (0.65, 0) -- (1, 0) -- (1, -0.75) -- (-1, -0.75) -- (-1, 0.5) -- (0, 0.5) -- (0, 0.15)
		(0, -0.15) -- (0, -0.5) -- (0.5, -0.5) -- (0.5, 0.75) -- (-1.5, 0.75) -- (-1.5, 0) -- (-1.15, 0);
	

	\draw [->, very thick, gray] (-0.6	, 0.3) arc (135 : 235 : {0.3 * sqrt(2)});

&

	\draw [->] (-0.5, 0) -- node [above] {RV move} (0.5, 0);
	
&

	
	\draw [rounded corners] (-1, 0.35) -- (-1, -0.75) -- (1, -0.75) -- (1, 0) -- (0.65, 0)
		(-1, 0.65) -- (-1, 0.75) -- (0.5, 0.75) -- (0.5, -0.5) -- (0, -0.5) -- (0, -0.15);
	
	
	\draw [fill = white] (-1, 0) circle (0.15);
	\draw [fill = white] (0, 0) circle (0.15);
	
	
	\draw [rounded corners] (0.35, 0) -- (-1.5, 0) -- (-1.5, 0.5) -- (0, 0.5) -- (0, 0.15);

	
	\draw [dashed] (-1.25, -0.25) rectangle (0.25, 0.25);
	
&

	\draw [->] (-0.5, 0) -- node [above] {2-3 move} (0.5, 0);
	
&

	\draw (0, 0.5) circle (0.15);
	\draw (0.75, 0) circle (0.15);
	\draw (0, -0.5) circle (0.15);
	
	\draw [rounded corners] (0, 0.5) -- (0, 1) -- (0.75, 1) -- (0.75, 0.65) (0.75, 0.35) -- (0.75, 0.15) (0.75, -0.15) -- (0.75, -0.35) (0.75, -0.65) --
			(0.75, -1) -- (0, -1) -- (0, 0.5)
		(0.15, 0) -- (1.6, 0) (1.9, 0) -- (2, 0) -- (2, -1.25) -- (-0.5, -1.25) -- (-0.5, -0.5) -- (-0.15, -0.5) (0.15, -0.5) -- (1.75, -0.5)
			-- (1.75, 1.5) -- (-0.5, 1.5) -- (-0.5, 1.4) (-0.5, 1.1) -- (-0.5, 0.5) -- (-0.15, 0.5) (0.15, 0.5) -- (1.25, 0.5) -- (1.25, 1.25) -- (-1, 1.25)
			-- (-1, 0) -- (-0.15, 0);

\\
};
\end{tikzpicture}
\caption{\label{bad-pach-end}The inconsistency of using the 2-3 move on the graph $3^\ell 5^\ell 1^-$. Using the RV move on the left-hand vertex in opposite directions, then acting with the Pachner move, gives two distinct graphs as a result.}
\end{figure}
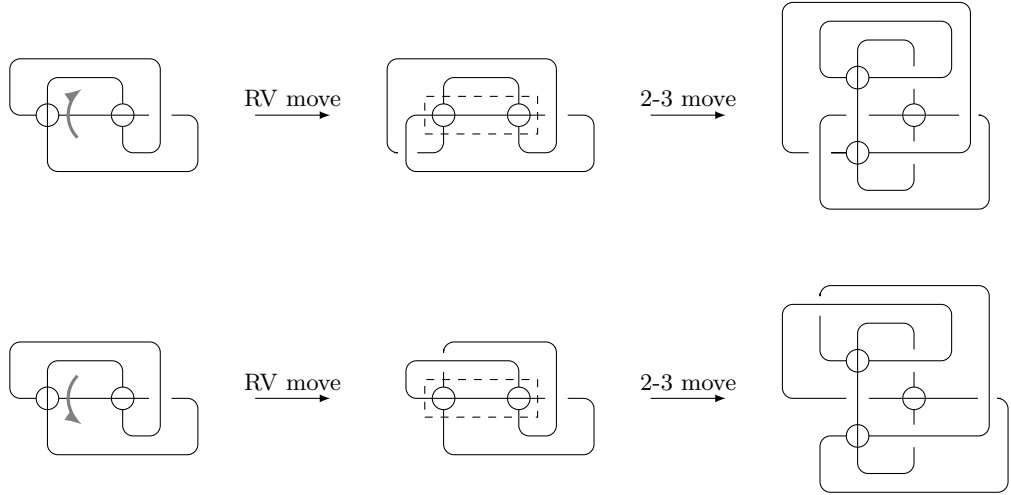

We want to ensure that the Pachner moves are applied consistently on all graphs. However, as defined up to this point, these graph moves give different graphs, depending on how the move is set up and carried out. As an example of what can go wrong, consider the graph diagram for the sequence $3^\ell 5^\ell 1^-$ shown in Figure \ref{bad-pach-start}. In the diagram shown there, the 2-3 move cannot be used on the edge inside the dashed box, since the two vertices the edge is incident to have opposite vertex states. However, this can be changed by using the RV graph move on one of the two vertices, to place both in the same state. Figure \ref{bad-pach-end} shows two variations of a RV move acting on the left-hand vertex, with opposite directions for the vertex rotation. Both of these change the state of that vertex to one where the 2-3 move can now be used on the edge inside the dashed box. Note, however, that by using the RV move in opposite directions, the final graphs after the Pachner move are different -- they do not have the same adjacency matrix, for example.  Another perspective on this -- which does not require using any graph moves (such as the RV move) -- is to say that the same abstract graph is represented in the middle pictures of Figure \ref{bad-pach-end} by two choices of the projective plane used to show the graph. Thus, the result of a 2-3 Pachner move on this graph depends on the projection plane used.

\begin{figure}[hbt]
\begin{tikzpicture}


	\draw (-1, 0) circle (0.15);
	\draw [fill = white] (0, 0) circle (0.15);


	\draw [rounded corners] (-0.85, 0) -- (0.35, 0) (0.65, 0) -- (1, 0) -- (1, -0.75) -- (-1, -0.75) -- (-1, 0.5) -- (0, 0.5) -- (0, 0.15)
		(0, -0.15) -- (0, -0.5) -- (0.5, -0.5) -- (0.5, 0.75) -- (-1.5, 0.75) -- (-1.5, 0) -- (-1.15, 0);
	
	
	\draw [dashed] (-1.25, -0.25) rectangle (0.25, 0.25);
	
\end{tikzpicture}
\caption{\label{bad-pach-start}The graph $3^\ell 5^\ell 1^-$, shown in a non-standard form so that the two vertices appear next to each other in the projection. We consider the 2-3 move acting on the vertices and common edge surrounded by the dashed line.}
\end{figure}
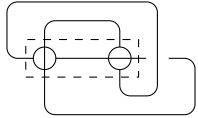

Obviously, this situation is not tenable: we expect that the Pachner moves, applied to the same subgraph, will have results independent of the graph diagram used -- recall that the Pachner moves on triangulation preserve the manifold under changes in the given discretization. To understand this situation, it is helpful to go back to generalized triangulations, and the Pachner moves defined on the dual tetrahedra. When constructing such a triangulation out of tetrahedral simplices, how the tetrahedra are glued together must be specified. In other words, when gluing together the faces of tetrahedra, one must know what the edges on one face correspond to those on the other. This information is not explicitly given when we turn to knotted graphs. The trouble comes when it is assumed that for adjacent vertices in the graph, there is a unique choice of gluing for the tetrahedra dual to these vertices. Recall the Pachner moves shown in Figure \ref{Pach-tet} for the dual triangulation. Focusing on the 2-3 move, it is crucial that the two original tetrahedra have faces consistently glued together; this is what can go wrong when looking at the graph picture dual to these simplices. In the graph picture, this situation would be two vertices sharing an incident edge. At this point, there is an implicit choice for the dual gluing of tetrahedra; the origin of the issue shown in Figure \ref{bad-pach-end} is that this choice is not consistent.

\begin{figure}[hbt]
\begin{tikzpicture}[> = latex]
\matrix[column sep = 1 cm, row sep = 0.5 cm]{

	\draw (-1.25, 0) node [left] {$B$} -- node [above] {$D$} (1.25, 0) node [right] {$B'$};

	\draw [fill = white] (-0.75, 0) circle (0.15);
	\draw [fill = white] (0.75, 0) circle (0.15);
	
	\draw (-0.75, 0.5) node [above] {$A$} -- (-0.75, -0.5) node [below] {$C$};
	\draw (0.75, 0.5) node [above] {$A'$} -- (0.75, -0.5) node [below] {$C'$};
	
&
	\draw [<->] (-0.5, 0) -- (0.5, 0);
&


	\draw [fill = gray!30] (-1.5, 1) node [above] {$c$} -- (-0.5, 0) node [right] {$b$} --
		(-1.5, -1) node [below] {$a$} -- (-2.5, 0) node [left] {$d$} -- cycle
		(-2.5, 0) -- (-0.5, 0);
	\draw [dashed] (-1.5, 1) -- (-1.5, -1);

	\draw [fill = gray!30] (1.5, 1) node [above] {$c'$} -- (2.5, 0) node [right] {$d'$} --
		(1.5, -1) node [below] {$a'$} -- (0.5, 0) node [left] {$b'$} -- cycle
		(0.5, 0) -- (2.5, 0);
	\draw [dashed] (1.5, 1) -- (1.5, -1);

\\
};
\end{tikzpicture}
\caption{\label{good-glue}Two vertices in the $\oplus$ state sharing an edge, and the corresponding dual tetrahedra. Unprimed vertices of the left dual tetrahedron are glued to their primed counterparts on the right-hand tetrahedron.}
\end{figure}
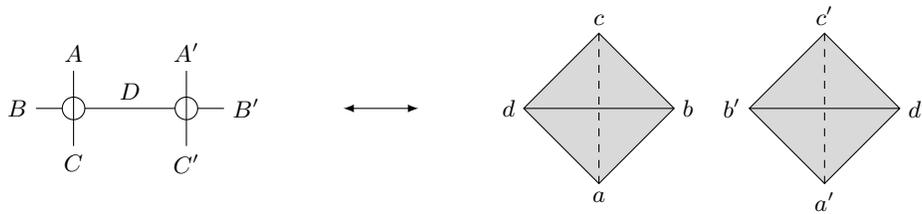

\begin{figure}[hbt]
\begin{tikzpicture}[> = latex]
\matrix[column sep = 1 cm, row sep = 0.5 cm]{

	\draw (-1.25, 0) node [left] {$B$} -- (0, 0) node [above] {$D$};
	\draw (0.75, 0.5) node [above] {$A'$} -- (0.75, -0.5) node [below] {$C'$};

	\draw [fill = white] (-0.75, 0) circle (0.15);
	\draw [fill = white] (0.75, 0) circle (0.15);
	
	\draw (-0.75, 0.5) node [above] {$A$} -- (-0.75, -0.5) node [below] {$C$};
	
	\draw (0, 0) -- (1.25, 0) node [right] {$B'$};
	
&
	\draw [<->] (-0.5, 0) -- (0.5, 0);
&


	\draw [fill = gray!30] (-1.5, 1) node [above] {$c$} -- (-0.5, 0) node [right] {$b$} --
		(-1.5, -1) node [below] {$a$} -- (-2.5, 0) node [left] {$d$} -- cycle
		(-2.5, 0) -- (-0.5, 0);
	\draw [dashed] (-1.5, 1) -- (-1.5, -1);

	\draw [fill = gray!30] (1.5, 1) node [above] {$c'$} -- (2.5, 0) node [right] {$d'$} --
		(1.5, -1) node [below] {$a'$} -- (0.5, 0) node [left] {$b'$} -- cycle
		(1.5, 1) -- (1.5, -1);
	\draw [dashed] (0.5, 0) -- (2.5, 0);
\\
};
\end{tikzpicture}
\caption{\label{bad-glue}Two vertices in unequal states sharing an edge $D$. Since the dual faces $abc$ and $a'b'c'$ have different alignments relative to the projection plane, there is no canonical choice for a gluing between the two faces.}
\end{figure}
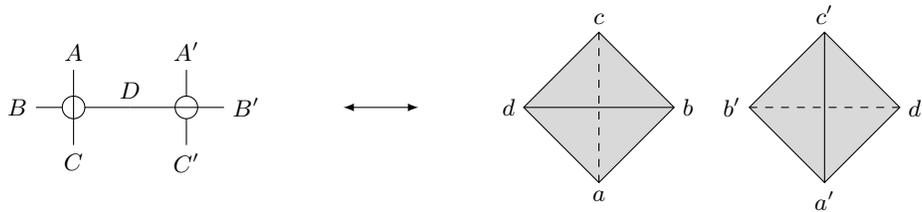

To see this, let us consider the possible cases for the two vertex states. In Figure \ref{good-glue}, two vertices in the $\oplus$ state share an edge are shown on the left-hand side of the picture; the matching dual tetrahedra are shown on the right side. The external edges of the subgraph have their own label, as do the vertices of the dual tetrahedra. As before, the edge $A$ on the left is dual to the triangular face $bcd$ on the right, i.e. the face without the same (lower-case) letter. The common edge $D$ between the two vertices in the graph then is dual to the faces $abc$ and $a'b'c'$. For the projection shown, the vertices $b$ and $b'$ point towards the viewer, while the edges $ac$ and $a'c'$ are located away from the viewer; both triangles $abc$ and $a'b'c'$ have the same orientation. In this picture, there is a natural gluing between these two faces, namely gluing the unprimed vertex label with its matching primed label, so that $a \leftrightarrow a'$, etc. Now we consider the situation where the two vertex states are not equal (as we saw in Figure \ref{bad-pach-end}). This case is shown in Figure \ref{bad-glue}. Unlike the previous case, there is no natural gluing between the two faces $abc$ and $a'b'c'$. This is because the faces $abc$ and $a'b'c'$ do not have the same orientation -- the face $abc$ has the vertex $b$ closest to the viewer, while the triangle $a'b'c'$ has the edge $a'c'$ closest. Note that we can now see what the RV move was accomplishing in the 2-3 move shown in Figure \ref{bad-pach-end}: in the dual picture, it serves to rotate the left-hand dual tetrahedra so that the common faces are now aligned. Either of the two ways of doing this can align the two dual faces, but they give different triangulations, and thus different graphs.


\section{Framed graphs}
\label{frame}

\subsection{Adding twists to edges}

In the discussion of the last section, we have seen that the gluing information needed for obtaining a consistent dual triangulation is currently missing for the knotted graph. This requires giving extra structure to the graph edges, indicating how the tetrahedra dual to the graph vertices are glued together. We do this by introducing {\it framed} edges for the graph~\cite{Mar-Pre-Sch08, Smo-Wan08, Wan07} (see also Burton and Pettersson~\cite{Bur-Pet14} in the context of combinatorial topology). Recall that a two-dimensional vector basis can be defined in the triangular cross-section of a framed edge. These framed edges can be twisted as they pass from one vertex to another; this twist is a rotation of the two non-tangent vectors in their common plane. For our purposes, these twists occur in units of $\pi/3$ radians, moving between the possible orientations for the faces of the tetrahedra dual to the graph vertices. Thus, from this point on, we will give the twist in multiples of $\pi/3$, i.e. a twist of $+1$ is that of $+\pi/3$ radians, while $-2$ is for $-2\pi/3$ radians. Positive and negative twists are defined using the right-hand rule: pointing the thumb of the right-hand towards the vertex, the fingers will curl in the direction of a positive twist. Note that this is independent of which of the two vertices incident to the edge are used -- positive twists will be the same for either choice. We will comment further on this definitional choice in Section \ref{3-line}.

Now we can interpret why certain situations gave consistency between the knotted graph and the dual tetrahedra. Up to this point, it has been tacitly assumed that all edges have a twist of zero. With the like vertex states in Figure \ref{good-glue}, for example, assuming a zero twist on the common edge $D$ led to the gluing of the dual faces $abc$ and $a'b'c'$. No twist is necessary, though, since the faces have the same orientation. However, this does not work out for the case of unlike vertex states in Figure \ref{bad-glue}. Here, a non-zero twist is needed to pass from the face $abc$ (with the vertex $b$ pointing towards the viewer) to the face $a'b'c'$ (with the edge $a'c'$ closest to the viewer). In this projection, the two faces are misaligned by $\pi/3$ radians, and the unframed edge gives no clue for how to perform the identification of dual vertices. Adding the appropriate twist to the edge resolves this. For example, placing a twist of $+1$ on the graph edge would then would identify the dual vertices $a \leftrightarrow b', b \leftrightarrow a', c \leftrightarrow c'$, while a $-1$ twist corresponds to the mapping $a \leftrightarrow a', b \leftrightarrow c', c \leftrightarrow b'$. More generally, we can correctly align dual faces if the graph edge has an even twist for like vertex states, and an odd twist for unlike vertex states.

\subsection{3-line graphs}
\label{3-line}

In order to show how to consistently apply the Pachner moves on graphs from the dual formulation on triangulations, Markopoulou and Pr\'emont-Schwarz~\cite{Mar-Pre-Sch08} introduced the concept of a 3-line graph. Consider a graph diagram for a  framed graph $G$. Then, we can create a {\it 3-line graph} for $G$ by replacing each framed edge by a leg of three lines. These lines will represent the gluing of the dual edges in any potential triangulation associated with the graph. Thus, each line in the 3-line graph can be labeled with the corresponding dual edge in the triangulation. Formulating the Pachner moves on the 3-line graphs allows us to work backwards, and determine the required twists on the graph edges.

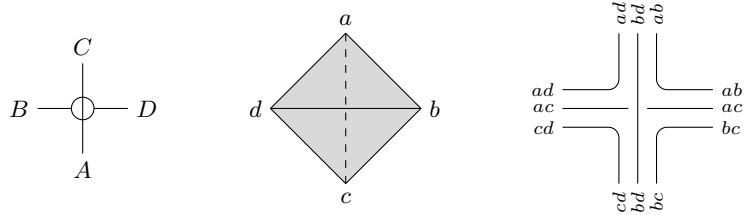
\begin{figure}[hbt]
\begin{tikzpicture}
\matrix[column sep = 1 cm]{


	\draw (-0.6, 0) node [left] {$B$} -- (0.6, 0) node [right] {$D$};
	\draw [fill = white] (0, 0) circle (0.15);
	\draw (0, -0.6) node [below] {$A$} -- (0, 0.6) node [above] {$C$};
	
	&


	\draw [fill = gray!30] (0, 1) node [above] {$a$} -- (1, 0) node [right] {$b$} --
		(0, -1) node [below] {$c$} -- (-1, 0) node [left] {$d$} -- cycle
		(-1, 0) -- (1, 0);
	\draw [dashed] (0, 1) -- (0, -1);

	&
	
	
	\begin{scope}[font = \scriptsize]

		\draw (0, -1) node [rotate = 90, left] {$bd$} -- (0, 1) node [rotate = 90, right] {$bd$}
			(-1, 0) node [left] {$ac$} -- (-0.125, 0) (0.125, 0) -- (1, 0) node [right] {$ac$};
		\draw [rounded corners] (-1, 0.25) node [left] {$ad$} -- (-0.25, 0.25) -- (-0.25, 1) node [rotate = 90, right] {$ad$}
			(-1, -0.25) node [left] {$cd$} -- (-0.25, -0.25) -- (-0.25, -1) node [rotate = 90, left] {$cd$}
			(1, 0.25) node [right] {$ab$} -- (0.25, 0.25) -- (0.25, 1) node [rotate = 90, right] {$ab$}
			(1, -0.25) node [right] {$bc$} -- (0.25, -0.25) -- (0.25, -1) node [rotate = 90, left] {$bc$};
		
	\end{scope}
\\
};

\end{tikzpicture}
\caption{\label{3-line-vert}Correspondence between a vertex in the $\oplus$ state, its dual tetrahedron, and the 3-line graph for the vertex.}
\end{figure} 

Consider the situation shown in Figure \ref{3-line-vert}. The three lines matching the incident graph edge $A$ in the 3-line graph give information about the dual edges $bc, bd,$ and $cd$ for the face $bcd$; there are similar matchings between the other incident edges and their corresponding lines. Looking at the tetrahedron, each edge appears in two different faces, so the 3-line graph matches the lines for these dual edges. For example, the line matching the edge $ad$ appears in the groups for the graph edges for $B$ and $C$, and so are connected together -- it is glued to one face as a part of the face $acd$, and another (not necessarily distinct) face as part of the face $abd$. For the lines of graph edges on opposite sides of the vertex, the vertex state determines the type of crossing. Thus, the line corresponding to $bd$ passes over that for $ac$, since the original vertex is in the $\oplus$ state. Looking at the dual tetrahedron, this is because the edge $bd$ faces the viewer, above the projection plane, while the edge $ac$ is below the plane, farthest from the viewer.

\begin{figure}[hbt]
\begin{tikzpicture}
\matrix[column sep = 2 cm]{


	\draw [fill = gray!30] (-1, 0) node [above left] {$a$}-- (1, 0) node [above right] {$b$} -- (0, {-sqrt(3)}) node [below] {$c$} -- cycle;
	

	\draw [fill = gray!30] (-1, {-sqrt(3) + 3}) node [below left] {$c'$} -- (0, 3) node [above] {$a'$} -- (1, {-sqrt(3) + 3}) node [below right] {$b'$} -- cycle;

	
	\draw [dashed] (-0.5, {-0.5 * sqrt(3)}) -- (-0.5, 0);
	\draw [dashed] (0.5, {-0.5 * sqrt(3)}) -- (0.5, 0);
	
	\draw (-0.5, 0) -- (-0.5, {-0.5 * sqrt(3) + 3});
	\draw (0.5, 0) -- (0.5, 0.25);
	\draw [rotate around = {90 : (0.5, 0.25)}] (0.5, 0.25) cos (1, 0.5) sin (1.5, 0.75);
	
	\draw (0, 0) -- (0, 0.25) (0.5, 1.25) -- (0.5, {-0.5 * sqrt(3) + 3});
	\draw [draw = white, double = black, double distance between line centers = 3 pt, line width = 2.6 pt, rotate around = {90 : (0, 0.25)}]
		(0, 0.25) cos (0.5, 0) sin (1, -0.25);
		
	
	\begin{scope}[->, > = latex, very thick]
	
		\draw (2, {-0.5 * sqrt(3) + 1}) -- (2, {-0.5 * sqrt(3) + 2});
		\draw (-2, {-0.5 * sqrt(3) + 2}) -- (-2, {-0.5 * sqrt(3) + 1});
	
	\end{scope}
	
	\begin{scope}[->, gray, very thick]
	
		\draw  (1.75, {-0.5 * sqrt(3) + 1.5}) arc (180 : 350 : 0.25);
		\draw (-1.75, {-0.5 * sqrt(3) + 1.5}) arc (0 : 170 : 0.25);
	
	\end{scope}
	
	
	\node at (0, -2.5) {$+1$ twist};

&


	\draw [fill = gray!30] (-1, 0) node [above left] {$a$}-- (1, 0) node [above right] {$b$} -- (0, {-sqrt(3)}) node [below] {$c$} -- cycle;
	

	\draw [fill = gray!30] (-1, {-sqrt(3) + 3}) node [below left] {$a'$} -- (0, 3) node [above] {$b'$} -- (1, {-sqrt(3) + 3}) node [below right] {$c'$} -- cycle;

	
	\draw [dashed] (-0.5, {-0.5 * sqrt(3)}) -- (-0.5, 0);
	\draw [dashed] (0.5, {-0.5 * sqrt(3)}) -- (0.5, 0);
	
	\draw (0.5, 0) -- (0.5, {-0.5 * sqrt(3) + 3});
	\draw (-0.5, 0) -- (-0.5, 0.25);
	\draw [rotate around = {90 : (-0.5, 0.25)}] (-0.5, 0.25) cos (0, 0) sin (0.5, -0.25);
	
	\draw (0, 0) -- (0, 0.25) (-0.5, 1.25) -- (-0.5, {-0.5 * sqrt(3) + 3});
	\draw [draw = white, double = black, double distance between line centers = 3 pt, line width = 2.6 pt, rotate around = {90 : (0, 0.25)}]
		(0, 0.25) cos (0.5, 0.5) sin (1, 0.75);
		
	
	\begin{scope}[->, > = latex, very thick]
	
		\draw (2, {-0.5 * sqrt(3) + 1}) -- (2, {-0.5 * sqrt(3) + 2});
		\draw (-2, {-0.5 * sqrt(3) + 2}) -- (-2, {-0.5 * sqrt(3) + 1});
	
	\end{scope}
	
	\begin{scope}[->, gray, very thick]
	
		\draw (2.25, {-0.5 * sqrt(3) + 1.5}) arc (360 : 170 : 0.25);
		\draw (-2.25, {-0.5 * sqrt(3) + 1.5}) arc (180 : 10 : 0.25);
	
	\end{scope}
	
	
	\node at (0, -2.5) {$-1$ twist};

\\
};

\end{tikzpicture}
\caption{\label{line-twist-map}How face rotations of the dual tetrahedron lead to braids in the legs of a 3-line graph. In each diagram, the legs extend between the far side of the face $abc$ to the near side of the face $a'b'c'$.}
\end{figure}
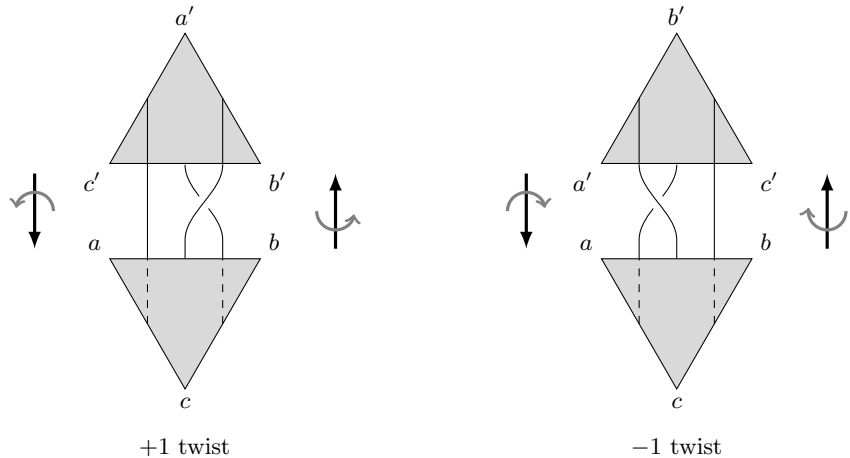

Once we have the correspondence between the original graph $G$ and its 3-line graph, we can discuss how the edge twists affect the lines of each leg in the 3-line graph. Because the twist will change the ordering of the lines, twists will give rise to braids in the lines of each leg. First, recall that the edge twist is defined using the right-hand rule around a vector pointing towards either vertex incident to the edge. Thus, as we pass from one vertex to another, the framed edge's triangular cross-section will rotate based on the given edge twist. To figure out how this applies to the legs of the 3-line graph, imagine that we move a copy of the dual face along the graph edge, and turn it based on the edge twist. This is what is shown in Figure \ref{line-twist-map}, which shows how the lines of a 3-line graph rotate as one passes along from one vertex to another.

Each part of Figure \ref{line-twist-map} shows the dual faces $abc$ and $a'b'c'$ to be glued together, and the lines for $ab$, $ac$, and $bc$. In both cases, the unprimed vertices are to be identified with their primed counterpart. The lines are drawn from the far side of the dual face $abc$ to the near side of the $a'b'c'$ face. Thus, the leg of the 3-line graph are drawn superimposed on the dual faces. For the $+1$ twist on the left, imagine that we move the bottom dual face $abc$ upward along the edge until it meets the face $a'b'c'$. Notice that we are looking from ``inside'' the dual tetrahedron with face $abc$ outward toward the face $a'b'c$; the projection plane is chosen so that edge $ab$ is above the plane, and edge $b'c'$ is below. The $+1$ twist will be a counterclockwise rotation along the edge. As face $abc$ moves upward towards face $a'b'c'$ and rotates $\pi/3$ radians in the counterclockwise direction, edge $ab$ will pass over edge $bc$ (as seen above the projection plane), so the corresponding lines are braided to reflect this, with the $ab$ line crossing over the $bc$ line. Line $ac$ is unaffected by the twist, and is not part of the braid. This gives the situation for a $+1$ twist, passing from a dual face with a single vertex $c$ below the projection plane to a face with a single vertex $a'$ above. When the opposite situation is required, then the face $abc$ rotates clockwise as it moves towards the face $a'b'c'$; the $-1$ twist is shown on the right-hand part of the figure.

One potentially confusing aspect to this definition of the twists is that the faces are rotated around the {\it outward} facing normal to the face, with positive being the counterclockwise direction. This appears to contradict the previously stated choice of the twist as using the {\it inward} facing normal to the vertex. However, as spelled out above, the picture is of transporting the dual face along the edge towards the other face, and rotating it as it goes along. This gives the $+1$ twist as a right-handed rotation of the face as it approaches the second face, so that this rotation is given by the vector pointing towards the second vertex. Thus, our definition of the twist is consistent with how the lines are braided in Figure \ref{line-twist-map}.

\begin{figure}[hbt]
\begin{tikzpicture}
	

	\draw [fill = gray!30] (-1, {-sqrt(3) - 3}) node [below left] {$c'$} -- (0, -3) node [left] {$a'$} -- (1, {-sqrt(3) - 3}) node [below right] {$b'$} -- cycle;


	\draw [fill = gray!30] (-1, 0) node [above left] {$a$}-- (1, 0) node [above right] {$b$} -- (0, {-sqrt(3)}) node [left] {$c$} -- cycle;
	

	\draw [fill = gray!30] (-1, {-sqrt(3) + 3}) node [below left] {$c'$} -- (0, 3) node [above] {$a'$} -- (1, {-sqrt(3) + 3}) node [below right] {$b'$} -- cycle;

	
	\draw [dashed] (-0.5, {-0.5 * sqrt(3)}) -- (-0.5, 0);
	\draw [dashed] (0.5, {-0.5 * sqrt(3)}) -- (0.5, 0);
	\draw [dashed] (0, -3) -- (0, {-sqrt(3) - 3});
	
	\draw (-0.5, 0) -- (-0.5, {-0.5 * sqrt(3) + 3});
	\draw (0.5, 0) -- (0.5, 0.25);
	\draw [rotate around = {90 : (0.5, 0.25)}] (0.5, 0.25) cos (1, 0.5) sin (1.5, 0.75);
	
	\draw (0, 0) -- (0, 0.25) (0.5, 1.25) -- (0.5, {-0.5 * sqrt(3) + 3});
	\draw [draw = white, double = black, double distance between line centers = 3 pt, line width = 2.6 pt, rotate around = {90 : (0, 0.25)}]
		(0, 0.25) cos (0.5, 0) sin (1, -0.25);
	
	\draw (-0.5, {-0.5 * sqrt(3)}) -- (-0.5, {-0.5 * sqrt(3) - 3});
	\draw (0, 0) -- (0, -1.75) (0, -2.75) -- (0, -3);
	\draw (0.5, {-0.5 * sqrt(3)}) -- (0.5, -1.75) (0.5, -2.75) -- (0.5, {-0.5 * sqrt(3) - 3});
	
	\draw [rotate around = {-90 : (0.5, -1.75)}] (0.5, -1.75) cos (1, -2) sin (1.5, -2.25);
	\draw [draw = white, double = black, double distance between line centers = 3 pt, line width = 2.6 pt, rotate around = {-90 : (0, -1.75)}]
		(0, -1.75) cos (0.5, -1.5) sin (1, -1.25);
		
	
	\node at (2.5, {-0.5 * sqrt(3) + 1.5}) {$+1$ twist};
	\node at (2.5, {-0.5 * sqrt(3) - 1.5}) {$-1$ twist};
		
\end{tikzpicture}
\caption{\label{line-twist-inv}Using the line twists shown in Figure \ref{line-twist-map} to demonstrate the $+1$ and $-1$ twists are inverses.}
\end{figure}
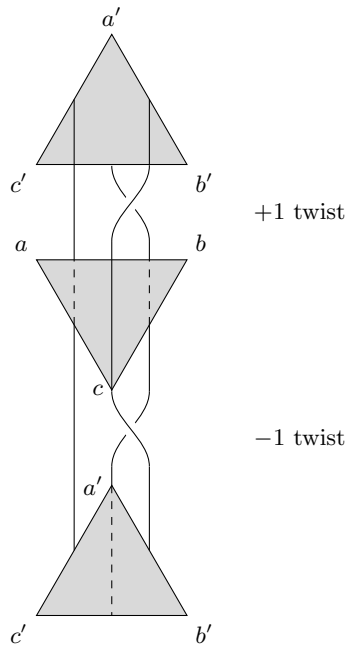

As one would expect, the $-1$ twist reverses the braiding of the $+1$ twist. One way to see this is to compose the two twists in the following way. Either twist diagram can be rotated around so that we are looking from the ``inside'' of the tetrahedron with the face $a'b'c'$ towards the face $abc$. Then, the two $abc$ faces can be combined together, and the twists combined into one; note that this is the same as composing braids in the braid group $B_3$ for three strands. An example of this is given in Figure \ref{line-twist-inv}. Here, the $+1$ twist is copied exactly as shown in the original definition of Figure \ref{line-twist-map}. On the other hand, the $-1$ twist is taken by starting with the $abc$ face, then moving down towards the $a'b'c'$ face -- the opposite orientation from that shown in Figure \ref{line-twist-map}. However, the braiding is exactly the same, with the line associated to the $ab$ dual edge passing over the line for the $ac$ dual edge. By following the lines throughout the entire illustration -- ignoring the superimposed dual faces -- it is easy to see it is equivalent to the three lines having no crossings at all.

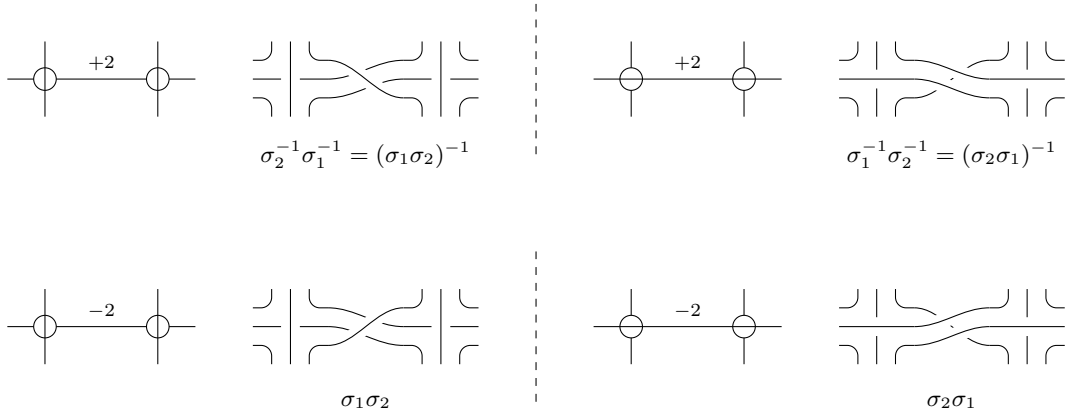
\begin{figure}[hbt]
\begin{tikzpicture}
\matrix[column sep = 0.75 cm, row sep = 1 cm]{


	\draw (-1.25, 0) -- node [above, font = \scriptsize] {$+2$} (1.25, 0);

	\draw [fill = white] (-0.75, 0) circle (0.15);
	\draw [fill = white] (0.75, 0) circle (0.15);
	
	\draw (-0.75, 0.5) -- (-0.75, -0.5);
	\draw (0.75, 0.5) -- (0.75, -0.5);
	
	&
	
	\draw (-1.5, 0) -- (-1.125, 0) (-0.875, 0) -- (-0.5, 0) (0.5, 0) -- (0.875, 0) (1.125, 0) -- (1.5, 0);
	
	\draw (-1, 0.5) -- (-1, -0.5);
	\draw (1, 0.5) -- (1, -0.5);
	
	\begin{scope}[rounded corners]
	
		\draw (-1.5, 0.25) -- (-1.25, 0.25) -- (-1.25, 0.5);
		\draw (-1.5, -0.25) -- (-1.25, -0.25) -- (-1.25, -0.5);
		
		\draw (-0.75, 0.5) -- (-0.75, 0.25) -- (-0.5, 0.25) (0.5, 0.25) -- (0.75, 0.25) -- (0.75, 0.5);
		\draw (-0.75, -0.5) -- (-0.75, -0.25) -- (-0.5, -0.25) (0.5, -0.25) -- (0.75, -0.25) -- (0.75, -0.5);
	
		\draw (1.5, 0.25) -- (1.25, 0.25) -- (1.25, 0.5);
		\draw (1.5, -0.25) -- (1.25, -0.25) -- (1.25, -0.5);
	
	\end{scope}
	
	\draw (-0.5, 0) cos (0, 0.125) sin (0.5, 0.25);
	\draw (-0.5, -0.25) cos (0, -0.125) sin (0.5, 0);
	\draw [draw = white, double = black, double distance between line centers = 3 pt, line width = 2.6 pt] (-0.5, 0.25) cos (0, 0) sin (0.5, -0.25);
	
	\node at (0, -1) {$\sigma_2 ^{-1} \sigma_1 ^{-1} = (\sigma_1 \sigma_2)^{-1}$};

	
	&
	
	\draw [dashed] (0, -1) -- (0, 1);
	
	&

	
	\draw (-0.75, 0.5) -- (-0.75, -0.5);
	\draw (0.75, 0.5) -- (0.75, -0.5);

	\draw [fill = white] (-0.75, 0) circle (0.15);
	\draw [fill = white] (0.75, 0) circle (0.15);

	\draw (-1.25, 0) -- node [above, font = \scriptsize] {$+2$} (1.25, 0);
	
	&
	
	\draw (-1.5, 0) -- (-0.5, 0) (0.5, 0) -- (1.5, 0);
	
	\draw (-1, 0.5) -- (-1, 0.125) (-1, -0.125) -- (-1, -0.5);
	\draw (1, 0.5) -- (1, 0.125) (1, -0.125) -- (1, -0.5);
	
	\begin{scope}[rounded corners]
	
		\draw (-1.5, 0.25) -- (-1.25, 0.25) -- (-1.25, 0.5);
		\draw (-1.5, -0.25) -- (-1.25, -0.25) -- (-1.25, -0.5);
		
		\draw (-0.75, 0.5) -- (-0.75, 0.25) -- (-0.5, 0.25) (0.5, 0.25) -- (0.75, 0.25) -- (0.75, 0.5);
		\draw (-0.75, -0.5) -- (-0.75, -0.25) -- (-0.5, -0.25) (0.5, -0.25) -- (0.75, -0.25) -- (0.75, -0.5);
	
		\draw (1.5, 0.25) -- (1.25, 0.25) -- (1.25, 0.5);
		\draw (1.5, -0.25) -- (1.25, -0.25) -- (1.25, -0.5);
	
	\end{scope}
	
	\draw (-0.5, -0.25) cos (0, 0) sin (0.5, 0.25);
	\draw [draw = white, double = black, double distance between line centers = 3 pt, line width = 2.6 pt] (-0.5, 0) cos (0, -0.125) sin (0.5, -0.25);
	\draw [draw = white, double = black, double distance between line centers = 3 pt, line width = 2.6 pt] (-0.5, 0.25) cos (0, 0.125) sin (0.5, 0);
	
	\node at (0, -1) {$\sigma_1 ^{-1} \sigma_2 ^{-1} = (\sigma_2 \sigma_1)^{-1}$};

	
\\


	\draw (-1.25, 0) -- node [above, font = \scriptsize] {$-2$} (1.25, 0);

	\draw [fill = white] (-0.75, 0) circle (0.15);
	\draw [fill = white] (0.75, 0) circle (0.15);
	
	\draw (-0.75, 0.5) -- (-0.75, -0.5);
	\draw (0.75, 0.5) -- (0.75, -0.5);
	
	&
	
	\draw (-1.5, 0) -- (-1.125, 0) (-0.875, 0) -- (-0.5, 0) (0.5, 0) -- (0.875, 0) (1.125, 0) -- (1.5, 0);
	
	\draw (-1, 0.5) -- (-1, -0.5);
	\draw (1, 0.5) -- (1, -0.5);
	
	\begin{scope}[rounded corners]
	
		\draw (-1.5, 0.25) -- (-1.25, 0.25) -- (-1.25, 0.5);
		\draw (-1.5, -0.25) -- (-1.25, -0.25) -- (-1.25, -0.5);
		
		\draw (-0.75, 0.5) -- (-0.75, 0.25) -- (-0.5, 0.25) (0.5, 0.25) -- (0.75, 0.25) -- (0.75, 0.5);
		\draw (-0.75, -0.5) -- (-0.75, -0.25) -- (-0.5, -0.25) (0.5, -0.25) -- (0.75, -0.25) -- (0.75, -0.5);
	
		\draw (1.5, 0.25) -- (1.25, 0.25) -- (1.25, 0.5);
		\draw (1.5, -0.25) -- (1.25, -0.25) -- (1.25, -0.5);
	
	\end{scope}
	
	\draw (-0.5, 0) cos (0, -0.125) sin (0.5, -0.25);
	\draw (-0.5, 0.25) cos (0, 0.125) sin (0.5, 0);
	\draw [draw = white, double = black, double distance between line centers = 3 pt, line width = 2.6 pt] (-0.5, -0.25) cos (0, 0) sin (0.5, 0.25);
	
	\node at (0, -1) {$\sigma_1 \sigma_2$};

	
	&
	
	\draw [dashed] (0, -1) -- (0, 1);
	
	&

	
	\draw (-0.75, 0.5) -- (-0.75, -0.5);
	\draw (0.75, 0.5) -- (0.75, -0.5);

	\draw [fill = white] (-0.75, 0) circle (0.15);
	\draw [fill = white] (0.75, 0) circle (0.15);

	\draw (-1.25, 0) -- node [above, font = \scriptsize] {$-2$} (1.25, 0);
	
	&
	
	\draw (-1.5, 0) -- (-0.5, 0) (0.5, 0) -- (1.5, 0);
	
	\draw (-1, 0.5) -- (-1, 0.125) (-1, -0.125) -- (-1, -0.5);
	\draw (1, 0.5) -- (1, 0.125) (1, -0.125) -- (1, -0.5);
	
	\begin{scope}[rounded corners]
	
		\draw (-1.5, 0.25) -- (-1.25, 0.25) -- (-1.25, 0.5);
		\draw (-1.5, -0.25) -- (-1.25, -0.25) -- (-1.25, -0.5);
		
		\draw (-0.75, 0.5) -- (-0.75, 0.25) -- (-0.5, 0.25) (0.5, 0.25) -- (0.75, 0.25) -- (0.75, 0.5);
		\draw (-0.75, -0.5) -- (-0.75, -0.25) -- (-0.5, -0.25) (0.5, -0.25) -- (0.75, -0.25) -- (0.75, -0.5);
	
		\draw (1.5, 0.25) -- (1.25, 0.25) -- (1.25, 0.5);
		\draw (1.5, -0.25) -- (1.25, -0.25) -- (1.25, -0.5);
	
	\end{scope}
	
	\draw (-0.5, 0.25) cos (0, 0) sin (0.5, -0.25);
	\draw [draw = white, double = black, double distance between line centers = 3 pt, line width = 2.6 pt] (-0.5, 0) cos (0, 0.125) sin (0.5, 0.25);
	\draw [draw = white, double = black, double distance between line centers = 3 pt, line width = 2.6 pt] (-0.5, -0.25) cos (0, -0.125) sin (0.5, 0);
	
	\node at (0, -1) {$\sigma_2 \sigma_1$};

	
\\
};

\end{tikzpicture}
\caption{\label{like-dict}Twist dictionary for like vertex states separated by an edge, and their associated 3-line graphs. When the twist is zero, the braid is the identity braid, with no crossings.}
\end{figure}

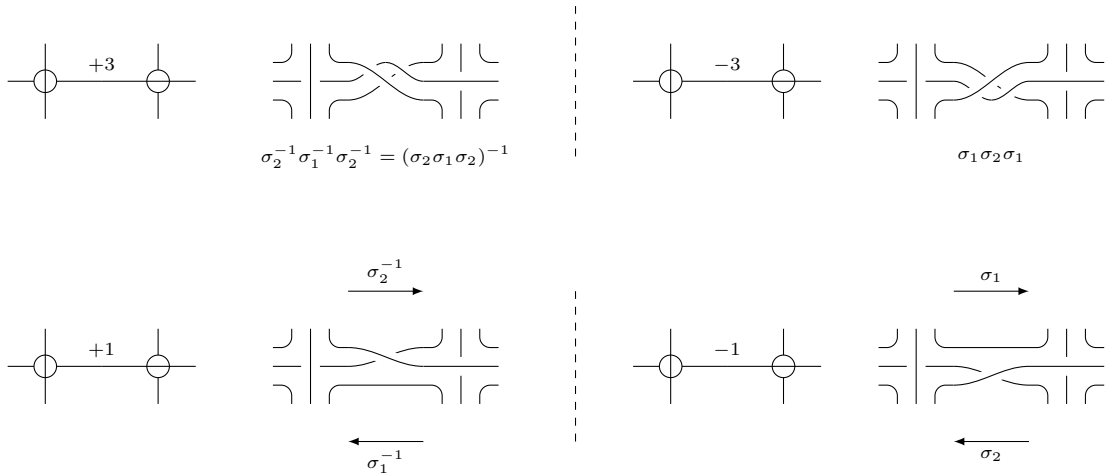
\begin{figure}[hbt]
\begin{tikzpicture}[> = latex, font = \scriptsize]
\matrix[column sep = 0.75 cm, row sep = 1 cm]{

	\draw (-1.25, 0) -- (0, 0);
	\draw (0.75, 0.5) -- (0.75, -0.5);

	\draw [fill = white] (-0.75, 0) circle (0.15);
	\draw [fill = white] (0.75, 0) circle (0.15);
	
	\draw (-0.75, 0.5) -- (-0.75, -0.5);
	
	\draw (0, 0) -- (1.25, 0);
	\node at (0, 0) [above] {$+3$};
	
	&
	
	\draw (-1.5, 0) -- (-1.125, 0) (-0.875, 0) -- (-0.5, 0) (0.5, 0) -- (1.5, 0);
	
	\draw (-1, 0.5) -- (-1, -0.5);
	\draw (1, 0.5) -- (1, 0.125) (1, -0.125) -- (1, -0.5);
	
	\begin{scope}[rounded corners]
	
		\draw (-1.5, 0.25) -- (-1.25, 0.25) -- (-1.25, 0.5);
		\draw (-1.5, -0.25) -- (-1.25, -0.25) -- (-1.25, -0.5);
		
		\draw (-0.75, 0.5) -- (-0.75, 0.25) -- (-0.5, 0.25) (0.5, 0.25) -- (0.75, 0.25) -- (0.75, 0.5);
		\draw (-0.75, -0.5) -- (-0.75, -0.25) -- (-0.5, -0.25) (0.5, -0.25) -- (0.75, -0.25) -- (0.75, -0.5);
	
		\draw (1.5, 0.25) -- (1.25, 0.25) -- (1.25, 0.5);
		\draw (1.5, -0.25) -- (1.25, -0.25) -- (1.25, -0.5);
	
	\end{scope}
	
	\draw (-0.5, 0) cos (-0.25, 0.125) sin (0, 0.25);
	
	\draw (-0.5, -0.25) cos (0, 0) sin (0.5, 0.25);
	\draw [draw = white, double = black, double distance between line centers = 3 pt, line width = 2.6 pt] (-0.5, 0.25) cos (0, 0) sin (0.5, -0.25);
	
	\draw [draw = white, double = black, double distance between line centers = 3 pt, line width = 2.6 pt] (0, 0.25) cos (0.25, 0.125) sin (0.5, 0);
	
	\node at (0, -1) {$\sigma_2 ^{-1} \sigma_1 ^{-1} \sigma_2 ^{-1} = (\sigma_2 \sigma_1 \sigma_2) ^{-1}$};
	
	
	&
	
	\draw [dashed] (0, -1) -- (0, 1);
	
	&


	\draw (-1.25, 0) -- (0, 0);
	\draw (0.75, 0.5) -- (0.75, -0.5);

	\draw [fill = white] (-0.75, 0) circle (0.15);
	\draw [fill = white] (0.75, 0) circle (0.15);
	
	\draw (-0.75, 0.5) -- (-0.75, -0.5);
	
	\draw (0, 0) -- (1.25, 0);
	\node at (0, 0) [above] {$-3$};
	
	&
	
	\draw (-1.5, 0) -- (-1.125, 0) (-0.875, 0) -- (-0.5, 0) (0.5, 0) -- (1.5, 0);
	
	\draw (-1, 0.5) -- (-1, -0.5);
	\draw (1, 0.5) -- (1, 0.125) (1, -0.125) -- (1, -0.5);
	
	\begin{scope}[rounded corners]
	
		\draw (-1.5, 0.25) -- (-1.25, 0.25) -- (-1.25, 0.5);
		\draw (-1.5, -0.25) -- (-1.25, -0.25) -- (-1.25, -0.5);
		
		\draw (-0.75, 0.5) -- (-0.75, 0.25) -- (-0.5, 0.25) (0.5, 0.25) -- (0.75, 0.25) -- (0.75, 0.5);
		\draw (-0.75, -0.5) -- (-0.75, -0.25) -- (-0.5, -0.25) (0.5, -0.25) -- (0.75, -0.25) -- (0.75, -0.5);
	
		\draw (1.5, 0.25) -- (1.25, 0.25) -- (1.25, 0.5);
		\draw (1.5, -0.25) -- (1.25, -0.25) -- (1.25, -0.5);
	
	\end{scope}
	
	\draw (-0.5, 0) cos (-0.25, -0.125) sin (0, -0.25);
	
	\draw (-0.5, 0.25) cos (0, 0) sin (0.5, -0.25);
	\draw [draw = white, double = black, double distance between line centers = 3 pt, line width = 2.6 pt] (-0.5, -0.25) cos (0, 0) sin (0.5, 0.25);
	
	\draw [draw = white, double = black, double distance between line centers = 3 pt, line width = 2.6 pt] (0, -0.25) cos (0.25, -0.125) sin (0.5, 0);
	
	\node at (0, -1) {$\sigma_1 \sigma_2 \sigma_1$};
	
	
\\


	\draw (-1.25, 0) -- (0, 0);
	\draw (0.75, 0.5) -- (0.75, -0.5);

	\draw [fill = white] (-0.75, 0) circle (0.15);
	\draw [fill = white] (0.75, 0) circle (0.15);
	
	\draw (-0.75, 0.5) -- (-0.75, -0.5);
	
	\draw (0, 0) -- (1.25, 0);
	\node at (0, 0) [above] {$+1$};
	
	&
	
	\draw (-1.5, 0) -- (-1.125, 0) (-0.875, 0) -- (-0.5, 0) (0.5, 0) -- (1.5, 0);
	
	\draw (-1, 0.5) -- (-1, -0.5);
	\draw (1, 0.5) -- (1, 0.125) (1, -0.125) -- (1, -0.5);
	
	\begin{scope}[rounded corners]
	
		\draw (-1.5, 0.25) -- (-1.25, 0.25) -- (-1.25, 0.5);
		\draw (-1.5, -0.25) -- (-1.25, -0.25) -- (-1.25, -0.5);
		
		\draw (-0.75, 0.5) -- (-0.75, 0.25) -- (-0.5, 0.25) (0.5, 0.25) -- (0.75, 0.25) -- (0.75, 0.5);
		\draw (-0.75, -0.5) -- (-0.75, -0.25) -- (0.75, -0.25) -- (0.75, -0.5);
	
		\draw (1.5, 0.25) -- (1.25, 0.25) -- (1.25, 0.5);
		\draw (1.5, -0.25) -- (1.25, -0.25) -- (1.25, -0.5);
	
	\end{scope}
	
	\draw (-0.5, 0) cos (0, 0.125) sin (0.5, 0.25);
	\draw [draw = white, double = black, double distance between line centers = 3 pt, line width = 2.6 pt] (-0.5, 0.25) cos (0, 0.125) sin (0.5, 0);
	
	\draw [->] (-0.5, 1) -- node [above] {$\sigma_2 ^{-1}$} (0.5, 1);
	\draw [->] (0.5, -1) -- node [below] {$\sigma_1 ^{-1}$} (-0.5, -1);

	
	&
	
	\draw [dashed] (0, -1) -- (0, 1);
	
	&


	\draw (-1.25, 0) -- (0, 0);
	\draw (0.75, 0.5) -- (0.75, -0.5);

	\draw [fill = white] (-0.75, 0) circle (0.15);
	\draw [fill = white] (0.75, 0) circle (0.15);
	
	\draw (-0.75, 0.5) -- (-0.75, -0.5);
	
	\draw (0, 0) -- (1.25, 0);
	\node at (0, 0) [above] {$-1$};
	
	&
	
	\draw (-1.5, 0) -- (-1.125, 0) (-0.875, 0) -- (-0.5, 0) (0.5, 0) -- (1.5, 0);
	
	\draw (-1, 0.5) -- (-1, -0.5);
	\draw (1, 0.5) -- (1, 0.125) (1, -0.125) -- (1, -0.5);
	
	\begin{scope}[rounded corners]
	
		\draw (-1.5, 0.25) -- (-1.25, 0.25) -- (-1.25, 0.5);
		\draw (-1.5, -0.25) -- (-1.25, -0.25) -- (-1.25, -0.5);
		
		\draw (-0.75, 0.5) -- (-0.75, 0.25) -- (0.75, 0.25) -- (0.75, 0.5);
		\draw (-0.75, -0.5) -- (-0.75, -0.25) -- (-0.5, -0.25) (0.5, -0.25) -- (0.75, -0.25) -- (0.75, -0.5);
	
		\draw (1.5, 0.25) -- (1.25, 0.25) -- (1.25, 0.5);
		\draw (1.5, -0.25) -- (1.25, -0.25) -- (1.25, -0.5);
	
	\end{scope}
	
	\draw (-0.5, 0) cos (0, -0.125) sin (0.5, -0.25);
	\draw [draw = white, double = black, double distance between line centers = 3 pt, line width = 2.6 pt] (-0.5, -0.25) cos (0, -0.125) sin (0.5, 0);
	
	\draw [->] (-0.5, 1) -- node [above] {$\sigma_1$} (0.5, 1);
	\draw [->] (0.5, -1) -- node [below] {$\sigma_2$} (-0.5, -1);
	
\\
};

\end{tikzpicture}
\caption{\label{unlike-dict}Twist dictionary for unlike vertex states separated by an edge, and their associated 3-line graphs. The arrows for the $\pm 1$ cases show the braid, as read from ``top to bottom''. Recall the braid identity $\sigma_1 \sigma_2 \sigma_1 = \sigma_2 \sigma_2 \sigma_1$ implies the $+3$ twist is the inverse of the $-3$ twist.}
\end{figure}

From here, we can turn to matching the edges between two graph vertices with the corresponding braided lines of the 3-line graph. We can only consistently do this matching if the graph edge, and the two vertices, give the correct gluing of two dual faces in the triangulation. The cases where this can be done are shown in Figure \ref{like-dict} when the two vertices project to the same vertex state, and Figure \ref{unlike-dict} when they do not. All other cases can be built up from these by induction; for example, a twist of $+4$ between two vertices in the $\oplus$ state would result from applying the braid shown in the upper left corner of Figure \ref{like-dict} twice.

\begin{figure}[hbt]
\begin{tikzpicture}
\matrix[column sep = 1 cm]{

	\draw [rotate around = {90 : (-0.25, -1)}] (-0.25, -1) cos (0.25, -1.125) sin (0.75, -1.25);
	\draw (0.25, -1) -- (0.25, 0);
	\draw [rotate around = {90 : (0, -1)}, draw = white, double = black, double distance between line centers = 3 pt, line width = 2.6 pt]
		(0, -1) cos (0.5, -0.875) sin (1, -0.75);
		
	\node at (0, -1.5) {$\sigma_1$};

&

	\draw [rotate around = {90 : (0, -1)}] (0, -1) cos (0.5, -0.875) sin (1, -0.75);
	\draw (0.25, -1) -- (0.25, 0);
	\draw [rotate around = {90 : (-0.25, -1)}, draw = white, double = black, double distance between line centers = 3 pt, line width = 2.6 pt]
		(-0.25, -1) cos (0.25, -1.125) sin (0.75, -1.25);
		
	\node at (0, -1.5) {$\sigma_1 ^{-1}$};

&

	\draw [rotate around = {90 : (0, -1)}] (0, -1) cos (0.5, -1.125) sin (1, -1.25);
	\draw (-0.25, -1) -- (-0.25, 0);
	\draw [rotate around = {90 : (0.25, -1)}, draw = white, double = black, double distance between line centers = 3 pt, line width = 2.6 pt]
		(0.25, -1) cos (0.75, -0.875) sin (1.25, -0.75);
		
	\node at (0, -1.5) {$\sigma_2$};

&

	\draw [rotate around = {90 : (0.25, -1)}] (0.25, -1) cos (0.75, -0.875) sin (1.25, -0.75);
	\draw (-0.25, -1) -- (-0.25, 0);
	\draw [rotate around = {90 : (0, -1)}, draw = white, double = black, double distance between line centers = 3 pt, line width = 2.6 pt]
		(0, -1) cos (0.5, -1.125) sin (1, -1.25);
		
	\node at (0, -1.5) {$\sigma_2 ^{-1}$};

\\
};
\end{tikzpicture}
\caption{\label{braids}Elements of the braid group $B_3$.}
\end{figure}
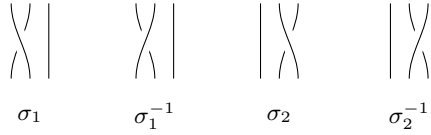

The edge twists for the knotted graph lead to braids in the associated 3-line graph. Thus, it is helpful to recall the definition of the braid group $B_3$, as an aid to describing the 3-line graphs. We will also see in Section \ref{gen-braid} that some of the twists lead to a natural extension of the braid group. $B_3$ is the group of all braids on three strands; these braids can be written in terms of the elements $\sigma_1$ and $\sigma_2$, along with their inverses. These elements are shown in Figure \ref{braids}. When a braid has multiple crossings, the elements are read from top to bottom. This method of reading the braid word matches the composition of the braids given in Figure \ref{line-twist-inv} for the $+1$ twist and its inverse $-1$ twist; these correspond to the $\sigma_2 ^{-1}$ and $\sigma_2$ braids, respectively. For all the cases in the twist dictionaries shown in Figure \ref{like-dict} and \ref{unlike-dict}, the matching braid words are also given.

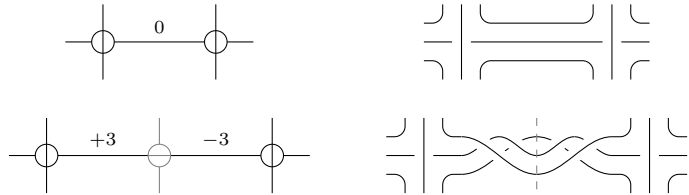
\begin{figure}[hbt]
\begin{tikzpicture}[font = \scriptsize]
\matrix[column sep = 1 cm, row sep = 0.5 cm]{


	\draw (-1.25, 0) -- node [above] {0} (1.25, 0);

	\draw [fill = white] (-0.75, 0) circle (0.15);
	\draw [fill = white] (0.75, 0) circle (0.15);
	
	\draw (-0.75, 0.5) -- (-0.75, -0.5);
	\draw (0.75, 0.5) -- (0.75, -0.5);
	
&
	
	\draw (-1.5, 0) -- (-1.125, 0) (-0.875, 0) -- (0.875, 0) (1.125, 0) -- (1.5, 0);
	
	\draw (-1, 0.5) -- (-1, -0.5);
	\draw (1, 0.5) -- (1, -0.5);
	
	\begin{scope}[rounded corners]
	
		\draw (-1.5, 0.25) -- (-1.25, 0.25) -- (-1.25, 0.5);
		\draw (-1.5, -0.25) -- (-1.25, -0.25) -- (-1.25, -0.5);
		
		\draw (-0.75, 0.5) -- (-0.75, 0.25) -- (0.75, 0.25) -- (0.75, 0.5);
		\draw (-0.75, -0.5) -- (-0.75, -0.25) -- (0.75, -0.25) -- (0.75, -0.5);
	
		\draw (1.5, 0.25) -- (1.25, 0.25) -- (1.25, 0.5);
		\draw (1.5, -0.25) -- (1.25, -0.25) -- (1.25, -0.5);
	
	\end{scope}

\\

	\draw (-2, 0) -- node [above, pos = 0.625] {$+3$} (0, 0) -- node [above, pos = 0.375] {$-3$} (2, 0);

	\draw [fill = white] (-1.5, 0) circle (0.15);
	\draw [fill = white] (1.5, 0) circle (0.15);
	\draw [gray, fill = white] (0, 0) circle (0.15);
	
	\draw (-1.5, 0.5) -- (-1.5, -0.5);
	\draw (1.5, 0.5) -- (1.5, -0.5);
	\draw [gray] (0, -0.5) -- (0, -0.15) (0, 0.15) -- (0, 0.5) (-0.15, 0) -- (0.15, 0);
	
&
	
	\draw (-2, 0) -- (-1.625, 0) (-1.375, 0) -- (-1, 0) (1, 0) -- (1.375, 0) (1.625, 0) -- (2, 0);
	
	\draw (-1.5, 0.5) -- (-1.5, -0.5);
	\draw (1.5, 0.5) -- (1.5, -0.5);
	
	\begin{scope}[rounded corners]
	
		\draw (-2, 0.25) -- (-1.75, 0.25) -- (-1.75, 0.5);
		\draw (-2, -0.25) -- (-1.75, -0.25) -- (-1.75, -0.5);
		
		\draw (-1.25, 0.5) -- (-1.25, 0.25) -- (-1, 0.25) (1, 0.25) -- (1.25, 0.25) -- (1.25, 0.5);
		\draw (-1.25, -0.5) -- (-1.25, -0.25) -- (-1, -0.25) (1, -0.25) -- (1.25, -0.25) -- (1.25, -0.5);
	
		\draw (2, 0.25) -- (1.75, 0.25) -- (1.75, 0.5);
		\draw (2, -0.25) -- (1.75, -0.25) -- (1.75, -0.5);
	
	\end{scope}
	
	
	\draw (-1, 0) cos (-0.75, 0.125) sin (-0.5, 0.25);
	
	\draw (-1, -0.25) cos (-0.5, 0) sin (0, 0.25);
	\draw [draw = white, double = black, double distance between line centers = 3 pt, line width = 2.6 pt] (-1, 0.25) cos (-0.5, 0) sin (0, -0.25);
	
	\draw [draw = white, double = black, double distance between line centers = 3 pt, line width = 2.6 pt] (-0.5, 0.25) cos (-0.25, 0.125) sin (0, 0);
	
	
	\draw (0, 0.25) cos (0.5, 0) sin (1, -0.25);
	
	\draw (0.5, 0.25) cos (0.75, 0.125) sin (1, 0);
	\draw [draw = white, double = black, double distance between line centers = 3 pt, line width = 2.6 pt] (0, 0) cos (0.25, 0.125) sin (0.5, 0.25);
	
	\draw [draw = white, double = black, double distance between line centers = 3 pt, line width = 2.6 pt] (0, -0.25) cos (0.5, 0) sin (1, 0.25);

	
	\draw [gray, dashed] (0, 0.5) -- (0, -0.5);

\\
};
\end{tikzpicture}
\caption{\label{3-twist}The 3-line graphs for twists $+3$ and $-3$ must be inverses, since we can take two like vertices, and place a ``virtual'' vertex of opposite state between them. Then the two twists should cancel to get the original zero twist edge. The braid for the $-3$ twist is different than shown in Figure \ref{unlike-dict}, where it is rotated $\pi$ radians from the situation shown here.} 
\end{figure}

Notice that, for Figure \ref{like-dict}, these words are the same regardless of whether you move from left to right, or right to left, along the leg in the 3-line graph. This is what you would expect, since the original graph is the same under a $\pi$ rotation. Since the $\pm 1$ twist case of unlike vertices does not have this symmetry, then the braid word depends on the direction traveled along the leg, as shown in the bottom two entries of Figure \ref{unlike-dict}. The arrows shown in those cases give the ``downward'' direction as the braid is read top to bottom. The braid identity $\sigma_1 \sigma_2 \sigma_1 = \sigma_2 \sigma_1 \sigma_2$ ensures that the $+3$ twist is the same braid word, when read in either direction along the graph edge; the same is true for the $-3$ twist (see also Figure \ref{3-twist}).

\begin{figure}[hbt]
\begin{tikzpicture}[> = latex, font = \scriptsize]
\matrix[column sep = 1 cm]{

	\draw (0, -2) node [left, rotate = 90] {$a'c'$} -- (0, 2) node [right, rotate = 90] {$ac$};
	
	\draw[rounded corners]
		(-1, -1.25) node [left] {$c'd'$} -- (-0.25, -1.25) -- (-0.25, -2) node [left, rotate = 90] {$c'd'$}
		(1, -1.25) node [right] {$a'd'$} -- (0.25, -1.25) -- (0.25, -2) node [left, rotate = 90] {$a'd'$}
		
		(-1, -0.75) node [left] {$b'c'$} -- (-0.25, -0.75) -- (-0.25, 0.75) -- (-1, 0.75) node [left] {$bc$}
		(1, -0.75) node [right] {$a'b'$} -- (0.25, -0.75) -- (0.25, 0.75) -- (1, 0.75) node [right] {$ab$}
		
		(-1, 1.25) node [left] {$cd$} -- (-0.25, 1.25) -- (-0.25, 2) node [right, rotate = 90] {$cd$}
		(1, 1.25) node [right] {$ad$} -- (0.25, 1.25) -- (0.25, 2) node [right, rotate = 90] {$ad$};
		
	\draw [draw = white, double = black, double distance between line centers = 3 pt, line width = 2.6 pt]
		(-1, -1) node [left] {$b'd'$} -- (1, -1) node [right] {$b'd'$};
	\draw [draw = white, double = black, double distance between line centers = 3 pt, line width = 2.6 pt]
		(-1, 1) node [left] {$bd$} -- (1, 1) node [right] {$bd$};

&

	\begin{scope}[->, font = \footnotesize]

		\draw (0, 0.25) -- node [above] {2-3 move} (1, 0.25);
		\draw (1, -0.25) -- node [below] {3-2 move} (0, -0.25);

	\end{scope}

&

	\draw (0, -2) node [left, rotate = 90] {$a'c'$} -- (0, 2) node [right, rotate = 90] {$ac$};
	
	\draw [rounded corners]
		(-0.25, -2) node [left, rotate = 90] {$c'd'$}-- (-0.25, 0.75) -- (-2.25, 0.75) -- (-2.25, 0.5) (-2.75, -0.5) -- (-2.75, -1.25) -- (-1.75, -1.25) -- (-1.75, -1.5)
		(0.25, -2) node [left, rotate = 90] {$a'd'$} -- (0.25, 0.75) -- (2.25, 0.75) -- (2.25, 0.5) (2.75, -0.5) -- (2.75, -1.25) -- (1.75, -1.25) -- (1.75, -1.5)
		(-2.5, -0.5) -- (-2.5, -1) -- (-1, -1)
		(1, -1) -- (2.5, -1) -- (2.5, -0.5)
		(-1, 1) -- (-2.5, 1) -- (-2.5, 0.5)
		(1, 1) -- (2.5, 1) -- (2.5, 0.5)
		(-0.25, 2) node [right, rotate = 90] {$cd$} -- (-0.25, 1.25) -- (-2.75, 1.25) -- (-2.75, 0.5)
		(0.25, 2) node [right, rotate = 90] {$ad$} -- (0.25, 1.25) -- (2.75, 1.25) -- (2.75, 0.5);
		
	\draw [rotate around = {-90 : (-2.5, 0.5)}] (-2.5, 0.5) cos (-2.25, 0.375) sin (-2, 0.25);
	\draw [rotate around = {-90 : (-2.25, 0.5)}] (-2.25, 0.5) cos (-1.75, 0.25) sin (-1.25, 0);
	\draw [rotate around = {-90 : (-1.75, -1.5)}] (-1.75, -1.5) cos (-1.25, -1.25) sin (-0.75, -1);
	
	\draw [rotate around = {90 : (-1.75, 1.25)}] (-1.75, 1.25) cos (-1.25, 1) sin (-0.75, 0.75);
	\draw [rotate around = {90 : (1.75, 1.25)}] (1.75, 1.25) cos (2.25, 1.5) sin (2.75, 1.75);
	
	\draw [rotate around = {-90 : (1.75, -1.5)}] (1.75, -1.5) cos (2.25, -1.75) sin (2.75, -2);
	\draw [rotate around = {-90 : (2.25, 0.5)}] (2.25, 0.5) cos (2.75, 0.75) sin (3.25, 1);
	\draw [rotate around = {-90 : (2.5, 0.5)}] (2.5, 0.5) cos (2.75, 0.625) sin (3, 0.75);
	
	\draw [rounded corners, draw = white, double = black, double distance between line centers = 3 pt, line width = 2.6 pt]
		(-1, -1) -- (1, -1)
		(-1, 1) -- (1, 1)
		(-1.5, -1.5) -- (-1.5, 1.25)
		(-1.25, -1.5) -- (-1.25, -1.25) -- (1.25, -1.25) -- (1.25, -1.5)
		(-2.25, -0.5) -- (-2.25, -0.75) -- (-1.75, -0.75) -- (-1.75, 1.3)							
		(-1.25, 1.25) -- (-1.25, -0.5) -- (-1.25, -0.75) -- (1.25, -0.75) -- (1.25, -0.5) -- (1.25, 1.25)
		(1.5, -1.5) -- (1.5, 1.25)
		(2.25, -0.5) -- (2.25, -0.75) -- (1.75, -0.75) -- (1.75, 1.3)							
	;
	
	\draw [rotate around = {90 : (-2.5, -0.5)}, draw = white, double = black, double distance between line centers = 3 pt, line width = 2.6 pt]
		(-2.5, -0.5) cos (-2.25, -0.375) sin (-2, -0.25);
	
	\draw [rotate around = {90 : (-2.25, -0.5)}, draw = white, double = black, double distance between line centers = 3 pt, line width = 2.6 pt]
		(-2.25, -0.5) cos (-1.75, -0.25) sin (-1.25, 0);

	\draw [rotate around = {-90 : (-1.5, -1.5)}, draw = white, double = black, double distance between line centers = 3 pt, line width = 2.6 pt]
		(-1.5, -1.5) cos (-1, -1.625) sin (-0.5, -1.75);
	\draw [rotate around = {-90 : (-1.25, -1.5)}, draw = white, double = black, double distance between line centers = 3 pt, line width = 2.6 pt]
		(-1.25, -1.5) cos (-0.75, -1.625) sin (-0.25, -1.75);

	\draw [rotate around = {-90 : (1.5, -1.5)}, draw = white, double = black, double distance between line centers = 3 pt, line width = 2.6 pt]
		(1.5, -1.5) cos (2, -1.375) sin (2.5, -1.25);
	\draw [rotate around = {-90 : (1.25, -1.5)}, draw = white, double = black, double distance between line centers = 3 pt, line width = 2.6 pt]
		(1.25, -1.5) cos (1.75, -1.375) sin (2.25, -1.25);
	
	\draw [rotate around = {90 : (-1.5, 1.25)}, draw = white, double = black, double distance between line centers = 3 pt, line width = 2.6 pt]
		(-1.5, 1.25) cos (-1, 1.375) sin (-0.5, 1.5);
	\draw [rotate around = {90 : (-1.25, 1.25)}, draw = white, double = black, double distance between line centers = 3 pt, line width = 2.6 pt]
		(-1.25, 1.25) cos (-0.75, 1.375) sin (-0.25, 1.5);
	
	\draw [rotate around = {90 : (1.25, 1.25)}, draw = white, double = black, double distance between line centers = 3 pt, line width = 2.6 pt]
		(1.25, 1.25) cos (1.75, 1.125) sin (2.25, 1);
	\draw [rotate around = {90 : (1.5, 1.25)}, draw = white, double = black, double distance between line centers = 3 pt, line width = 2.6 pt]
		(1.5, 1.25) cos (2, 1.125) sin (2.5, 1);
	
	\draw [rotate around = {90 : (2.25, -0.5)}, draw = white, double = black, double distance between line centers = 3 pt, line width = 2.6 pt]
		(2.25, -0.5) cos (2.75, -0.75) sin (3.25, -1);
	
	\draw [rotate around = {90 : (2.5, -0.5)}, draw = white, double = black, double distance between line centers = 3 pt, line width = 2.6 pt]
		(2.5, -0.5) cos (2.75, -0.625) sin (3, -0.75);
		
	\node at (-1.75, -2.5) [left, rotate = 90] {$b'c'$};
	\node at (-1.5, -2.5) [left, rotate = 90] {$b'd'$};
	\node at (-1.25, -2.5) [left, rotate = 90] {$c'd'$};
		
	\node at (1.25, -2.5) [left, rotate = 90] {$a'd'$};
	\node at (1.5, -2.5) [left, rotate = 90] {$b'd'$};
	\node at (1.75, -2.5) [left, rotate = 90] {$a'b'$};
		
	\node at (-1.75, 2.25) [right, rotate = 90] {$bc$};
	\node at (-1.5, 2.25) [right, rotate = 90] {$bd$};
	\node at (-1.25, 2.25) [right, rotate = 90] {$cd$};
		
	\node at (1.25, 2.25) [right, rotate = 90] {$ad$};
	\node at (1.5, 2.25) [right, rotate = 90] {$bd$};
	\node at (1.75, 2.25) [right, rotate = 90] {$ab$};

\\
};

\end{tikzpicture}
\caption{\label{2-3-line}The 3-line graph version of 2-3 move and its inverse.}
\end{figure}
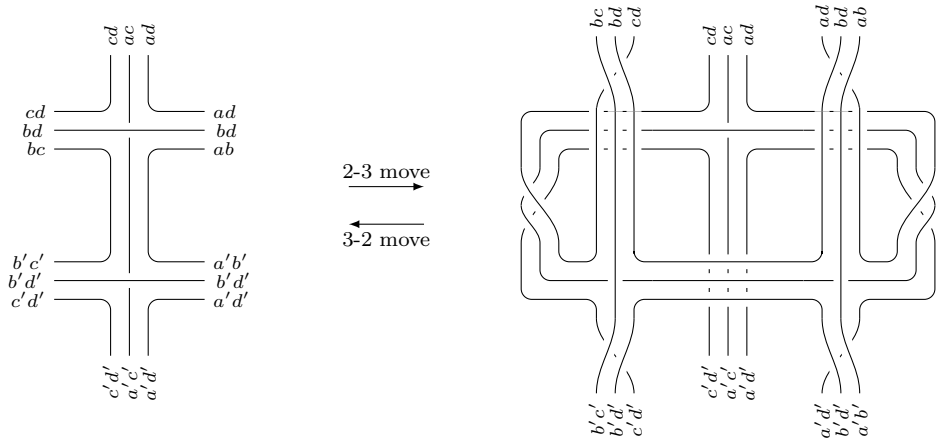

\begin{figure}[hbt]
\begin{tikzpicture}[> = latex, font = \scriptsize]
\matrix[column sep = 1 cm]{

	\draw (-1, 0) node [left] {$ac$} -- (1, 0) node [right] {$ac$};
	
	\begin{scope}[rounded corners]
	
		\draw (1, 0.25) node [right] {$ad$} -- (0.25, 0.25) -- (0.25, 1) node [right, rotate = 90] {$ad$};
		\draw (1, -0.25) node [right] {$cd$} -- (0.25, -0.25) -- (0.25, -1) node [left, rotate = 90] {$cd$};
	
		\draw (-1, 0.25) node [left] {$ab$} -- (-0.25, 0.25) -- (-0.25, 1) node [right, rotate = 90] {$ab$};
		\draw (-1, -0.25) node [left] {$bc$} -- (-0.25, -0.25) -- (-0.25, -1) node [left, rotate = 90] {$bc$};
		
	\end{scope}
	
	\draw [draw = white, double = black, double distance between line centers = 3 pt, line width = 2.6 pt] (0, -1) node [left, rotate = 90] {$bd$}
		-- (0, 1) node [right, rotate = 90] {$bd$};

&

	\begin{scope}[->, font = \footnotesize]

		\draw (0, 0.25) -- node [above] {1-4 move} (1, 0.25);
		\draw (1, -0.25) -- node [below] {4-1 move} (0, -0.25);

	\end{scope}

&


	\draw (-2.5, 0) node [left] {$ac$} -- (2.5, 0) node [right] {$ac$};
	
	\begin{scope}[rounded corners]
	
		
		\draw (-1.25, 1.25) -- (1.25, 1.25) -- (1.25, 0.25) -- (-1.25, 0.25) -- cycle;
		\draw (-1.25, -1.25) -- (1.25, -1.25) -- (1.25, -0.25) -- (-1.25, -0.25) -- cycle;
	
	
		\draw (2.5, 0.25) node [right] {$ad$} -- (1.75, 0.25) -- (1.75, 1) -- (1.25, 1.5) -- (0.25, 1.5) -- (0.25, 2) node [right, rotate = 90] {$ad$};
		\draw (2.5, -0.25) node [right] {$cd$} -- (1.75, -0.25) -- (1.75, -1) -- (1.25, -1.5) -- (0.25, -1.5) -- (0.25, -2) node [left, rotate = 90] {$cd$};
	
		\draw (-2.5, 0.25) node [left] {$ab$} -- (-1.75, 0.25) -- (-1.75, 1) -- (-1.25, 1.5) -- (-0.25, 1.5) -- (-0.25, 2) node [right, rotate = 90] {$ab$};
		\draw (-2.5, -0.25) node [left] {$bc$} -- (-1.75, -0.25) -- (-1.75, -1) -- (-1.25, -1.5) -- (-0.25, -1.5) -- (-0.25, -2) node [left, rotate = 90] {$bc$};
		
		
		\draw [draw = white, double = black, double distance between line centers = 3 pt, line width = 2.6 pt]
			(1.5, 1) -- (1.5, -1) -- (0.25, -1) -- (0.25, 1) -- cycle;
		\draw [draw = white, double = black, double distance between line centers = 3 pt, line width = 2.6 pt]
			(-1.5, 1) -- (-1.5, -1) -- (-0.25, -1) -- (-0.25, 1) -- cycle;
	
	\end{scope}
	
	
	\draw [draw = white, double = black, double distance between line centers = 3 pt, line width = 2.6 pt] (0, -2) node [left, rotate = 90] {$bd$}
		-- (0, 2) node [right, rotate = 90] {$bd$};

\\
};

\end{tikzpicture}
\caption{\label{1-4-line}The 3-line graph version of the 1-4 move and its inverse.}
\end{figure}
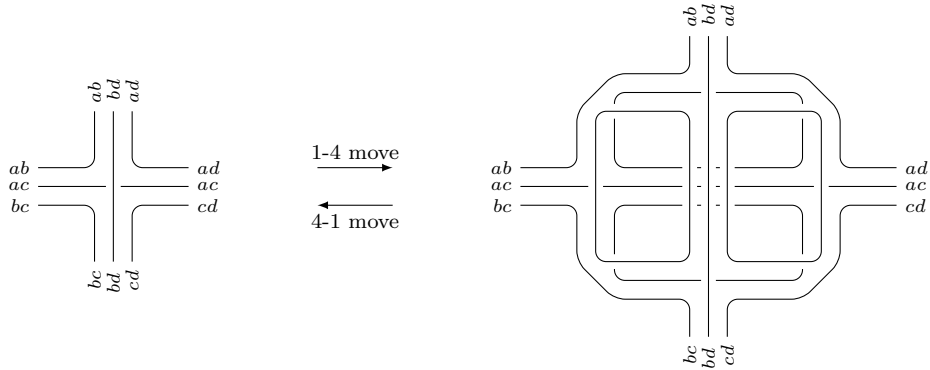

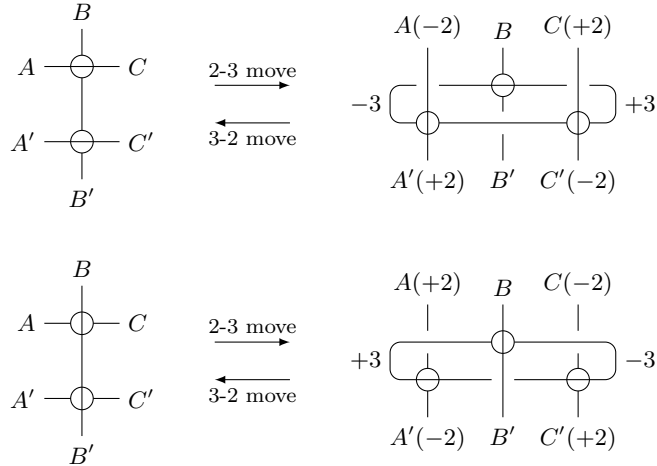
\begin{figure}[hbt]
\centering
\begin{tikzpicture}[> = latex]
\matrix[column sep = 0.5 cm, row sep = 0.5 cm]{


	\draw (0, -1) node [below] {$B'$} -- (0, 1) node [above] {$B$};

	\draw [fill = white] (0, -0.5) circle (0.15);
	\draw [fill = white] (0, 0.5) circle (0.15);

	\draw (0.5, -0.5) node [right] {$C'$} -- (-0.5, -0.5) node [left] {$A'$};
	\draw (0.5, 0.5) node [right] {$C$} -- (-0.5, 0.5) node [left] {$A$};

&

	\begin{scope}[->, font = \footnotesize]

		\draw (0, 0.25) -- node [above] {2-3 move} (1, 0.25);
		\draw (1, -0.25) -- node [below] {3-2 move} (0, -0.25);

	\end{scope}

&

	
	\draw (0, -0.75) node [below] {$B'$} -- (0, -0.4) (0, -0.1) -- (0, 0.75) node [above] {$B$};
	\draw (0, 0.25) [fill = white] circle (0.15);
	\draw [rounded corners] (0, 0.25) -- (0.85, 0.25) (1.15, 0.25) -- (1.5, 0.25) -- node [right] {$+3$} (1.5, -0.25) -- (1.15, -0.25)
		(0.85, -0.25) -- (-0.85, -0.25) (-1.15, -0.25) -- (-1.5, -0.25) -- node [left] {$-3$} (-1.5, 0.25) -- (-1.15, 0.25) (-0.85, 0.25) -- (0, 0.25);
	
	\draw (-1, -0.25) [fill = white] circle (0.15);
	\draw (1, -0.25) [fill = white] circle (0.15);
	
	\draw (-1, 0.75) node [above] {$A (-2)$} -- (-1, -0.75) node [below] {$A' (+2)$};
	\draw (1, 0.75) node [above] {$C (+2)$} -- (1, -0.75) node [below] {$C' (-2)$};

\\


	\draw (0.5, -0.5) node [right] {$C'$} -- (-0.5, -0.5) node [left] {$A'$};
	\draw (0.5, 0.5) node [right] {$C$} -- (-0.5, 0.5) node [left] {$A$};

	\draw [fill = white] (0, -0.5) circle (0.15);
	\draw [fill = white] (0, 0.5) circle (0.15);

	\draw (0, -1) node [below] {$B'$} -- (0, 1) node [above] {$B$};

&

	\begin{scope}[->, font = \footnotesize]

		\draw (0, 0.25) -- node [above] {2-3 move} (1, 0.25);
		\draw (1, -0.25) -- node [below] {3-2 move} (0, -0.25);

	\end{scope}

&

	
	\draw (-1, 0.75) node [above] {$A (+2)$} -- (-1, 0.4) (-1, 0.1) -- (-1, -0.75) node [below] {$A' (-2)$};
	\draw (1, 0.75) node [above] {$C (-2)$} -- (1, 0.4) (1, 0.1) -- (1, -0.75) node [below] {$C' (+2)$};
	
	\draw (-1, -0.25) [fill = white] circle (0.15);
	\draw (1, -0.25) [fill = white] circle (0.15);
	
	\draw [rounded corners] (0.15, -0.25) -- (1.5, -0.25) -- node [right] {$-3$} (1.5, 0.25) -- (0.15, 0.25) (-0.15, 0.25) -- (-1.5, 0.25) --
		node [left] {$+3$} (-1.5, -0.25) -- (-0.15, -0.25);
	
	\draw (0, 0.25) [fill = white] circle (0.15);
	\draw (0, -0.75) node [below] {$B'$} -- (0, 0.75) node [above] {$B$};

\\
};
\end{tikzpicture}
\caption{\label{2-3-twist}The 2-3 and 3-2 Pachner graph moves, with appropriate twists on the edges, coming from those shown in the 3-line graph of Figure \ref{2-3-line}.}
\end{figure}

\begin{figure}[hbt]
\begin{tikzpicture}[> = latex]
\matrix[column sep = 0.25 cm, row sep = 0.5 cm]{


	\draw (-0.5, 0) node [left] {$A$} -- (0.5, 0) node [right] {$C$};
	\draw [fill = white] (0, 0) circle (0.15);
	\draw (0, -0.5) node [below] {$D$} -- (0, 0.5) node [above] {$B$};

&

	\begin{scope}[->, font = \footnotesize]

		\draw (0, 0.25) -- node [above] {1-4 move} (1, 0.25);
		\draw (1, -0.25) -- node [below] {4-1 move} (0, -0.25);

	\end{scope}

&


	\draw (-1, 0) node [left] {$A$} -- (1, 0) node [right] {$C$};

	\draw [fill = white] (-0.5, 0) circle (0.15);
	\draw [fill = white] (0, 0.5) circle (0.15);
	\draw [fill = white] (0.5, 0) circle (0.15);
	\draw [fill = white] (0, -0.5) circle (0.15);

	\draw [rounded corners] (-0.15, -0.5) -- (-0.5, -0.5) node [below = 0.25 em, left = 0.25 em] {$-1$} --
		(-0.5, 0.5) node [above = 0.25 em, left = 0.25 em] {$+1$} -- (-0.15, 0.5)
		(0.15, -0.5) -- (0.5, -0.5) node [below = 0.25 em, right = 0.25 em] {$+1$} --
		(0.5, 0.5) node [above = 0.25 em, right = 0.25 em] {$-1$} -- (0.15, 0.5);

	\draw [draw = white, double = black, double distance between line centers = 3 pt, line width = 2.6 pt] (0, -0.35) -- (0, 0.35);

	\draw (0, 1) node [above] {$B$} -- (0, 0.35);
	\draw (0, -0.35) -- (0, -1) node [below] {$D$};

&
	\draw [dashed] (0, -1) -- (0, 1);
&


	\draw (0, -0.5) node [below] {$D$} -- (0, 0.5) node [above] {$B$};
	\draw [fill = white] (0, 0) circle (0.15);
	\draw (-0.5, 0) node [left] {$A$} -- (0.5, 0) node [right] {$C$};s

&

	\begin{scope}[->, font = \footnotesize]

		\draw (0, 0.25) -- node [above] {1-4 move} (1, 0.25);
		\draw (1, -0.25) -- node [below] {4-1 move} (0, -0.25);

	\end{scope}

&


	\draw (0, 1) node [above] {$B$} -- (0, -1) node [below] {$D$};

	\draw [fill = white] (-0.5, 0) circle (0.15);
	\draw [fill = white] (0, 0.5) circle (0.15);
	\draw [fill = white] (0.5, 0) circle (0.15);
	\draw [fill = white] (0, -0.5) circle (0.15);

	\draw [rounded corners] (-0.5, -0.15) -- (-0.5, -0.5) node [below = 0.25 em, left = 0.25 em] {$+1$} --
		(0.5, -0.5) node [below = 0.25 em, right = 0.25 em] {$-1$} -- (0.5, -0.15)
		(-0.5, 0.15) -- (-0.5, 0.5) node [above = 0.25 em, left = 0.25 em] {$-1$} -- (0.5, 0.5) node [above = 0.25 em, right = 0.25 em] {$+1$}
		-- (0.5, 0.15);

	\draw [draw = white, double = black, double distance between line centers = 3 pt, line width = 2.6 pt] (-0.35, 0) -- (0.35, 0);

	\draw (-1, 0) node [left] {$A$} -- (-0.35, 0);
	\draw (0.35, 0) -- (1, 0) node [right] {$C$};

\\
};
\end{tikzpicture}
\caption{\label{1-4-twist}The 1-4 and 4-1 Pachner graph moves, with appropriate twists on the edges, coming from those shown in the 3-line graph of Figure \ref{1-4-line}.}
\end{figure}
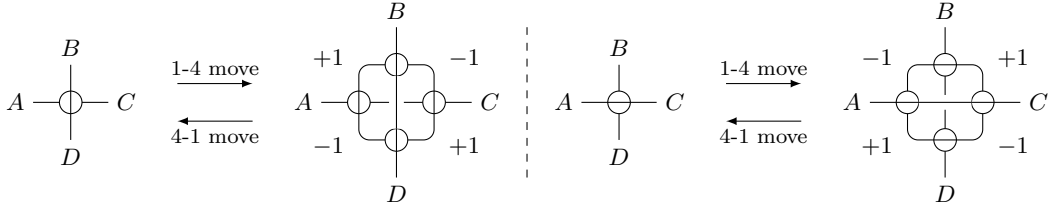

Now we consider the 3-line graph versions of the Pachner moves. Since the lines in the 3-line graph correspond to the edges in the dual triangulation, we return to the original Pachner moves, given in Figure \ref{Pach-tet}. From there, the identifications of the dual edges can be read off. In addition, the necessary rotations of the dual faces is determined; this gives how the matching lines are braided, using the dictionary between graph edge twists and leg braids. The resulting equivalent subgraphs for the 3-line graph are shown as Figure \ref{2-3-line} for the 2-3 move and its inverse 3-2 move, and Figure \ref{1-4-line} for the 1-4 and 4-1 moves. From the leg braidings of these graphs, the twists on the graph edges are found; these give Figure \ref{2-3-twist} for the 2-3 move, while the 1-4 move is drawn in Figure \ref{1-4-twist}. There is an assumption that the edge common to the two vertices before the 2-3 move has a twist of zero; this can easily be obtained by any appropriate RV moves to eliminate a pre-existing twist on this graph edge. It is also worthwhile to note that for the closed cycles after the 1-4 and 2-3 moves, the sum of the edge twists along either cycle is zero. However, there are non-zero twists on the individual edges required for the 3-2 move to take place; this is different than the requirement stated by Smolin and Wan~\cite{Smo-Wan08} for no twist on any internal edge. Yet it can be seen from the 3-line graph that these twists are necessary, for otherwise the line identifications would be incorrect. An example of this is that $cd$ and $c'd'$ dual edges would be distinct before a 2-3 move, but glued together afterward, if no internal twists were present.

One important consideration is that the subgraphs before and after the Pachner move retain their symmetry under reflections, including the twists on the framed edges. If we ignore these twists for a moment, it is easy to see that this holds true for the 1-4 and 4-1 moves in Figure \ref{1-4-twist} -- the resulting graphs are invariant under both horizontal and vertical reflections. The sign of each twist will also change under a reflection, so these graphs are indeed symmetric. A little more work is required to show the same for the 2-3 and 3-2 moves. The invariance under a reflection about a vertical axis through the center of each graph in Figure \ref{2-3-twist} is relatively easy to see, with the edge twists changing sign under the reflection. However, any symmetry under reflection across a horizontal axis through the graph after the 2-3 move is not as obvious. In particular, although the $\pm 2$ would change sign to give the correct behavior, so would the $\pm 3$ twists, meaning these latter twists no longer match the twists on the loop edges before the reflection. Thus, there is an apparent choice in how the $\pm 3$ twists are placed on the internal edges, depending on which initial vertex is assigned as that with the unprimed incident edges. The answer to this issue is in how the vertices are shifted by graph moves to the same positions as before the reflection. The two $\oplus$ vertices are both moved via the RIV move, but this results in two curls in the edges of the central loop (exactly those with the $\pm 3$ twists). If the edges were unframed, these are removable by subsequent RI moves. This cannot be done for framed edges, as shown in Section \ref{frame-Reid}; it is this obstruction that answers the seeming issue of a lack of reflection invariance.

\subsection{Generalized Reidemeister moves on framed graphs}
\label{frame-Reid}

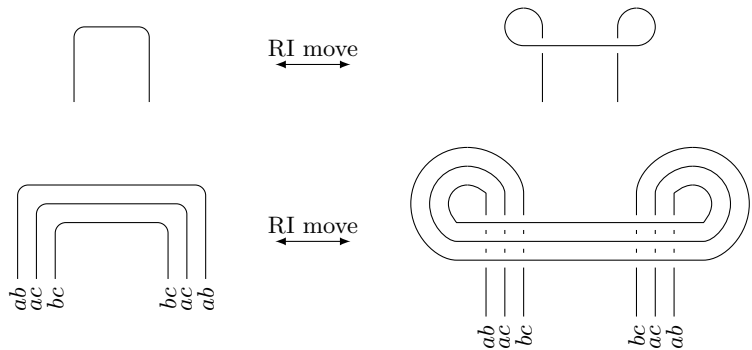
\begin{figure}[hbt]
\begin{tikzpicture}[> = latex]
\matrix[column sep = 0.5 cm, row sep = 0.5 cm]{

	\draw [rounded corners] (-0.5, 0) -- (-0.5, 1) -- (0.5, 1) -- (0.5, 0);
	
&

	\draw [<->] (-0.5, 0.5) -- node [above] {RI move} (0.5, 0.5);

&

	\draw (-0.5, 0) -- (-0.5, 1) arc (0 : 90 : 0.25);
	\draw (0.5, 0) -- (0.5, 1) arc (180 : 90 : 0.25);

	\draw [draw = white, double = black, double distance between line centers = 3 pt, line width = 2.6 pt] (-0.75, 1.25) arc (90 : 270 : 0.25)
		-- (0.75, 0.75) arc (-90 : 90 : 0.25);
	
\\

	\draw [rounded corners]
		(-1.25, 0) node [left, rotate = 90] {$ab$} -- (-1.25, 1.25) -- (1.25, 1.25) -- (1.25, 0) node [left, rotate = 90] {$ab$}
		(-1, 0) node [left, rotate = 90] {$ac$} -- (-1, 1) -- (1, 1) -- (1, 0) node [left, rotate = 90] {$ac$}
		(-0.75, 0) node [left, rotate = 90] {$bc$} -- (-0.75, 0.75) -- (0.75, 0.75) -- (0.75, 0) node [left, rotate = 90] {$bc$};

&
	
	\draw [<->] (-0.5, 0.5) -- node [above] {RI move} (0.5, 0.5);

&

	\draw [rounded corners]
		(-1.25, -0.5) node [left, rotate = 90] {$ab$} -- (-1.25, 1) arc (0 : 90 : 0.25)
		(1.25, -0.5) node [left, rotate = 90] {$ab$} -- (1.25, 1) arc (180 : 90 : 0.25)
		(-1, -0.5) node [left, rotate = 90] {$ac$} -- (-1, 1) arc (0 : 90 : 0.5)
		(1, -0.5) node [left, rotate = 90] {$ac$} -- (1, 1) arc (180 : 90 : 0.5)
		(-0.75, -0.5) node [left, rotate = 90] {$bc$} -- (-0.75, 1) arc (0 : 90 : 0.75)
		(0.75, -0.5) node [left, rotate = 90] {$bc$} -- (0.75, 1) arc (180 : 90 : 0.75);
		
	\draw [rounded corners, draw = white, double = black, double distance between line centers = 3 pt, line width = 2.6 pt]
		(-1.5, 1.25) arc (90 : 270 : 0.25) -- (1.5, 0.75) arc (-90 : 90 : 0.25)
		(-1.5, 1.5) arc (90 : 270 : 0.5) -- (1.5, 0.5) arc (-90 : 90 : 0.5)
		(-1.5, 1.75) arc (90 : 270 : 0.75) -- (1.5, 0.25) arc (-90 : 90 : 0.75); 

\\
};
\end{tikzpicture}
\caption{\label{3-line-RI}The modified Reidemeister RI move for framed graphs, and its associated 3-line graph version.}
\end{figure}

At this point, we consider how the Reidemeister moves act on framed graphs, and their associated 3-line graphs. Most of the moves remain the same for framed graphs as for the unframed case. Of the moves that do change, the first case to consider is that of the RI move. The original RI move (shown in Figure \ref{Reid-move}) allowed for the removal of a single curl in the edge. With a framed graph, this is only possible when there are two such curls, where the edge curls in opposite directions, as shown in Figure \ref{3-line-RI}. One way to see the necessity of this is the following. If one looks at the 3-line graph after the RI move in Figure \ref{3-line-RI}, and focuses purely on one of the curls, this portion of the leg is equivalent to a twist of $\pm 6$. This can be seen if one repeatedly uses the (original) RI and RIII moves on the lines of the leg to undo the curl of the leg. Thus, the two curls must turn in opposite directions to cancel out these twists. As mentioned in the context of the 2-3 move, this modified RI move is necessary to ensure there is no ambiguity under the Pachner move.

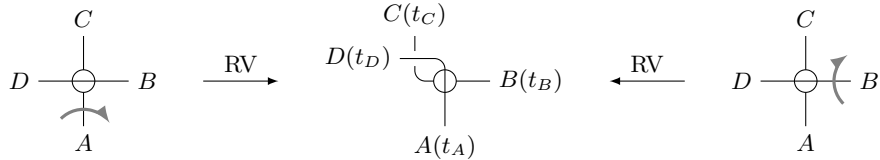
\begin{figure}[hbt]
\begin{tikzpicture}[> = latex]
\matrix[column sep = 0.5 cm]{


	\draw (0, -0.6) node [below] {$A$} -- (0, 0.6) node [above] {$C$};
	\draw [fill = white] (0, 0) circle (0.15);
	\draw (-0.6, 0) node [left] {$D$} -- (0.6, 0) node [right] {$B$};

	\draw [->, very thick, gray] (-0.3, -0.5) arc (135 : 35 : {0.3 * sqrt(2)});
	
&

	\draw [->] (-0.5, 0) -- node [above] {RV} (0.5, 0);

&


	\draw [rounded corners] (0.6, 0) node [right] {$B (t_B)$} -- (-0.4, 0) -- (-0.4, 0.2) (-0.4, 0.4) -- (-0.4, 0.6) node [above] {$C (t_C)$};
	\draw [fill = white] (0, 0) circle (0.15);
	\draw [rounded corners] (0, -0.6) node [below] {$A (t_A)$} -- (0, 0.3) -- (-0.6, 0.3) node [left] {$D (t_D)$};

	
&

	\draw [<-] (-0.5, 0) -- node [above] {RV} (0.5, 0);
	
&


	\draw (0, -0.6) node [below] {$A$} -- (0, 0.6) node [above] {$C$};
	\draw [fill = white] (0, 0) circle (0.15);
	\draw (-0.6, 0) node [left] {$D$} -- (0.6, 0) node [right] {$B$};

	\draw [->, very thick, gray] (0.5, -0.3) arc (225 : 125 : {0.3 * sqrt(2)});
	
\\
};
\end{tikzpicture}
\caption{\label{two-cases}Two RV moves that give the same final graph from the same starting vertex. The numbers $t_i$ give the change due to the vertex rotation in the twist for each edge $i = A, B, C, D$.}
\end{figure}

The more complicated move is the RV move. Here, the rotation of a given vertex around one of its edges will result in twists on all incident edges. The final twists are given incorrectly by Wan~\cite{Wan07}, so it is worthwhile to spend some time explaining the correct form of the graph move. First, we motivate the final form of the twists. Consider the case of the vertices shown in Figure \ref{two-cases}. The original vertices are on the left and right sides of the diagram, with the final vertex in the middle after the RV move. In both situations, we start from, and end with, the same graph vertex; the figure shows the two possible ways of doing this with a vertex rotation. In terms of the dual triangulation, the equal vertices mean that the initial and final dual tetrahedra are also the same, so that the faces of this tetrahedron must be rotated the same amount. This implies that the change $t_i (i = A, B, C, D)$ in the twists on the graph edges incident to the vertex also change identically. Additionally, the vertex is rotated around either edge $A$ or $B$ by the same rotation amount and direction. Therefore, these two edges must change their twist by the same amount, so $t_A = t_B$.

\begin{figure}
\begin{tikzpicture}[> = latex]
\matrix[column sep = 1 cm]{

	
	\begin{scope}[dotted]
	
		\draw (-1, -1, -1) -- (-1, -1, 1) -- (-1, 1, 1) -- (-1, 1, -1) -- cycle;
		\draw (1, -1, -1) -- (1, -1, 1) -- (1, 1, 1) -- (1, 1, -1) -- cycle;
		\draw (-1, -1, -1) -- (1, -1, -1);
		\draw (-1, -1, 1) -- (1, -1, 1);
		\draw (-1, 1, 1) -- (1, 1, 1);
		\draw (-1, 1, -1) -- (1, 1, -1);
	
	\end{scope}

	
	\draw [fill = white] (-1, -1, -1) circle (3 pt) node [left = 0.25 em] {$a'$};
	\draw [fill = white] (-1, 1, 1) circle (3 pt) node [left = 0.25 em] {$b'$};
	\draw [fill = white] (1, 1, -1) circle (3 pt) node [right = 0.25 em] {$c'$};
	\draw [fill = white] (1, -1, 1) circle (3 pt) node [right = 0.25 em] {$d'$};
	
	
	\begin{scope}[thick]
	
		\draw (-1, -1, 1) -- (-1, 1, -1) -- (1, -1, -1) -- (1, 1, 1) -- cycle;
		\draw (-1, -1, 1) -- (1, -1, -1) -- (1, 1, 1) -- (-1, 1, -1) -- cycle;
		
	\end{scope}

	
	\filldraw [gray!50, opacity = 0.5] (-1, 1, -1) -- (-1, -1, 1) -- (1, -1, -1) -- (1, 1, 1) -- cycle;

	
	\filldraw (-1, -1, 1) circle (3 pt) node [left = 0.25 em] {$b$};
	\filldraw (-1, 1, -1) circle (3 pt) node [left = 0.25 em] {$a$};
	\filldraw (1, -1, -1) circle (3 pt) node [right = 0.25 em] {$d$};
	\filldraw (1, 1, 1) circle (3 pt) node [right = 0.25 em] {$c$};
	
	
	\draw [dashed] (-2, 0, 0) node [left] {rot. axis} -- (2, 0, 0);
	\filldraw (-1, 0, 0) circle (2 pt);
	\filldraw (1, 0, 0) circle (2 pt);
	
	
	\draw [yshift = -3 cm] (135 : 0.5) node [above left] {$C$} -- (315 : 0.5) node [below right] {$A$};
	\draw [fill = white] (0, -3) circle (0.15);
	\draw [yshift = -3 cm] (45 : 0.5) node [above right] {$B$} -- (225 : 0.5) node [below left] {$D$};
	
&
	\draw [->] (-0.5, 0) -- node [above] {RV move} (0.5, 0);
&
	
	
	\filldraw [gray!50, opacity = 0.5] (-1, -1, -1) -- (-1, 1, 1) -- (1, 1, -1) -- (1, -1, 1) -- cycle;

	
	\begin{scope}[dotted]
	
		\draw (-1, -1, -1) -- (-1, -1, 1) -- (-1, 1, 1) -- (-1, 1, -1) -- cycle;
		\draw (1, -1, -1) -- (1, -1, 1) -- (1, 1, 1) -- (1, 1, -1) -- cycle;
		\draw (-1, -1, -1) -- (1, -1, -1);
		\draw (-1, -1, 1) -- (1, -1, 1);
		\draw (-1, 1, 1) -- (1, 1, 1);
		\draw (-1, 1, -1) -- (1, 1, -1);
	
	\end{scope}

	
	\begin{scope}[thick]
	
		\draw (-1, -1, -1) -- (-1, 1, 1) -- (1, -1, 1) -- (1, 1, -1) -- cycle;
		\draw (-1, -1, -1) -- (1, -1, 1) -- (1, 1, -1) -- (-1, 1, 1) -- cycle;
	
	\end{scope}


	\filldraw (-1, -1, -1) circle (3 pt) node [left = 0.25 em] {$a'$};
	\filldraw (-1, 1, 1) circle (3 pt) node [left = 0.25 em] {$b'$};
	\filldraw (1, 1, -1) circle (3 pt) node [right = 0.25 em] {$c'$};
	\filldraw (1, -1, 1) circle (3 pt) node [right = 0.25 em] {$d'$};
	
	\draw [fill = white] (-1, -1, 1) circle (3 pt) node [left = 0.25 em] {$b$};
	\draw [fill = white] (-1, 1, -1) circle (3 pt) node [left = 0.25 em] {$a$};
	\draw [fill = white] (1, -1, -1) circle (3 pt) node [right = 0.25 em] {$d$};
	\draw [fill = white] (1, 1, 1) circle (3 pt) node [right = 0.25 em] {$c$};
	
	
	\draw [yshift = -3 cm] (45 : 0.5) node [above right] {$B'$} -- (225 : 0.5) node [below left] {$C'$};
	\draw [fill = white] (0, -3) circle (0.15);
	\draw [yshift = -3 cm] (135 : 0.5) node [above left] {$D'$} -- (315 : 0.5) node [below right] {$A'$};

\\
};

\end{tikzpicture}
\caption{\label{tet-rot}Rotation of the tetrahedron about the line passing through the midpoints of the edges $ab$ and $cd$, which are the centers of the cubic faces $aa'bb'$ and $cc'dd'$, respectively. Each vertex $v$ in the left-hand picture moves to the vertex $v'$ after the rotation, in the right illustration.}
\end{figure}
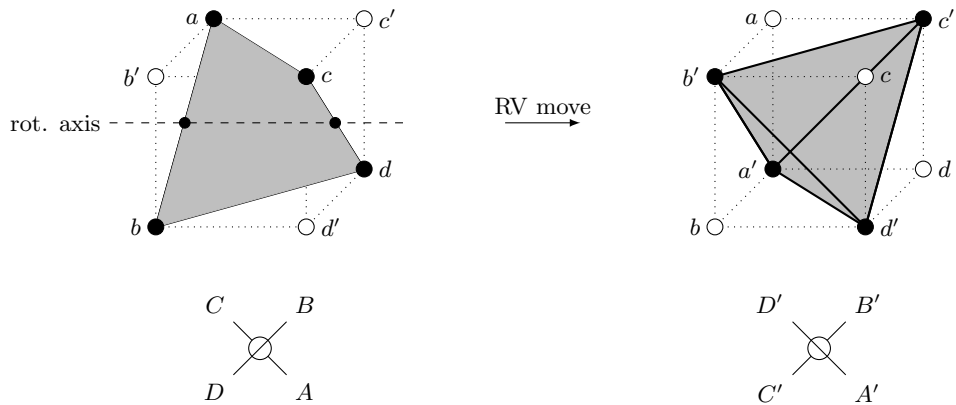

To find the actual twist values on each edge, it is useful to have a geometric picture of the vertex rotation; so we consider the tetrahedron dual to the vertex. Our method will be based on the corners of a cube, and the vertices of the dual tetrahedron will be chosen from these corners. A cube has eight corners, which we can place at the coordinates $(\pm 1, \pm 1, \pm 1)$. We shall consider these points to be adjacent if they differ only by one coordinate. The corners can then be grouped together into two sets, each of which contain four points, none of whom are adjacent to each other. If we connect these points in each set together, we obtain a tetrahedron. Thus, there are two possible tetrahedra constructible from the corners of the cube, as given in Figure \ref{tet-rot}. By taking the projection plane to be parallel to the sides $ab'cc'$ and $a'cd'd$ of the cube, we can associate a dual graph vertex to each tetrahedron, shown below the cubes. The unprimed corners of the cube form the dual tetrahedron before the RV move, and the primed points are used for the final tetrahedron. Remember that each edge for the vertex is dual to the face which does not feature the same letter; thus, for example, edge $A$ is dual to the face $bcd$.

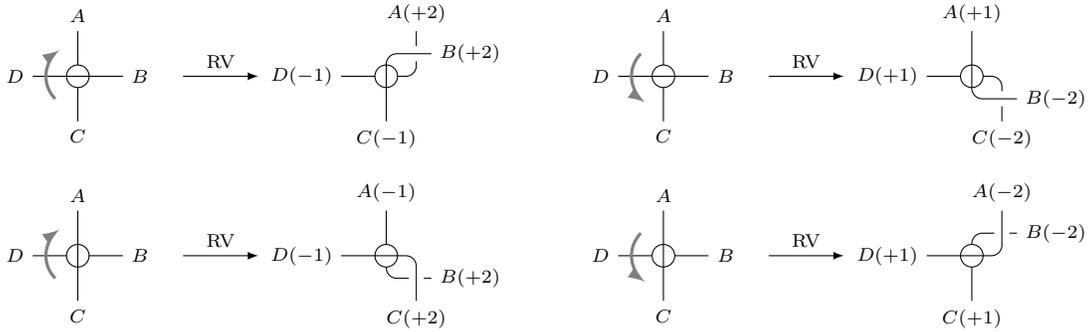
\begin{figure}[hbt]
\centering
\begin{tikzpicture}[> = latex, font = \scriptsize]
\matrix[column sep = 1 cm, row sep = 0.25 cm]{


	\draw (0, -0.6) node [below] {$C$} -- (0, 0.6) node [above] {$A$};
	\draw [fill = white] (0, 0) circle (0.15);
	\draw (-0.6, 0) node [left] {$D$} -- (0.6, 0) node [right] {$B$};

	\draw [->, very thick, gray] (-0.3, -0.3) arc (225 : 125 : {0.3 * sqrt(2)});

	\draw [->] (1.4, 0) -- node [above] {RV} (2.4, 0);


	\draw [rounded corners] (3.5, 0) node [left] {$D (-1)$} -- (4.5, 0) -- (4.5, 0.2) (4.5, 0.4) -- (4.5, 0.6) node [above] {$A (+2)$};
	\draw [fill = white] (4.1, 0) circle (0.15);
	\draw [rounded corners] (4.1, -0.6) node [below] {$C (-1)$} -- (4.1, 0.3) -- (4.7, 0.3) node [right] {$B (+2)$};

&


	\draw (0, -0.6) node [below] {$C$} -- (0, 0.6) node [above] {$A$};
	\draw [fill = white] (0, 0) circle (0.15);
	\draw (-0.6, 0) node [left] {$D$} -- (0.6, 0) node [right] {$B$};

	\draw [->, very thick, gray] (-0.3, 0.3) arc (135 : 235 : {0.3 * sqrt(2)});

	\draw [->] (1.4, 0) -- node [above] {RV} (2.4, 0);


	\draw [rounded corners] (3.5, 0) node [left] {$D (+1)$} -- (4.5, 0) -- (4.5, -0.2) (4.5, -0.4) -- (4.5, -0.6) node [below] {$C (-2)$};
	\draw [fill = white] (4.1, 0) circle (0.15);
	\draw [rounded corners] (4.1, 0.6) node [above] {$A (+1)$} -- (4.1, -0.3) -- (4.7, -0.3) node [right] {$B (-2)$};

\\


	\draw (-0.6, 0) node [left] {$D$} -- (0.6, 0) node [right] {$B$};
	\draw [fill = white] (0, 0) circle (0.15);
	\draw (0, -0.6) node [below] {$C$} -- (0, 0.6) node [above] {$A$};

	\draw [->, very thick, gray] (-0.3, -0.3) arc (225 : 125 : {0.3 * sqrt(2)});

	\draw [->] (1.4, 0) -- node [above] {RV} (2.4, 0);


	\draw [rounded corners] (4.1, 0.6) node [above] {$A (-1)$} -- (4.1, -0.3) -- (4.4, -0.3) (4.6, -0.3) -- (4.7, -0.3) node [right] {$B (+2)$};
	\draw [fill = white] (4.1, 0) circle (0.15);
	\draw [rounded corners] (3.5, 0) node [left] {$D (-1)$} -- (4.5, 0) -- (4.5, -0.6) node [below] {$C (+2)$};

&


	\draw (-0.6, 0) node [left] {$D$} -- (0.6, 0) node [right] {$B$};
	\draw [fill = white] (0, 0) circle (0.15);
	\draw (0, -0.6) node [below] {$C$} -- (0, 0.6) node [above] {$A$};

	\draw [->, very thick, gray] (-0.3, 0.3) arc (135 : 235 : {0.3 * sqrt(2)});

	\draw [->] (1.4, 0) -- node [above] {RV} (2.4, 0);


	\draw [rounded corners] (4.1, -0.6) node [below] {$C (+1)$} -- (4.1, 0.3) -- (4.4, 0.3) (4.6, 0.3) -- (4.7, 0.3) node [right] {$B (-2)$};
	\draw [fill = white] (4.1, 0) circle (0.15);
	\draw [rounded corners] (3.5, 0) node [left] {$D (+1)$} -- (4.5, 0) -- (4.5, 0.6) node [above] {$A (-2)$};

\\
};

\end{tikzpicture}
\caption{\label{RV-move-twist}The RV move for rotating a vertex around one of its incident edges, based on its current vertex state, with the appropriate twists included.}
\end{figure}

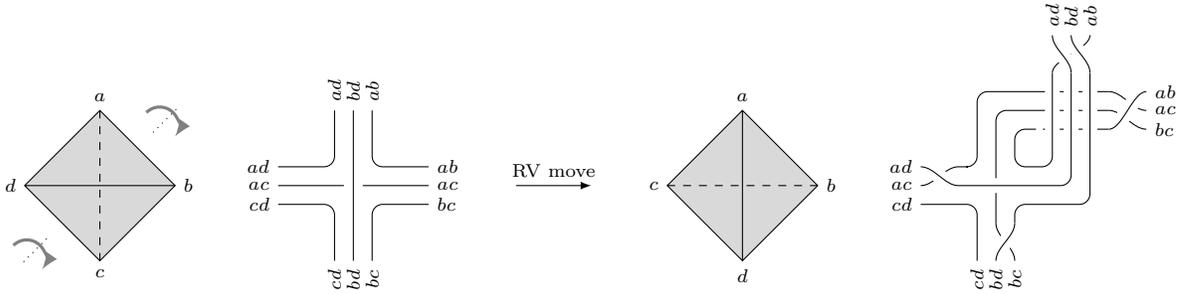
\begin{figure}[hbt]
\begin{tikzpicture}[> = latex, font = \scriptsize]
\matrix[column sep = 0.5 cm]{


	\draw [fill = gray!30] (0, 1) node [above] {$a$} -- (1, 0) node [right] {$b$} --
		(0, -1) node [below] {$c$} -- (-1, 0) node [left] {$d$} -- cycle
		(-1, 0) -- (1, 0);
	\draw [dashed] (0, 1) -- (0, -1);

	\draw [dotted] (45 : 1) -- (45 : 1.5) (225 : 1) -- (225 : 1.5);
	\draw [->, very thick, gray] ({1.125 * cos(45) + 0.25 * cos(135)}, {1.125 * sin(45) + 0.25 * sin(135)}) arc (135 : -35 : 0.25);
	\draw [->, very thick, gray] ({1.375 * cos(225) + 0.25 * cos(135)}, {1.375 * sin(225) + 0.25 * sin(135)}) arc (135 : -35 : 0.25);

	&
	

	\draw (0, -1) node [rotate = 90, left] {$bd$} -- (0, 1) node [rotate = 90, right] {$bd$}
		(-1, 0) node [left] {$ac$} -- (-0.125, 0) (0.125, 0) -- (1, 0) node [right] {$ac$};
	\draw [rounded corners] (-1, 0.25) node [left] {$ad$} -- (-0.25, 0.25) -- (-0.25, 1) node [rotate = 90, right] {$ad$}
		(-1, -0.25) node [left] {$cd$} -- (-0.25, -0.25) -- (-0.25, -1) node [rotate = 90, left] {$cd$}
		(1, 0.25) node [right] {$ab$} -- (0.25, 0.25) -- (0.25, 1) node [rotate = 90, right] {$ab$}
		(1, -0.25) node [right] {$bc$} -- (0.25, -0.25) -- (0.25, -1) node [rotate = 90, left] {$bc$};
		
	&
	
	\draw [->] (0, 0) -- node [above] {RV move} (1, 0);
	
	&


	\draw [fill = gray!30] (0, 1) node [above] {$a$} -- (1, 0) node [right] {$b$} --
		(0, -1) node [below] {$d$} -- (-1, 0) node [left] {$c$} -- cycle
		(0, 1) -- (0, -1);
	\draw [dashed] (-1, 0) -- (1, 0);
	
	&
	
	
	
	\draw (-1, 0) node [left] {$ac$} cos (-0.75, 0.125) sin (-0.5, 0.25) -- (-0.375, 0.25) [rounded corners] -- (-0.25, 0.25) -- (-0.25, 1.25)
		-- (1.5, 1.25);
	\draw (1.5, 1.25) cos (1.75, 1.125) sin (2, 1) node [right] {$ac$};
	
	
	\draw [rounded corners] (-1, -0.25) node [left] {$cd$} -- (-0.25, -0.25) -- (-0.25, -1) node [rotate = 90, left] {$cd$};


	\draw [rotate around = {-90 : (0, -0.5)}] (0, -0.5) cos (0.25, -0.375) sin (0.5, -0.25) node [rotate = 90, left] {$bc$};
	\draw [rounded corners] (0, -0.5) -- (0, 1) -- (1.5, 1);
	\draw (1.5, 1) cos (1.75, 0.875) sin (2, 0.75) node [right] {$bc$};
	
	
	\draw [rotate around = {90 : (0.75, 1.5)}] (0.75, 1.5) cos (1, 1.25) sin (1.25, 1) node [rotate = 90, right] {$ab$};
	\draw (1.5, 0.75) -- (0.625, 0.75);
	
	
	\draw [draw = white, double = black, double distance between line centers = 3 pt, line width = 2.6 pt, rotate around = {-90 : (0.25, -0.5)}]
		(0.25, -0.5) cos (0.5, -0.625) sin (0.75, -0.75);
	\node at (0, -1) [rotate = 90, left] {$bd$};
	\draw [draw = white, double = black, double distance between line centers = 3 pt, line width = 2.6 pt, rounded corners] (0.25, -0.5) -- (0.25, -0.25)
		-- (1.25, -0.25) -- (1.25, 1.5);
	\draw [draw = white, double = black, double distance between line centers = 3 pt, line width = 2.6 pt, rotate around = {90 : (1.25, 1.5)}]
		(1.25, 1.5) cos (1.5, 1.625) sin (1.75, 1.75);
	\node at (1, 2) [rotate = 90, right] {$bd$};
		
	
	\node at (-1, 0.25) [left] {$ad$};
	\draw [draw = white, double = black, double distance between line centers = 3 pt, line width = 2.6 pt] (-1, 0.25) cos (-0.75, 0.125) sin (-0.5, 0)
		-- (0.5, 0) [rounded corners] -- (1, 0) -- (1, 1.5);
	\draw [draw = white, double = black, double distance between line centers = 3 pt, line width = 2.6 pt, rotate around = {90 : (1, 1.5)}]
		(1, 1.5) cos (1.25, 1.625) sin (1.5, 1.75);
	\node at (0.75, 2) [rotate = 90, right] {$ad$};
	
	
	\node at (2, 1.25) [right] {$ab$};
	\draw [draw = white, double = black, double distance between line centers = 3 pt, line width = 2.6 pt] (0.75, 1.5) -- (0.75, 0.8175)
		[rounded corners] -- (0.75, 0.25) -- (0.25, 0.25) -- (0.25, 0.75) -- (0.533, 0.75);
	\draw [draw = white, double = black, double distance between line centers = 3 pt, line width = 2.6 pt] (2, 1.25) cos (1.75, 1) sin (1.5, 0.75);

\\
};

\end{tikzpicture}
\caption{\label{3-line-RV}RV move for a vertex, and the associated 3-line graphs.}
\end{figure}

Now we consider the rotation of the tetrahedron around the axis shown, which passes through the centers of the lines $ab$ and $cd$. Again, this is done so that each tetrahedral vertex moves from the unprimed to the primed corner of the cube; this is the necessary rotation to take one vertex state to the other. The amount of rotation for each tetrahedral face gives the corresponding twist for each dual graph edge. Recall that a counterclockwise rotation, as seen around the normal vector pointing towards the tetrahedron center, is defined as a positive rotation. Looking towards the center of the cube, the face $bcd$ is rotated $\pi/3$ radians counterclockwise -- the triangular face originally has a single vertex $c$ at the top, and an edge $bd$ along the bottom; afterwards, edge $b'c'$ is along the top, and vertex $d'$ points downward. A similar process happens for face $acd$, starting with the single vertex $d$ pointing downward, and rotating to get the edge $a'd'$ along the bottom. With identical rotations of the faces, the dual edges $A$ and $B$ have twists of $+1$ after the RV move. The other two faces get twists of $-2\pi/3$. For example, the face $abd$ goes from having the vertex $a$ pointing upwards to the same vertex on the lower left corner. Thus, after the RV move the dual edges $C$ and $D$ have twists of $-2$. By going through the various possible rotations of the dual tetrahedron in this manner, we can find the resulting twists on the edges incident to the vertex. These are shown in Figure \ref{RV-move-twist}. From the twists on the graph edges, we can see the effect on the associated 3-line graph, with one case given in Figure \ref{3-line-RV}. As in the situation using the cube to define the two tetrahedra in Figure \ref{tet-rot}, this shows the effect of a rotation around the midpoints of the edges $ab$ and $cd$ of the dual tetrahedron. Thus, the final 3-line graph is the result of a $+1$ twist around either the $A$ or $B$ edges; all other situations can be obtained from this one by the appropriate rotation or reflection.

\subsection{An invariant of framed 4-regular graphs}
\label{invariant}

Up to now, the 3-line graph has been a device to keep track of the gluings in the triangulation dual to the graph. However, we can also consider it as exactly what it looks like -- namely, the diagram for a link. This associates a link with each knotted 4-regular graph. The utility of this link will be that it gives rise to an invariant associated with the original graph, not only under the graph Reidemeister moves but also the Pachner moves as well. We now motivate this statement.

First, look at the 3-line graph versions of the RI (Figure \ref{3-line-RI}) and RV (Figure \ref{3-line-RV}) moves. If these are considered as acting on a link, then one can use the Reidemeister moves on the lines as edges themselves; note that this includes the original RI move for unframed edges. Then, with these graph moves acting on the link, it is not difficult to show the links coming from the 3-line graph remain the same before and after each equivalence move. Althought the other three Reidemeister moves on graphs have not been shown in a 3-line graph version, the same equivalence holds for them as well. This proves the 3-line graph, taken as a link, is invariant under the graph Reidemeister moves. Note this means any graph diagram of a knotted graph gives the same link.

However, we can go further, and find an invariant under the graph Pachner moves as well. Unfortunately, this is not simply the 3-line graph taken as a link. The reason is that both the 1-4 and 2-3 moves introduce additional unknot components to the link. Recall that each line in the 3-line graph corresponds to an edge in the dual triangulation. Looking back at the Pachner moves acting on the dual tetrahedra, shown in Figure \ref{Pach-tet}, the 2-3 move introduces one new dual edge, while the 1-4 move creates four new dual edges. Thus, in Figure \ref{1-4-twist}, one can see there are four additional unknots after the 1-4 move -- two above the $ac$ line, and two below the $bd$ line. In Figure \ref{2-3-twist}, there is a single new unknot, given by the middle strand in the loop around the center, after the 2-3 move. All of these created unknots are unlinked with the rest of the link. On the flip side, the 3-2 move deletes one unlinked unknot, while the 4-1 move removes four. Therefore, the number of unknot components of the link associated with a knotted graph does not remain constant after a Pachner move is performed on the original graph.

This being said, the rest of the link is equivalent after the Pachner move. In particular, if the added unknots are ignored in Figures \ref{2-3-twist} and \ref{1-4-twist}, the link coming from the 3-line graph is the same before and after each move. Again, this is shown by using the knot Reidemeister moves on the link. Thus, it is useful to define the link $L(G)$ as the link obtained from the 3-line graph corresponding to the graph $G$, where all unknotted unknot components are removed. Based on this discussion above, we have now shown the following.

\begin{lemma}
	The link $L(G)$ obtained from the 3-line graph of a graph diagram of a graph $G$ is a topological invariant under both the Reidemeister moves and the Pachner moves on the original graph diagram.
\end{lemma}

Unlike the invariants of knotted 4-regular graphs introduced in Paper I, this new invariant $L(G)$ is preserved under the graph Pachner moves, so we can now look at the behavior of graphs under these moves. Most important is the question of which other graphs can be reached from one starting graph by a finite set of Pachner moves. Remember that for the dual triangulations, the Pachner moves were sufficient to go between any two triangulations of the same manifold. We can restate this theorem in a manner that will aid us later for the knotted graph case. For a choice of manifold $M$, Burton~\cite{Bur11} defines the {\it Pachner graph} ${\mathscr P} (M)$ as follows. The nodes of the graph are generalized triangulations of the manifold $M$, up to isomorphisms of the tetrahedra labeling; this means triangulations with a different number of tetrahedra, or different gluing structure, comprise different nodes. There is an edge between two nodes if there is a Pachner move that takes one triangulation to the other. Phrased in this manner, the theorem that any two triangulations of the same manifold are isomorphic under a finite sequence of Pachner moves is the same as saying the Pachner graph ${\mathscr P} (M)$ is connected.

For those knotted graphs where we can define a 3-line graph without issue -- i.e. the edge twists match the states of the vertices the graph edge is incident to -- then, in many cases, a manifold can be associated with the graph. There are additional conditions on the dual edge identifications to give a manifold from the 3-line graph, given in more detail by Burton~\cite{Bur11}. However, even with these possible obstructions, there are many spatial graphs whose dual is a triangulation of some manifold. This gives a natural classification of knotted graphs, and allows us to define a graph ${\mathscr P}_G (M)$ for knotted 4-regular graphs, analogous to the Pachner graph for triangulations. The nodes of ${\mathscr P}_G (M)$ are now graphs dual to the triangulation for the manifold $M$, and its edges are the graph analogues of the Pachner moves. Note that all graph diagrams for the same graph are included in the same node, e.g. any two graph diagrams joined by a finite list of Reidemester moves. The expectation is that the different embeddings of graphs in ${\mathscr P}_G (M)$ for a fixed $M$ prevent one from moving between any two graphs in the same ${\mathscr P}_G (M)$; in other words, the graph ${\mathscr P}_G (M)$ does not consist of a single connected component. We now show this is true, for the case of the 3-sphere $S^3$.

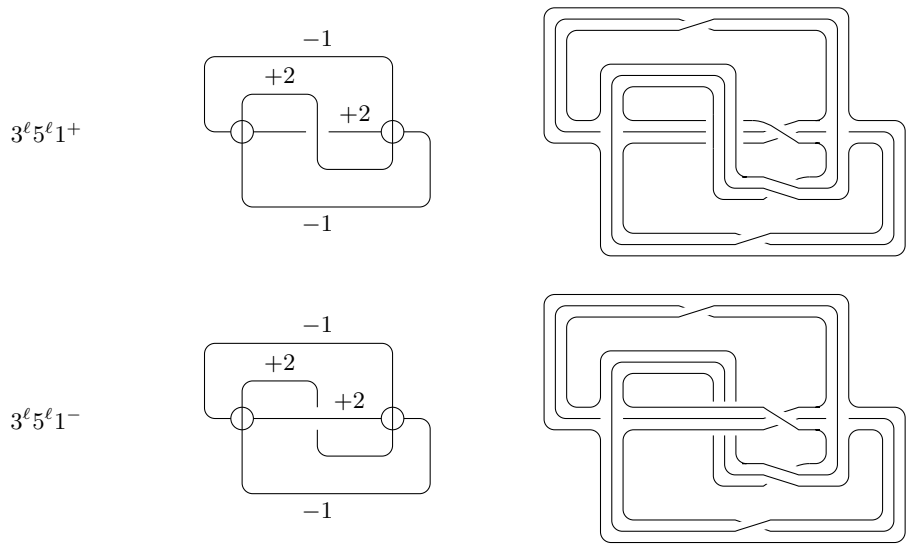
\begin{figure}[hbt]
\begin{tikzpicture}
\matrix[column sep = 1.5 cm, row sep = 0.5 cm]{

	\node at (0, 0) {$3^\ell 5^\ell 1^+$};

&


	\draw (-1, 0) circle (0.15);
	\draw (1, 0) circle (0.15);
	
	\draw [rounded corners] (-0.15, 0) -- (-0.85, 0) (-1.15, 0) -- (-1.5, 0) -- (-1.5, 1) -- node [above, pos = 0.6] {$-1$} (1, 1) -- (1, -0.5)
		-- (0, -0.5) -- (0, 0.5) -- node [above] {$+2$} (-1, 0.5) -- (-1, -1) -- node [below, pos = 0.4] {$-1$} (1.5, -1) -- (1.5, 0) -- (1.15, 0)
		(0.85, 0) -- node [above] {$+2$} (0.15, 0);

&
	
	
	\begin{scope}[scale = 0.75]
	
	\draw [rounded corners]
		(1.5, -0.8) cos (1, -1) sin (0.5, -1.2) -- (-0.2, -1.2) -- (-0.2, -0.3) (-0.2, 0.3) -- (-0.2, 0.8) -- (-1.8, 0.8) -- (-1.8, 0.2)
			-- (0.5, 0.2)	
		(-2.2, 0) -- (-3, 0) -- (-3, 2) -- (-1, 2) cos (-0.5, 1.9) sin (0, 1.8) -- (1.8, 1.8) -- (1.8, 0.2) --
			(1.5, 0.2) cos (1, 0.1) sin (0.5, 0) -- (-1.8, 0)			
		(1, -2.2) -- (3.2, -2.2) -- (3.2, 0.2) -- (2.2, 0.2) -- (2.2, 2.2) -- (-3.2, 2.2) -- (-3.2, -0.2) -- (-2.2, -0.2) -- (-2.2, -2.2) -- cycle 
		(1.8, 0) -- (1.5, 0) cos (1, -0.1) sin (0.5, -0.2) -- (-1.8, -0.2) -- (-1.8, -1.8) -- (0, -1.8) cos (0.5, -1.9) sin (1, -2) -- (3, -2) -- (3, 0)
			-- (2.2, 0) 
		;
	
	\draw [rounded corners, draw = white, double = black, double distance between line centers = 3 pt, line width = 2.6 pt]
		(0, 0.3) -- (0, 1) -- (-2, 1) -- (-2, -2) -- (0, -2) cos (0.5, -1.9) sin (1, -1.8) -- (2.8, -1.8)
		-- (2.8, -0.2) -- (2.2, -0.2) -- (2.2, -1.2) -- (1.5, -1.2) cos (1, -1.1) sin (0.5, -1) -- (0, -1) -- cycle			
		(0.2, 0.3) -- (0.2, 1.2) -- (-2.2, 1.2) -- (-2.2, 0.2) -- (-2.8, 0.2) -- (-2.8, 1.8) -- (-1, 1.8) cos (-0.5, 1.9) sin (0, 2) -- (2, 2)
		-- (2, -1) -- (1.5, -1) cos (1, -0.9) sin (0.5, -0.8) -- (0.2, -0.8) -- cycle 
		(-0.2, -0.3) -- (-0.2, 0.3) (0.5, 0.2) cos (1, 0) sin (1.5, -0.2) -- (1.8, -0.2)
		-- (1.8, -0.8) -- (1.5, -0.8) 
		;
		
	\end{scope}

\\

	\node at (0, 0) {$3^\ell 5^\ell 1^-$};

&


	\draw (-1, 0) circle (0.15);
	\draw (1, 0) circle (0.15);
	
	\draw [rounded corners] (0, 0.15) -- (0, 0.5) -- node [above] {$+2$} (-1, 0.5) -- (-1, -1) -- node [below, pos = 0.4] {$-1$} (1.5, -1) --
		(1.5, 0) -- (1.15, 0) (0.85, 0) -- node [above, pos = 0.25] {$+2$} (-0.85, 0) (-1.15, 0) -- (-1.5, 0) -- (-1.5, 1) --
		node [above, pos = 0.6] {$-1$} (1, 1) -- (1, -0.5) -- (0, -0.5) -- (0, -0.15);
	
&
	
	
	\begin{scope}[scale = 0.75]
	
	\draw [rounded corners]
		(1.5, -0.8) cos (1, -1) sin (0.5, -1.2) -- (-0.2, -1.2) -- (-0.2, -0.3)	
		(-2.2, 0) -- (-3, 0) -- (-3, 2) -- (-1, 2) cos (-0.5, 1.9) sin (0, 1.8) -- (1.8, 1.8) -- (1.8, 0.2) --
			(1.5, 0.2) cos (1, 0.1) sin (0.5, 0) -- (-1.8, 0)			
		(1, -2.2) -- (3.2, -2.2) -- (3.2, 0.2) -- (2.2, 0.2) -- (2.2, 2.2) -- (-3.2, 2.2) -- (-3.2, -0.2) -- (-2.2, -0.2) -- (-2.2, -2.2) -- cycle 
		(1.8, 0) -- (1.5, 0) cos (1, -0.1) sin (0.5, -0.2) -- (-1.8, -0.2) -- (-1.8, -1.8) -- (0, -1.8) cos (0.5, -1.9) sin (1, -2) -- (3, -2) -- (3, 0)
			-- (2.2, 0) 
		;
	
	\draw [rounded corners, draw = white, double = black, double distance between line centers = 3 pt, line width = 2.6 pt]
		(0, 0.3) -- (0, 1) -- (-2, 1) -- (-2, -2) -- (0, -2) cos (0.5, -1.9) sin (1, -1.8) -- (2.8, -1.8)
		-- (2.8, -0.2) -- (2.2, -0.2) -- (2.2, -1.2) -- (1.5, -1.2) cos (1, -1.1) sin (0.5, -1) -- (0, -1) -- (0, -0.3)			
		(0.2, 0.3) -- (0.2, 1.2) -- (-2.2, 1.2) -- (-2.2, 0.2) -- (-2.8, 0.2) -- (-2.8, 1.8) -- (-1, 1.8) cos (-0.5, 1.9) sin (0, 2) -- (2, 2)
		-- (2, -1) -- (1.5, -1) cos (1, -0.9) sin (0.5, -0.8) -- (0.2, -0.8) -- (0.2, -0.3) 
		(-0.2, 0.3) -- (-0.2, 0.8) -- (-1.8, 0.8) -- (-1.8, 0.2) -- (0.5, 0.2) cos (1, 0) sin (1.5, -0.2) -- (1.8, -0.2)
		-- (1.8, -0.8) -- (1.5, -0.8) 
		;
		
	\end{scope}
	
\\
};
\end{tikzpicture}
\caption{\label{3-sph-graphs}Graph diagrams for the sequences $3^\ell 5^\ell 1^+$ and $3^\ell 5^\ell 1^-$, and their corresponding 3-line graphs. The only difference between the graph diagrams is the type of crossing in the middle of the diagram.}
\end{figure}

Specifically, we look at the graphs $3^\ell 5^\ell 1^-$ and $3^\ell 5^\ell 1^+$; the details of how to show these graphs are associated with the triangulation of a 3-sphere are given in the Appendix. The duality of the graphs with a triangulation of $S^3$ requires the edge twists shown in Figure \ref{3-sph-graphs}, where both the graph diagrams and the associated 3-line graphs, are given for each sequence. Both of the 3-line graphs have six distinct loops, corresponding to the six distinct edges in the dual triangulation. If the two graphs are connected in the Pachner graph ${\mathscr P}_G (S^3)$, then they must have the same modified link $L(G)$; we demonstrate that this is not the case. First, look at the 3-line graph for the sequence $3^\ell 5^\ell 1^+$. Its 3-line graph, shown in the top row of Figure \ref{3-sph-graphs}, is composed of six unlinked unknots. The easiest way to see this is to repeatedly locating loops that have no crossings at all, or else all of its crossings are the same type, and then removing these loops to simplify the link. This shows that $L(3^\ell 5^\ell 1^+) = \varnothing$. On the other hand, the 3-line graph for the $3^\ell 5^\ell 1^-$ sequence is a more complicated link. In fact, only one unlinked unknot is present, with the rest of the 3-line graph forming a link of five components. Since the unknots involved in the latter link have non-zero linking numbers, $L(3^\ell 5^\ell 1^-) \neq \varnothing$; this can also be verified using the Jones polynomial. This implies the following result.

\begin{proposition}
	The Pachner graph ${\mathscr P}_G (S^3)$ is not a connected graph.
\end{proposition}
The relative ease of finding two graphs, both dual to the same triangulation of the 3-sphere, but not connected by any finite sequence of Pachner moves, leads to the conjecture that this is a general condition. In other words, it is likely that the Pachner graph ${\mathscr P}_G (M)$ is not a connected graph for any manifold $M$.


\section{An extension of 3-line graphs}
\label{gen-braid}

Up to this point, we have considered twists on edges such that this information is compatible with the dual triangulation. In other words, the twist on the graph edge can be unambiguously related to the gluing of the dual faces of tetrahedra. From these edge twists, a 3-line graph can be associated with the original graph. However, the presence of a single twist where this is not the case derails this process, since it is not obvious how to go from the graph to the 3-line graph in this situation. This issue is now explored. Suppose we are given a graph with generic twists on the edges. Where the twists are compatible with the dual tetrahedra, we define the 3-line graph as before. For the edges were this is not possible, for the moment we place a marker on the 3-line graph to represent those twists, with the possible choices shown in Figure \ref{improp-mark}. In that figure, the base cases for incompatible twists are given; all other such twists can be built from these objects, in combination with the dictionary given in Figures \ref{like-dict} and \ref{unlike-dict}. 3-line graphs where at least one of these markers has been placed is called an {\it extended} 3-line graph. The objects $\hat{\alpha}$ and $\hat{\beta}$ are for two vertices in the same state. When the two vertices are not in the same state, it is important to show which vertex state is on which side of the edge. Thus, for $|\hat{\omega}$, the $\oplus$ vertex is to the left of the edge, while for $\hat{\omega}|$, it is to the right. Note that one can go from one of the latter two diagrams to the other by a $\pi$ rotation of the diagram. Similarly, the $\hat{\beta}$ diagram is simply the $\hat{\alpha}$ picture, seen from the opposite side of the projecting plane. We now study the properties that these algebraic objects must have to give a consistent extension of 3-line graphs for all possible edge twists.

\begin{figure}[hbt]
\begin{tikzpicture}[> = latex]
\matrix[column sep = 1 cm, row sep = 0.5 cm]{

	\draw (-1.25, 0) -- node [above] {$+1$} (1.25, 0);

	\draw [fill = white] (-0.75, 0) circle (0.15);
	\draw [fill = white] (0.75, 0) circle (0.15);
	
	\draw (-0.75, 0.5) -- (-0.75, -0.5);
	\draw (0.75, 0.5) -- (0.75, -0.5);
	
&

	\draw [<->] (-0.5, 0) -- (0.5, 0);

&

	\draw (-1.5, 0) -- (1.5, 0);
	
	\begin{scope}[rounded corners]
	
		\draw (-1.25, -0.5) -- (-1.25, -0.25) -- (-1.5, -0.25);
		\draw (-1.25, 0.5) -- (-1.25, 0.25) -- (-1.5, 0.25);
		
		\draw (-0.75, -0.5) -- (-0.75, -0.25) -- (0.75, -0.25) -- (0.75, -0.5);
		\draw (-0.75, 0.5) -- (-0.75, 0.25) -- (0.75, 0.25) -- (0.75, 0.5);
		
		\draw (1.25, -0.5) -- (1.25, -0.25) -- (1.5, -0.25);
		\draw (1.25, 0.5) -- (1.25, 0.25) -- (1.5, 0.25);
	
	\end{scope}

	\draw [draw = white, double = black, double distance between line centers = 3 pt, line width = 2.6 pt] (-1, -0.5) -- (-1, 0.5);
	\draw [draw = white, double = black, double distance between line centers = 3 pt, line width = 2.6 pt] (1, -0.5) -- (1, 0.5);
	
	\draw [fill = gray!50] (-0.25, -0.5) rectangle (0.25, 0.5);
	\node at (0, 0) {$\hat{\alpha}$};

\\	
	
	\draw (-0.75, 0.5) -- (-0.75, -0.5);
	\draw (0.75, 0.5) -- (0.75, -0.5);

	\draw [fill = white] (-0.75, 0) circle (0.15);
	\draw [fill = white] (0.75, 0) circle (0.15);

	\draw (-1.25, 0) -- node [above] {$+1$} (1.25, 0);
	
&

	\draw [<->] (-0.5, 0) -- (0.5, 0);

&

	\draw (-1, -0.5) -- (-1, 0.5);
	\draw (1, -0.5) -- (1, 0.5);
	
	\begin{scope}[rounded corners]
	
		\draw (-1.25, -0.5) -- (-1.25, -0.25) -- (-1.5, -0.25);
		\draw (-1.25, 0.5) -- (-1.25, 0.25) -- (-1.5, 0.25);
		
		\draw (-0.75, -0.5) -- (-0.75, -0.25) -- (0.75, -0.25) -- (0.75, -0.5);
		\draw (-0.75, 0.5) -- (-0.75, 0.25) -- (0.75, 0.25) -- (0.75, 0.5);
		
		\draw (1.25, -0.5) -- (1.25, -0.25) -- (1.5, -0.25);
		\draw (1.25, 0.5) -- (1.25, 0.25) -- (1.5, 0.25);
	
	\end{scope}

	\draw [draw = white, double = black, double distance between line centers = 3 pt, line width = 2.6 pt] (-1.5, 0) -- (1.5, 0);
	
	\draw [fill = gray!50] (-0.25, -0.5) rectangle (0.25, 0.5);
	\node at (0, 0) {$\hat{\beta}$};
	
\\

	\draw (-1.25, 0) -- (0, 0) node [above] {$0$};
	\draw (0.75, 0.5) -- (0.75, -0.5);

	\draw [fill = white] (-0.75, 0) circle (0.15);
	\draw [fill = white] (0.75, 0) circle (0.15);
	
	\draw (-0.75, 0.5) -- (-0.75, -0.5);
	
	\draw (0, 0) -- (1.25, 0);
	
&

	\draw [<->] (-0.5, 0) -- (0.5, 0);

&
	\draw (1, -0.5) -- (1, 0.5);
	\draw [draw = white, double = black, double distance between line centers = 3 pt, line width = 2.6 pt] (-1.5, 0) -- (1.5, 0);
	\draw [draw = white, double = black, double distance between line centers = 3 pt, line width = 2.6 pt]  (-1, -0.5) -- (-1, 0.5);
	
	\begin{scope}[rounded corners]
	
		\draw (-1.25, -0.5) -- (-1.25, -0.25) -- (-1.5, -0.25);
		\draw (-1.25, 0.5) -- (-1.25, 0.25) -- (-1.5, 0.25);
		
		\draw (-0.75, -0.5) -- (-0.75, -0.25) -- (0.75, -0.25) -- (0.75, -0.5);
		\draw (-0.75, 0.5) -- (-0.75, 0.25) -- (0.75, 0.25) -- (0.75, 0.5);
		
		\draw (1.25, -0.5) -- (1.25, -0.25) -- (1.5, -0.25);
		\draw (1.25, 0.5) -- (1.25, 0.25) -- (1.5, 0.25);
	
	\end{scope}

	
	\draw [fill = gray!50] (-0.25, -0.5) rectangle (0.25, 0.5);
	\node at (0, 0) {$|\hat{\omega}$};

\\

	\draw (1.25, 0) -- (0, 0) node [above] {$0$};
	\draw (-0.75, 0.5) -- (-0.75, -0.5);

	\draw [fill = white] (-0.75, 0) circle (0.15);
	\draw [fill = white] (0.75, 0) circle (0.15);
	
	\draw (0.75, 0.5) -- (0.75, -0.5);
	
	\draw (0, 0) -- (-1.25, 0);
	
&

	\draw [<->] (-0.5, 0) -- (0.5, 0);

&

	\draw (-1, -0.5) -- (-1, 0.5);
	\draw [draw = white, double = black, double distance between line centers = 3 pt, line width = 2.6 pt] (-1.5, 0) -- (1.5, 0);
	\draw [draw = white, double = black, double distance between line centers = 3 pt, line width = 2.6 pt]  (1, -0.5) -- (1, 0.5);
	
	\begin{scope}[rounded corners]
	
		\draw (-1.25, -0.5) -- (-1.25, -0.25) -- (-1.5, -0.25);
		\draw (-1.25, 0.5) -- (-1.25, 0.25) -- (-1.5, 0.25);
		
		\draw (-0.75, -0.5) -- (-0.75, -0.25) -- (0.75, -0.25) -- (0.75, -0.5);
		\draw (-0.75, 0.5) -- (-0.75, 0.25) -- (0.75, 0.25) -- (0.75, 0.5);
		
		\draw (1.25, -0.5) -- (1.25, -0.25) -- (1.5, -0.25);
		\draw (1.25, 0.5) -- (1.25, 0.25) -- (1.5, 0.25);
	
	\end{scope}

	\draw [fill = gray!50] (-0.25, -0.5) rectangle (0.25, 0.5);
	\node at (0, 0) {$\hat{\omega}|$};

\\
};
\end{tikzpicture}
\caption{\label{improp-mark}The improper twists can be replaced by algebraic objects $\hat{\alpha}, \hat{\beta}, |\hat{\omega},$ and $\hat{\omega}|$. Note that $\hat{\alpha}$ and $\hat{\beta}$ show the same situation from opposite sides of the projecting plane, while $|\hat{\omega}$ and $\hat{\omega}|$ are related by a $\pi$ rotation of the projection plane.}
\end{figure}
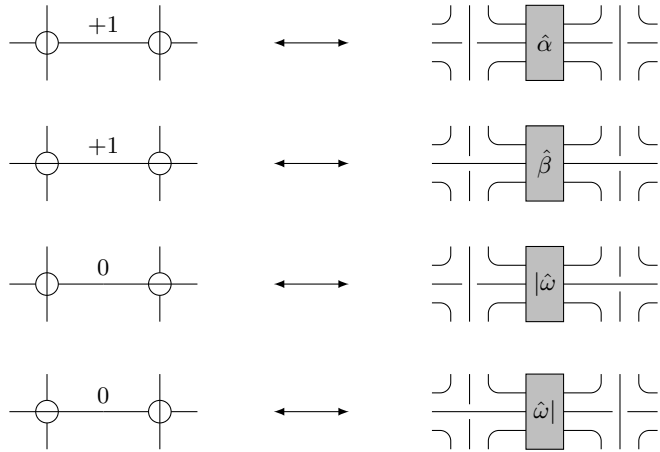

\begin{figure}[hbt]
\begin{tikzpicture}[> = latex]
\matrix[column sep = 0.75 cm, row sep = 1 cm]{

	\draw (-1.25, 0) -- (0, 0) node [above] {$0$};
	\draw (0.75, 0.5) -- (0.75, -0.5);

	\draw [fill = white] (-0.75, 0) circle (0.15);
	\draw [fill = white] (0.75, 0) circle (0.15);
	
	\draw (-0.75, 0.5) -- (-0.75, -0.5);
	
	\draw (0, 0) -- (1.25, 0);

&
	\draw (1, -0.5) -- (1, 0.5);
	\draw [draw = white, double = black, double distance between line centers = 3 pt, line width = 2.6 pt] (-1.5, 0) -- (1.5, 0);
	\draw [draw = white, double = black, double distance between line centers = 3 pt, line width = 2.6 pt]  (-1, -0.5) -- (-1, 0.5);
	
	\begin{scope}[rounded corners]
	
		\draw (-1.25, -0.5) -- (-1.25, -0.25) -- (-1.5, -0.25);
		\draw (-1.25, 0.5) -- (-1.25, 0.25) -- (-1.5, 0.25);
		
		\draw (-0.75, -0.5) -- (-0.75, -0.25) -- (0.75, -0.25) -- (0.75, -0.5);
		\draw (-0.75, 0.5) -- (-0.75, 0.25) -- (0.75, 0.25) -- (0.75, 0.5);
		
		\draw (1.25, -0.5) -- (1.25, -0.25) -- (1.5, -0.25);
		\draw (1.25, 0.5) -- (1.25, 0.25) -- (1.5, 0.25);
	
	\end{scope}

	
	\draw [fill = gray!50] (-0.25, -0.5) rectangle (0.25, 0.5);
	\node at (0, 0) {$|\hat{\omega}$};

\\
	
	\draw (1.5, 0.5) -- (1.5, -0.5);
	\draw [fill = white] (1.5, 0) circle (0.15);

	\draw (-2, 0) -- node [above, pos = 0.625] {$+1$} (0, 0) -- node [above, pos = 0.375] {$-1$} (2, 0);

	\draw [fill = white] (-1.5, 0) circle (0.15);
	\draw [gray, fill = white] (0, 0) circle (0.15);
	
	\draw (-1.5, 0.5) -- (-1.5, -0.5);
	\draw [gray] (0, -0.5) -- (0, 0.5);
	
&
	
	\draw (-2, 0) -- (-1.625, 0) (-1.375, 0) -- (0, 0) (1, 0) -- (2, 0);
	
	\draw (-1.5, 0.5) -- (-1.5, -0.5);
	\draw (1.5, 0.5) -- (1.5, 0.125) (1.5, -0.125) -- (1.5, -0.5);
	
	\begin{scope}[rounded corners]
	
		\draw (-2, 0.25) -- (-1.75, 0.25) -- (-1.75, 0.5);
		\draw (-2, -0.25) -- (-1.75, -0.25) -- (-1.75, -0.5);
		
		\draw (-1.25, 0.5) -- (-1.25, 0.25) -- (1.25, 0.25) -- (1.25, 0.5);
		\draw (-1.25, -0.5) -- (-1.25, -0.25)  -- (0, -0.25) (1, -0.25) -- (1.25, -0.25) -- (1.25, -0.5);
	
		\draw (2, 0.25) -- (1.75, 0.25) -- (1.75, 0.5);
		\draw (2, -0.25) -- (1.75, -0.25) -- (1.75, -0.5);
	
	\end{scope}
	
	\draw [fill = gray!50] (-0.875, -0.5) rectangle (-0.375, 0.5);
	\node at (-0.625, 0) {$\hat{\alpha}$};
	
	\draw (0, 0) cos (0.5, -0.125) sin (1, -0.25);
	\draw [draw = white, double = black, double distance between line centers = 3 pt, line width = 2.6 pt] (0, -0.25) cos (0.5, -0.125) sin (1, 0);
	
	
	\draw [gray, dashed] (0, 0.5) -- (0, -0.5);

\\
	
	\draw (1.5, 0.5) -- (1.5, -0.5);
	\draw [fill = white] (1.5, 0) circle (0.15);

	\draw (-2, 0) -- node [above, pos = 0.625] {$-1$} (0, 0) -- node [above, pos = 0.375] {$+1$} (2, 0);

	\draw [fill = white] (-1.5, 0) circle (0.15);
	\draw (-1.5, 0.5) -- (-1.5, -0.5);
	
	\draw [gray] (0, 0.5) -- (0, -0.5);
	\draw [gray, fill = white] (0, 0) circle (0.15);
	\draw [gray] (-0.15, 0) -- (0.15, 0);
	
&
	
	\draw (-2, 0) -- (-1.625, 0) (-1.375, 0) -- (-1, 0) (0, 0) -- (0.375, 0) (0.875, 0) -- (2, 0);
	
	\draw (-1.5, 0.5) -- (-1.5, -0.5);
	\draw (1.5, 0.5) -- (1.5, 0.125) (1.5, -0.125) -- (1.5, -0.5);
	
	\begin{scope}[rounded corners]
	
		\draw (-2, 0.25) -- (-1.75, 0.25) -- (-1.75, 0.5);
		\draw (-2, -0.25) -- (-1.75, -0.25) -- (-1.75, -0.5);
		
		\draw (-1.25, 0.5) -- (-1.25, 0.25) -- (1.25, 0.25) -- (1.25, 0.5);
		\draw (-1.25, -0.5) -- (-1.25, -0.25) -- (-1, -0.25) (0, -0.25) -- (1.25, -0.25) -- (1.25, -0.5);
	
		\draw (2, 0.25) -- (1.75, 0.25) -- (1.75, 0.5);
		\draw (2, -0.25) -- (1.75, -0.25) -- (1.75, -0.5);
	
	\end{scope}
	
	\draw [fill = gray!50] (0.875, -0.5) rectangle (0.375, 0.5);
	\node at (0.625, 0) {$\hat{\beta}$};
	
	\draw (-1, 0) cos (-0.5, -0.125) sin (0, -0.25);
	\draw [draw = white, double = black, double distance between line centers = 3 pt, line width = 2.6 pt] (-1, -0.25) cos (-0.5, -0.125) sin (0, 0);
	
	
	\draw [gray, dashed] (0, 0.5) -- (0, -0.5);

\\
};
\end{tikzpicture}
\caption{\label{trip-rel}A relationship between the three algebraic objects $\hat{\alpha}, \hat{\beta},$ and $|\hat{\omega}$; all three of the 3-line subgraphs shown to the right must be equivalent.}
\end{figure}
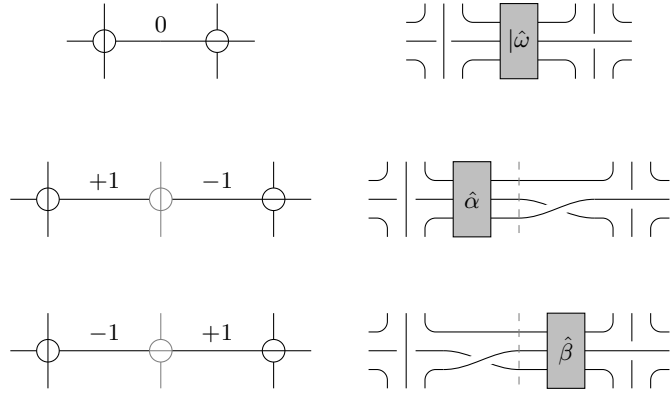

\begin{figure}[hbt]
\begin{tikzpicture}[> = latex]
\matrix[column sep = 0.75 cm, row sep = 1 cm]{

	\draw (-1.25, 0) -- node [above, font = \scriptsize] {$+2$} (1.25, 0);

	\draw [fill = white] (-0.75, 0) circle (0.15);
	\draw [fill = white] (0.75, 0) circle (0.15);
	
	\draw (-0.75, 0.5) -- (-0.75, -0.5);
	\draw (0.75, 0.5) -- (0.75, -0.5);
	
	&
	
	\draw (-1.5, 0) -- (-1.125, 0) (-0.875, 0) -- (-0.5, 0) (0.5, 0) -- (0.875, 0) (1.125, 0) -- (1.5, 0);
	
	\draw (-1, 0.5) -- (-1, -0.5);
	\draw (1, 0.5) -- (1, -0.5);
	
	\begin{scope}[rounded corners]
	
		\draw (-1.5, 0.25) -- (-1.25, 0.25) -- (-1.25, 0.5);
		\draw (-1.5, -0.25) -- (-1.25, -0.25) -- (-1.25, -0.5);
		
		\draw (-0.75, 0.5) -- (-0.75, 0.25) -- (-0.5, 0.25) (0.5, 0.25) -- (0.75, 0.25) -- (0.75, 0.5);
		\draw (-0.75, -0.5) -- (-0.75, -0.25) -- (-0.5, -0.25) (0.5, -0.25) -- (0.75, -0.25) -- (0.75, -0.5);
	
		\draw (1.5, 0.25) -- (1.25, 0.25) -- (1.25, 0.5);
		\draw (1.5, -0.25) -- (1.25, -0.25) -- (1.25, -0.5);
	
	\end{scope}
	
	\draw (-0.5, 0) cos (0, 0.125) sin (0.5, 0.25);
	\draw (-0.5, -0.25) cos (0, -0.125) sin (0.5, 0);
	\draw [draw = white, double = black, double distance between line centers = 3 pt, line width = 2.6 pt] (-0.5, 0.25) cos (0, 0) sin (0.5, -0.25);

\\

	\draw (-2, 0) -- node [above, pos = 0.625] {$+1$} (0, 0) -- node [above, pos = 0.375] {$+1$} (2, 0);

	\draw [fill = white] (-1.5, 0) circle (0.15);
	\draw [fill = white] (1.5, 0) circle (0.15);
	\draw [gray, fill = white] (0, 0) circle (0.15);
	
	\draw (-1.5, 0.5) -- (-1.5, -0.5);
	\draw (1.5, 0.5) -- (1.5, -0.5);
	\draw [gray] (0, -0.5) -- (0, 0.5);
	
&
	
	\draw (-2, 0) -- (-1.625, 0) (-1.375, 0) -- (1.375, 0) (1.625, 0) -- (2, 0);
	
	\draw (-1.5, 0.5) -- (-1.5, -0.5);
	\draw (1.5, 0.5) -- (1.5, -0.5);
	
	\begin{scope}[rounded corners]
	
		\draw (-2, 0.25) -- (-1.75, 0.25) -- (-1.75, 0.5);
		\draw (-2, -0.25) -- (-1.75, -0.25) -- (-1.75, -0.5);
		
		\draw (-1.25, 0.5) -- (-1.25, 0.25) -- (1.25, 0.25) -- (1.25, 0.5);
		\draw (-1.25, -0.5) -- (-1.25, -0.25)  -- (1.25, -0.25) -- (1.25, -0.5);
	
		\draw (2, 0.25) -- (1.75, 0.25) -- (1.75, 0.5);
		\draw (2, -0.25) -- (1.75, -0.25) -- (1.75, -0.5);
	
	\end{scope}
	
	\draw [fill = gray!50] (-0.875, -0.5) rectangle (-0.375, 0.5);
	\node at (-0.625, 0) {$\hat{\alpha}$};

	\draw [fill = gray!50] (0.375, -0.5) rectangle (0.875, 0.5);
	\node at (0.625, 0) {$\hat{\alpha}$};
	
	
	\draw [gray, dashed] (0, 0.5) -- (0, -0.5);

\\
};
\end{tikzpicture}
\caption{\label{2-twist}Writing a twist of $+2$ in terms of the algebraic object $\hat{\alpha}$.}
\end{figure}
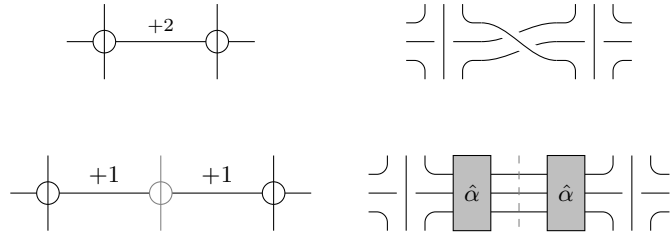

\begin{figure}[hbt]
\begin{tikzpicture}[> = latex]
\matrix[column sep = 0.75 cm, row sep = 1 cm]{

	\draw (-1.25, 0) -- node [above, font = \scriptsize] {$-1$} (1.25, 0);

	\draw [fill = white] (-0.75, 0) circle (0.15);
	\draw [fill = white] (0.75, 0) circle (0.15);
	
	\draw (-0.75, 0.5) -- (-0.75, -0.5);
	\draw (0.75, 0.5) -- (0.75, -0.5);
	
	&
	
	\draw (-1.5, 0) -- (-1.125, 0) (-0.875, 0) -- (0.875, 0) (1.125, 0) -- (1.5, 0);
	
	\draw (-1, 0.5) -- (-1, -0.5);
	\draw (1, 0.5) -- (1, -0.5);
	
	\begin{scope}[rounded corners]
	
		\draw (-1.5, 0.25) -- (-1.25, 0.25) -- (-1.25, 0.5);
		\draw (-1.5, -0.25) -- (-1.25, -0.25) -- (-1.25, -0.5);
		
		\draw (-0.75, 0.5) -- (-0.75, 0.25) -- (0.75, 0.25) -- (0.75, 0.5);
		\draw (-0.75, -0.5) -- (-0.75, -0.25) -- (0.75, -0.25) -- (0.75, -0.5);
	
		\draw (1.5, 0.25) -- (1.25, 0.25) -- (1.25, 0.5);
		\draw (1.5, -0.25) -- (1.25, -0.25) -- (1.25, -0.5);
	
	\end{scope}
	
	\draw [fill = gray!50] (-0.375, -0.5) rectangle (0.375, 0.5);
	\node at (0, 0) {$\hat{\alpha}^{-1}$};
	
\\

	\draw (-2, 0) -- node [above, pos = 0.625] {$-2$} (0, 0) -- node [above, pos = 0.375] {$+1$} (2, 0);

	\draw [fill = white] (-1.5, 0) circle (0.15);
	\draw [fill = white] (1.5, 0) circle (0.15);
	\draw [gray, fill = white] (0, 0) circle (0.15);
	
	\draw (-1.5, 0.5) -- (-1.5, -0.5);
	\draw (1.5, 0.5) -- (1.5, -0.5);
	\draw [gray] (0, -0.5) -- (0, 0.5);
	
&
	
	\draw (-2, 0) -- (-1.625, 0) (-1.375, 0) -- (-1, 0) (0, 0) -- (0.375, 0) (0.875, 0) -- (1.375, 0) (1.625, 0) -- (2, 0);
	
	\draw (-1.5, 0.5) -- (-1.5, -0.5);
	\draw (1.5, 0.5) -- (1.5, -0.5);
	
	\begin{scope}[rounded corners]
	
		\draw (-2, 0.25) -- (-1.75, 0.25) -- (-1.75, 0.5);
		\draw (-2, -0.25) -- (-1.75, -0.25) -- (-1.75, -0.5);
		
		\draw (-1.25, 0.5) -- (-1.25, 0.25) -- (-1, 0.25) (0, 0.25) -- (0.375, 0.25) (0.875, 0.25) -- (1.25, 0.25) -- (1.25, 0.5);
		\draw (-1.25, -0.5) -- (-1.25, -0.25) -- (-1, -0.25) (0, -0.25) -- (0.375, -0.25) (0.875, -0.25) -- (1.25, -0.25) -- (1.25, -0.5);
	
		\draw (2, 0.25) -- (1.75, 0.25) -- (1.75, 0.5);
		\draw (2, -0.25) -- (1.75, -0.25) -- (1.75, -0.5);
	
	\end{scope}

	\draw [fill = gray!50] (0.375, -0.5) rectangle (0.875, 0.5);
	\node at (0.625, 0) {$\hat{\alpha}$};
	
	\draw (-1, 0.25) cos (-0.5, 0) sin (0, -0.25);
	\draw [draw = white, double = black, double distance between line centers = 3 pt, line width = 2.6 pt] (-1, 0) cos (-0.5, 0.125) sin (0, 0.25);
	\draw [draw = white, double = black, double distance between line centers = 3 pt, line width = 2.6 pt] (-1, -0.25) cos (-0.5, -0.125) sin (0, 0);
	
	
	\draw [gray, dashed] (0, 0.5) -- (0, -0.5);

\\

	\draw (-2, 0) -- node [above, pos = 0.625] {$+1$} (0, 0) -- node [above, pos = 0.375] {$-2$} (2, 0);

	\draw [fill = white] (-1.5, 0) circle (0.15);
	\draw [fill = white] (1.5, 0) circle (0.15);
	\draw [gray, fill = white] (0, 0) circle (0.15);
	
	\draw (-1.5, 0.5) -- (-1.5, -0.5);
	\draw (1.5, 0.5) -- (1.5, -0.5);
	\draw [gray] (0, -0.5) -- (0, 0.5);
	
&
	
	\draw (-2, 0) -- (-1.625, 0) (-1.375, 0) -- (-0.875, 0) (-0.375, 0) -- (0, 0) (1, 0) -- (1.375, 0) (1.625, 0) -- (2, 0);
	
	\draw (-1.5, 0.5) -- (-1.5, -0.5);
	\draw (1.5, 0.5) -- (1.5, -0.5);
	
	\begin{scope}[rounded corners]
	
		\draw (-2, 0.25) -- (-1.75, 0.25) -- (-1.75, 0.5);
		\draw (-2, -0.25) -- (-1.75, -0.25) -- (-1.75, -0.5);
		
		\draw (-1.25, 0.5) -- (-1.25, 0.25) -- (-0.875, 0.25) (-0.375, 0.25) -- (0, 0.25) (1, 0.25) -- (1.25, 0.25) -- (1.25, 0.5);
		\draw (-1.25, -0.5) -- (-1.25, -0.25) -- (-0.875, -0.25) (-0.375, -0.25) -- (0, -0.25) (1, -0.25) -- (1.25, -0.25) -- (1.25, -0.5);
	
		\draw (2, 0.25) -- (1.75, 0.25) -- (1.75, 0.5);
		\draw (2, -0.25) -- (1.75, -0.25) -- (1.75, -0.5);
	
	\end{scope}
	
	\draw [fill = gray!50] (-0.875, -0.5) rectangle (-0.375, 0.5);
	\node at (-0.625, 0) {$\hat{\alpha}$};
	
	\draw (0, 0.25) cos (0.5, 0) sin (1, -0.25);
	\draw [draw = white, double = black, double distance between line centers = 3 pt, line width = 2.6 pt] (0, 0) cos (0.5, 0.125) sin (1, 0.25);
	\draw [draw = white, double = black, double distance between line centers = 3 pt, line width = 2.6 pt] (0, -0.25) cos (0.5, -0.125) sin (1, 0);
	
	
	\draw [gray, dashed] (0, 0.5) -- (0, -0.5);

\\
};
\end{tikzpicture}
\caption{\label{-1-twist}Writing a $-1$ twist in terms of a $-2$ twist and a $+1$ twist, in two different ways.}
\end{figure}
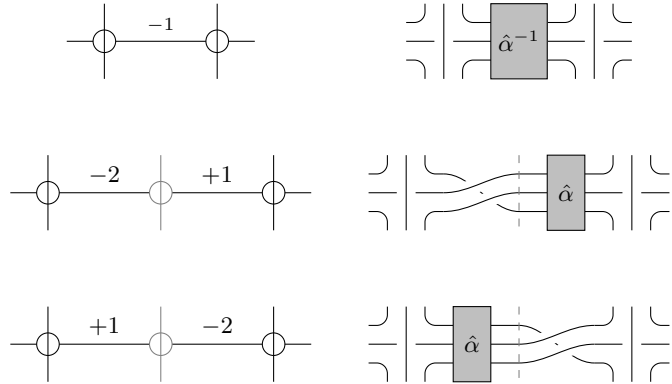

At the moment, we do not know whether these markers correspond to braids of three lines. From the comments just made, the answer of whether $\hat{\beta}$ is a braid will be the same as for $\hat{\alpha}$, and $\hat{\omega}|$ the same as $|\hat{\omega}$, from symmetry operations acting on the graph. In fact, we can go further, and write relations between all four objects. These are shown in Figure \ref{trip-rel}. The basic idea is that the same edge twist can be written in multiple ways. The top diagram in Figure \ref{trip-rel} shows a zero twist on an edge incident to vertices of opposite states. However, we can also represent this as two twists, with a ``virtual'' vertex in between; these are the cases shown in the middle and bottom of Figure \ref{trip-rel}. Suppose we take the middle case, and start from the left-hand $\oplus$ vertex. Then, moving left to right along the edge, the zero edge twist can be seen in the triangulation picture as taking the face dual to the edge, twisting it in one direction through $\pi/3$ radians, but then undoing this twist as one moves further along the edge before gluing it to the tetrahedron dual to the right-hand vertex. We can pretend we are gluing the actual faces to an imaginary face between the two real dual faces; this virtual vertex is shown in gray between the two actual vertices. The bottom diagram in Figure \ref{trip-rel} would be the same idea, just reversing the direction of the intermediate twists. When this is done, the fact that all of these versions must be equivalent gives relations between the algebraic objects. If we read the braid from left to right along the edge, then we obtain
\begin{equation}
\label{bar-omega}
	|\hat{\omega} = \hat{\alpha} \sigma_1 = \sigma_1 \hat{\beta}
\end{equation}
while if we read right to left, we have
\begin{equation}
\label{omega-bar}
	\hat{\omega}| = \sigma_2 \hat{\alpha} = \hat{\beta} \sigma_2
\end{equation}
Note that moving along the edge in two different ways, we have two relations between $\hat{\alpha}$ and $\hat{\beta}$; we will return to this point momentarily. We now have a way to write any of the four objects in terms of the others. In particular, using (\ref{bar-omega}) and (\ref{omega-bar}) to solve for $\hat{\beta}, | \hat{\omega},$ and $\hat{\omega}$ in terms of $\hat{\alpha}$, we get
\begin{equation}
\label{bw-dict}
	\hat{\beta} = \sigma_1 ^{-1} \hat{\alpha} \sigma_1 = \sigma_2 \hat{\alpha} \sigma_2^{-1} \qquad
	|\hat{\omega} = \hat{\alpha} \sigma_1 \qquad
	\hat{\omega}| = \sigma_2 \hat{\alpha}
\end{equation}

We now focus on finding the properties of $\hat{\alpha}$, with all other algebraic objects defined using these relations (\ref{bw-dict}). Using the same approach as before, we can use equivalent ways of writing a $+2$ twist between two $\oplus$ vertices, as in Figure \ref{2-twist}, to obtain
\begin{equation}
\label{a-sq-eqn}
	(\hat{\alpha})^2 = \sigma_2 ^{-1} \sigma_1 ^{-1} = (\sigma_1 \sigma_2)^{-1}
\end{equation}
In Figure \ref{-1-twist}, we look at the operator ${\hat{\alpha}}^{-1}$, which is the algebraic object associated with a $-1$ twist. Although we do not show it explicitly here, it is easy to show that ${\hat{\alpha}}^{-1}$ is, in fact, the inverse of $\hat{\alpha}$, in the sense that ${\hat{\alpha}}^{-1} \hat{\alpha} = \hat{\alpha} {\hat{\alpha}}^{-1} = \mathbb{1}_3$, with $\mathbb{1}_3$ the identity braid of three strands. Turning back to Figure \ref{-1-twist}, we get that
\[
	{\hat{\alpha}}^{-1} = \sigma_2 ^{-1} \sigma_1 ^{-1} \hat{\alpha} = \hat{\alpha} \sigma_2 ^{-1} \sigma_1 ^{-1}
\]
Not only does this give a relation ${\hat{\alpha}}^{-1}$, but also shows that $\hat{\alpha}$ commutes with $\sigma_2 ^{-1} \sigma_1 ^{-1} = (\sigma_1 \sigma_2)^{-1}$. This is why there is no problem with the two definitions of $\hat{\beta}$ in terms of $\hat{\alpha}$; from (\ref{bw-dict}), we have that
\[
	\sigma_1^{-2} \sigma_1 ^{-1} \hat{\alpha} \sigma_1 \sigma_2 = (\sigma_1 \sigma_2)^{-1} \hat{\alpha} (\sigma_1 \sigma_2)
\]
and the commutativity of $\hat{\alpha}$ and $\sigma_1 \sigma_2$ means the above is simply equal to $\hat{\alpha}$, giving a consistent definition of $\hat{\beta}$. This also means that the direction one moves along any given edge does not matter, since the same information is obtained up to the relationship between $\hat{\alpha}$ and $\sigma_1 \sigma_2$.

We now show that none of these algebraic objects corresponds to a braid of three strands. Specifically, we do this by showing that this cannot be done for $\hat{\alpha}$; this will imply that it is impossible for $\hat{\beta}, |\hat{\omega}$, and $\hat{\omega}|$ due to the relations (\ref{bw-dict}). We start with the relation (\ref{a-sq-eqn}) for $(\hat{\alpha})^2$, and work with the Burau representation $\rho: B_3 \to GL(3, \mathbb{Z}[t, t^{-1}])$ between the group $B_3$ of three-strand braids, and $3 \times 3$ matrices whose entries are Laurent polynomials in the abstract variable $t$. We use proof by contradiction, and show that any square root of the matrix $\rho((\sigma_1 \sigma_2)^{-1})$ is not in the image of the map $\rho$. Start with the matrices $\rho(\sigma_1)$ and $\rho(\sigma_2)$, given by
\[
	\rho(\sigma_1) = 
	\left[
		\begin{array}{ccc}
			1 - t 	& t 	& 0	\\
			1	& 0	& 0	\\
			0	& 0	& 1	\\
		\end{array}
	\right]
	\qquad
	\rho(\sigma_2) = 
	\left[
		\begin{array}{ccc}
			1 	& 0 		& 0	\\
			0	& 1 - t	& t	\\
			0	& 1		& 0	\\
		\end{array}
	\right]
\]
Multiplying these two matrices together, then finding the inverse, results in
\[
	\rho((\sigma_1 \sigma_2)^{-1}) =
	\left[
		\begin{array}{ccc}
			0 		& 1 				& 0			\\
			0		& 0				& 1			\\
			t^{-2}	& t^{-1} - t^{-2}	& 1 - t^{-1}	\\
		\end{array}
	\right]
\]
Let $z = \exp(2 \pi i / 3)$, so $z^3 = 1$. Then, we can write $\rho((\sigma_1 \sigma_2)^{-1})$ in terms of an eigendecomposition $\rho((\sigma_1 \sigma_2)^{-1}) = PDP^{-1}$, where
\[
	P =
	\left[
		\begin{array}{ccc}
			1 	& z t^2 	& (z t)^2	\\
			1	& z^2 t	& z t		\\
			1	& 1		& 1		\\
		\end{array}
	\right]
	\qquad
	D = 
	\left[
		\begin{array}{ccc}
			1 	& 0 		& 0			\\
			0	& z t^{-1}	& 0			\\
			0	& 0		&z^2 t^{-1}	\\
		\end{array}
	\right]
\]
To find the square root of $\rho((\sigma_1 \sigma_2)^{-1})$, we define a matrix $M$, where
\[
	M = PD^{1/2}P^{-1}
\]
There are $2^3 = 8$ choices for the matrix $D^{1/2}$, given by
\[
	D^{1/2} (\epsilon_i) = 
	\left[
		\begin{array}{ccc}
			\epsilon_1 	& 0 					& 0					\\
			0		& \epsilon_2 z^2 t^{-1/2}	& 0					\\
			0		& 0					& \epsilon_3 z t^{-1/2}	\\
		\end{array}
	\right]
\]
where $\epsilon_i = \pm 1, i = 1, 2, 3$, and we have used the fact that $(z^2)^2 = z^4 = z$. However, for any choice of $\epsilon_i$, the elements of $M$ will be outside the Laurent polynomials $\mathbb{Z} [t, t^{-1}]$, because of the (inverse) square roots of $t$ appearing in $D^{1/2}$. To show a specific example, we look at the entry in the first row and column of $M$. This is given by
\[
	M_{11} = \frac{1}{|P|} [\epsilon_1 (z^2 - z) t + (\epsilon_3 - \epsilon_2) t^{3/2} + (\epsilon_2 z - \epsilon_3 z^2) t^{5/2}]
\]
with the determinant $|P| = (z^2 - z)(t + t^2 + t^3)$. There will always be a $t^{5/2}$ term in the numerator of the entry $M_{11}$, possibly with a complex coefficient; a similar story occurs for the other entries of $M$. Thus, $M$ is outside the image of $\rho$, and so there is no braid in $B_3$ which satifies the original definition (\ref{a-sq-eqn}) of $\hat{\alpha}$. All of the algebraic objects $\hat{\alpha}, \hat{\beta}, |\hat{\omega},$ and $\hat{\omega}|$ are related to each other by (\ref{bw-dict}), so if one of them cannot be matched to a braid of three strands, neither can any of the others. This gives the following result.

\begin{proposition}
	None of the algebraic elements $\hat{\alpha}, \hat{\beta}, |\hat{\omega}$ or $\hat{\omega}|$ can be represented by a braid on three strands.
\end{proposition}

If the algebraic object $\hat{\alpha}$ could be written in terms of an element of three-strand braids $B_3$, then this would have meant the 3-line graph would have an equivalent formulation as a closed braid of $\ge 3$ strands. This comes from the theorem by Alexander\cite{Ale23} that any link has this alternative representation. The fact that $\hat{\alpha}$ cannot be written as a braid seems to obstruct this process when this algebraic object is placed on the extended 3-line graph. However, these 3-line graphs can still be written as a generalization of a closed braid, if the algebraic object $\hat{\alpha}$ included in the definition of a generalized braid group. In particular, suppose we start with the group $B_n$ of braids with $n$ strands. If we label the strands $\{1, 2, \cdots, n\}$, then the element $\sigma_k$, $1 \le k < n$, is a crossing of the $k$ strand over the $(k + 1)$ strand, while its inverse $\sigma_k ^{-1}$ is a crossing of the $(k + 1)$ strand over the $k$ strand. Now we add additional elements $\alpha_k$, $1 \le k < (n - 2)$, to represent the algebraic object coming from the extended 3-line graph. Then, the generalized braid group ${\widetilde B_n}$ is defined on $n$ strands such that the following properties are satisfied:
\begin{enumerate}

	\item $\sigma_i \sigma_j = \sigma_j \sigma_i$, for $|i - j| \le 2$
	\item $\sigma_i \sigma_{i + 1} \sigma_i = \sigma_{i + 1} \sigma_i \sigma_{i + 1}$
	\item $\alpha_i ^2 = \sigma_i \sigma_{i + 1}$
	\item $\alpha_i \sigma_i \sigma_{i + 1} = \sigma_i \sigma_{i + 1} \alpha_i$

\end{enumerate}
The first two relations are those that define the original braid group $B_n$; the last two give the conditions necessary to include the additional elements $\alpha_i$.

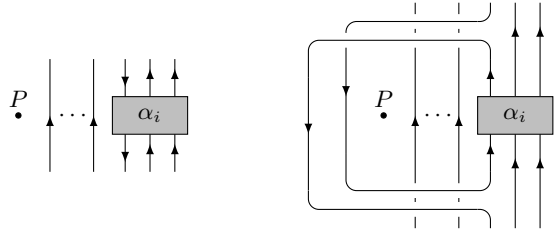
\begin{figure}[hbt]
\begin{tikzpicture}[> = latex]
\matrix[column sep = 1.5 cm]{

	
	\draw [fill = gray!50] (-0.5, -0.25) rectangle (0.5, 0.25);
	\node at (0, 0) {$\alpha_i$};
	
	\begin{scope}[decoration = {markings, mark = at position 0.5 with {\arrow{latex}}}]
	
		\draw [postaction = {decorate}] (-1.33, -0.75) -- (-1.33, 0.75);
		\draw [postaction = {decorate}] (-0.75, -0.75) -- (-0.75, 0.75);
		
	\end{scope}
	
	\node at (-1, 0) {$\cdots$};
	
	\begin{scope}[decoration = {markings, mark = at position 0.75 with {\arrow{latex}}}]
		
		\draw [postaction = {decorate}] (-0.33, 0.75) -- (-0.33, 0.25);
		\draw [postaction = {decorate}] (-0.33, -0.25) -- (-0.33, -0.75);
	
		\draw [postaction = {decorate}] (0, -0.75) -- (0, -0.25);
		\draw [postaction = {decorate}] (0, 0.25) -- (0, 0.75);
	
		\draw [postaction = {decorate}] (0.33, -0.75) -- (0.33, -0.25);
		\draw [postaction = {decorate}] (0.33, 0.25) -- (0.33, 0.75);
	
	\end{scope}
	
	\filldraw (-1.75, 0) circle (1 pt) node [above] {$P$};
	
&

	
	\begin{scope}[decoration = {markings, mark = at position 0.5 with {\arrow{latex}}}]
	
		\draw [postaction = {decorate}] (-1.33, -1.5) -- (-1.33, 1.5);
		\draw [postaction = {decorate}] (-0.75, -1.5) -- (-0.75, 1.5);
		
		\draw [postaction = {decorate}] (-2.25, 1) -- (-2.25, -0.5);
		\draw [postaction = {decorate}] (-2.75, 0.5) -- (-2.75, -1);
		
	\end{scope}
	
	\node at (-1, 0) {$\cdots$};
	
	\draw [rounded corners] (-0.33, 1.5) -- (-0.33, 1.25) -- (-0.5, 1.25) (-0.33, -1.5) -- (-0.33, -1.25) -- (-0.5, -1.25);

	\draw [draw = white, double = black, double distance between line centers = 3 pt, line width = 2.6 pt, rounded corners]
		(-0.5, 1.25) -- (-2.25, 1.25) -- (-2.25, 1) (-2.25, -0.5) -- (-2.25, -1) -- (-0.33, -1) -- (-0.33, -0.75);
	\draw [draw = white, double = black, double distance between line centers = 3 pt, line width = 2.6 pt, rounded corners]
		(-0.5, -1.25) -- (-2.75, -1.25) -- (-2.75, -1) (-2.75, 0.5) -- (-2.75, 1) -- (-0.33, 1) -- (-0.33, 0.75);

	
	\begin{scope}[decoration = {markings, mark = at position 0.75 with {\arrow{latex}}}]
		
		\draw [postaction = {decorate}] (-0.33, -0.75) -- (-0.33, -0.25);
		\draw [postaction = {decorate}] (-0.33, 0.25) -- (-0.33, 0.75);
	
		\draw [postaction = {decorate}] (0, -1.5) -- (0, -0.25);
		\draw [postaction = {decorate}] (0, 0.25) -- (0, 1.5);
	
		\draw [postaction = {decorate}] (0.33, -1.5) -- (0.33, -0.25);
		\draw [postaction = {decorate}] (0.33, 0.25) -- (0.33, 1.5);
	
	\end{scope}
	
	\filldraw (-1.75, 0) circle (1 pt) node [above] {$P$};
	
	\draw [fill = gray!50] (-0.5, -0.25) rectangle (0.5, 0.25);
	\node at (0, 0) {$\alpha_i$};

\\
};
\end{tikzpicture}
\caption{\label{trick}Using the method of Alexander to show how the element $\alpha_i$ can be included in a closed braid.}
\end{figure}

As stated previously, Alexander proved that it is possible for any knot or link to be written as a closed braid -- in other words, a braid whose matching edges at the top and bottom of the braid are joined together without crossing. The essence of the proof is to choose a point $P$ not on the link, and put an orientation on all components of the link. When one moves along the link with the orientation, the motion should appear at point $P$ as a rotation in the same direction. If there are any arcs of the link where this is not the case, the link can be modified by essentially moving the arc to the other side of $P$. Then, moving along the displaced arc now appears as a rotation with the correct direction. With the new generalized group ${\widetilde B_n}$, this can be done with the new object $\alpha_i$ as well. Consider a 3-edge graph and put an orientation on all edges of this graph. This orientation can be chosen to pass through any locations of $\alpha_i$ by assuming, without loss of generality, the identity from one side of the algebraic object to the other. This can always be done, since it is possible to do if $\alpha_i$ were not present; the addition of the algebraic object does not change the orientations along the edges. An example of this is shown on the left-hand side of Figure \ref{trick}. To make the presentation of the proof simpler, the orientations have been chosen so that all but one travel in the same direction around the axis point $P$, but the ideas here can be extended to any possible case. Most of the edges rotate around $P$ in a counter-clockwise direction. To deal with those that do not, we can perform repeated RII moves until the arc is on the opposite side of the axis $P$. In addition, the direction of the orientation for the left-hand strand incident to $\alpha_i$ is reversed, so that all three strands pass through the algebraic object in a counter-clockwise direction, as seen from $P$. The final result is shown on the right-hand side of Figure \ref{trick}. Note that it does not matter which edge is over the other, since the final closed braid will not be unique. By using this process repeatedly, any extended 3-line graph can be written as a closed generalized braid.


\section{Discussion}
\label{conclude}

This paper demonstrates that the graph equivalence of the Pachner moves can be consistently used on knotted 4-regular graphs with framed edges having certain twist values. This combination ensures there is enough information with the graph to potentially find a consistent dual triangulation. In particular, the edge twists must match the vertex states of the two vertices the edge is incident to, so the dual faces of the triangulation may be glued together uniquely. The triangulation then allows a possible interpretation as a manifold. This machinery also means the result of a Pachner move on the graph depends only on the original graph and the move used; there is no issue about how the first graph was manipulated to set up for the given move. In developing these ideas, a new invariant of knotted 4-regular framed graphs was found. This invariant was able to show that the Pachner graph $\mathscr{P}_G (S^3)$ of all knotted graphs dual to triangulations of the 3-sphere is not connected. Furthermore, the case of edge twists not matching their adjacent vertices was handled with the definition of a new algebraic object $\alpha$. Much like the closed braids of knot theory, with the addition of this new object, the graph diagram of a knotted 4-regular framed graph can be associated with a closed generalized braid.

With the tools developed in this paper, we can comment on the behavior of knotted 4-regular framed graphs under the Pachner moves. One consequence comes from the fact that the space of graphs is not connected by these moves. This was seen explicitly in the case of the Pachner graph $\mathscr{P}_G (S^3)$; however, it is natural to think this is true for the Pachner graph of any manifold. Indeed, since the duality between graphs and triangulations extends to pseudomanifolds as well, we conjecture this disconnectedness is a generic situation. This property has implications for any physical theory using a configuration space based on knotted graphs evolving under the Pachner moves. In particular, this gives rise to superselection sectors in the theory, suggesting there exist conserved quantities which identify these sectors. At the moment, the only such known quantity is the link invariant $L(G)$ for the graph $G$, but others may be discovered, including those utilizing the closed generalized braids coming from the 3-line graph. This suggests that the manner the graph is embedded is an important consideration for conservation laws, providing a way to include topological defects such as geons without changing the properties of the dual manifold at all. An example of these may be those twists which do not match the vertex states the edge is incident to, prohibiting the use of a Pachner move involving the edge. None of the graph moves change this matching between edge twists and vertex states for edges adjacent to the subgraph modified by the move, so a twist mismatch would be preserved on the edge in question.

A related issue is the overdetermined nature of the twists on the framed edges, in reference to the dual manifold. Specifically, the identifications of dual faces in the triangulation is not altered by changing any of the edge twists by a multiple of $2\pi$ radians. This leads to the question of which is the more fundamental object -- the knotted graph or its dual manifold (assuming it exists). If one gives primacy to the dual manifolds, then one approach to the associated graph states would be to use equivalence classes of graphs that give the same gluing information for the dual triangulation. Two graphs $G_1$ and $G_2$ would be in the same class if they have the same adjacency matrix, differing only by the twists on the edges. With this choice, we can readily identify edges in $G_1$ with their counterpart in $G_2$. For any such identified edge, the twist for that edge in $G_1$ must differ from the twist in $G_2$ by a multiple of $2\pi$ radians -- i.e. a multiple of 6 in the abbreviated notation. For this choice, the embedding would be arbitrary, so that many DT sequences would appear in the same class, differing by the crossing types. As we have already seen in the derivation given with Figure~\ref{3-sph-graphs}, the link invariant $L(G)$ would no longer be identical for all members of the equivalence class. Only invariants based on the manifold itself -- in graph terms, on the adjacency matrix and the edge twists mod $2\pi$ -- would identify elements of the class. This choice would lose the possibility of conserved quantities based on the graph embedding or twist information.

\begin{figure}[hbt]
\begin{tikzpicture}
\matrix[column sep = 1.5 cm, row sep = 0.5 cm]{


	\draw (-1, 0) circle (0.15);
	\draw (1, 0) circle (0.15);
	
	\draw [rounded corners] (-0.15, 0) -- (-0.85, 0) (-1.15, 0) -- (-1.5, 0) -- (-1.5, 1) -- node [above, pos = 0.6] {$-1$} (1, 1) -- (1, -0.5)
		-- (0, -0.5) -- (0, 0.5) -- node [above] {$+2$} (-1, 0.5) -- (-1, -1) -- node [below, pos = 0.4] {$-1$} (1.5, -1) -- (1.5, 0) -- (1.15, 0)
		(0.85, 0) -- node [above] {$-4$} (0.15, 0);

&
	
	
	\begin{scope}[scale = 0.75]
	
	\draw [rounded corners]
		(2, -0.8) cos (1.5, -1) sin (1, -1.2) -- (-0.2, -1.2) -- (-0.2, -0.3) (-0.2, 0.3) -- (-0.2, 0.8) -- (-1.8, 0.8) -- (-1.8, 0.2)
			-- (0.5, 0.2) cos (1.5, 0) sin (2.5, -0.2) -- (2.8, -0.2) -- (2.8, -0.8) -- (2, -0.8)
		(-2.2, 0) -- (-3, 0) -- (-3, 2) -- (-0.5, 2) cos (0, 1.9) sin (0.5, 1.8) -- (2.8, 1.8) -- (2.8, 0.2) -- (2.5, 0.2)
		(1.5, 0.2) cos (2, 0.1) sin (2.5, 0) -- (2.8, 0) (-1.8, 0) -- (0.5, 0) cos (1, -0.1) sin (1.5, -0.2)	 
		(1.5, -2.2) -- (4.2, -2.2) -- (4.2, 0.2) -- (3.2, 0.2) -- (3.2, 2.2) -- (-3.2, 2.2) -- (-3.2, -0.2) -- (-2.2, -0.2) -- (-2.2, -2.2) -- cycle 
		(0.5, -0.2) -- (-1.8, -0.2) -- (-1.8, -1.8) -- (0.75, -1.8) cos (1.25, -1.9) sin (1.75, -2) -- (4, -2) -- (4, 0)
			-- (3.2, 0) 
		;
	
	\draw [rounded corners, draw = white, double = black, double distance between line centers = 3 pt, line width = 2.6 pt]
		(0, 0.3) -- (0, 1) -- (-2, 1) -- (-2, -2) -- (0.75, -2) cos (1.25, -1.9) sin (1.75, -1.8) -- (3.8, -1.8)
		-- (3.8, -0.2) -- (3.2, -0.2) -- (3.2, -1.2) -- (2, -1.2) cos (1.5, -1.1) sin (1, -1) -- (0, -1) -- cycle			
		(0.2, 0.3) -- (0.2, 1.2) -- (-2.2, 1.2) -- (-2.2, 0.2) -- (-2.8, 0.2) -- (-2.8, 1.8) -- (-0.5, 1.8) cos (0, 1.9) sin (0.5, 2) -- (3, 2)
		-- (3, -1) -- (2, -1) cos (1.5, -0.9) sin (1, -0.8) -- (0.2, -0.8) -- cycle 
		(-0.2, -0.3) -- (-0.2, 0.3)
		
		(0.5, -0.2) cos (1, 0) sin (1.5, 0.2)
		(1.5, -0.2) cos (2, 0) sin (2.5, 0.2)
		;
		
	\end{scope}

\\
};
\end{tikzpicture}
\caption{\label{3-sph-graph-mod}Graph diagram for the sequence $3^\ell 5^\ell 1^+$, and its corresponding 3-line graph, where the twist of the middle edge has been changed by $-6$ from the case shown in Figure \ref{3-sph-graphs}.}
\end{figure}
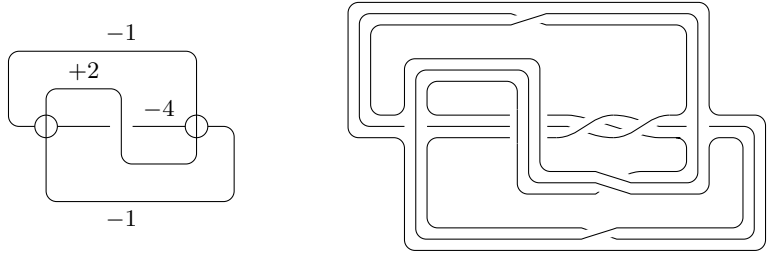

An intermediate possibility is to consider equivalence classes with the same adjacency matrix and embedding, but again restrict the values of edge twists only mod $2\pi$. Now, elements of each class would have the same DT sequence, although there would be many classes matched with the same manifold -- again, see the cases given in Figure \ref{3-sph-graphs} for two graphs giving gluing information for the 3-sphere. This choice allows graphs in the same class to be identified by their DT sequence, but not the invariant $L(G)$. The issue is that, in terms of the knotted graphs, changing any edge twist by a multiple of $2\pi$ will not give the same graph in general, or even equivalence under the Pachner moves. In the latter case, most Pachner moves commute with the operation of changing any edge twist by $2 \pi n$, but this is not true for the 2-3 move when the internal edge of the graph has its twist modified. That situation would generically give unequal link invariants $L(G_1)$ and $L(G_2)$, so that invariant could not be used to show two graphs are in the same equivalence class. An example of this would be to compare two versions of the graph $3^\ell 5^\ell 1^+$, the first with the same twists as seen in the top row of Figure \ref{3-sph-graphs}, and the second case as in Figure \ref{3-sph-graph-mod}. In the latter, the middle edge has been changed by $-2\pi$. Considering the 3-line graph in Figure \ref{3-sph-graph-mod}, there are three unlinked unknots in the diagram -- the outer loop, and the two loops with overcrossings on the top and bottom crossings, respectively. However, the remaining three loops are linked together in a single link, named $6^3 _3$ in Rolfsen notation. Since this is not the empty link, this graph is not Pachner-equivalent to the same graph with a twist of $+2$ on the central edge, which had $L(G) = \varnothing$.

The above makes the case that one must use all properties of knotted 4-regular graphs in order to get conserved properties that depend on the graph themselves. This is in distinction with the viewpoint taken by Markopoulou and Pr\'emont-Schwarz~\cite{}, who look at the situation from the viewpoint of the manifold instead of the graph. However, this is reasonable, considering they focus on 3-regular framed graphs, which are always associated with a manifold, unlilke the knotted 4-regular graphs. Although at this stage, the case is merely a plausibility argument, it does point to having graph states in a physical theory where a manifold description is insufficient. Obviously further research on these questions is required. On a more speculative note, this does add an avenue for including particles as graph defects, reminiscent of the suggestion by Bilson-Thompson and collaborators~\cite{braid}. In particular, the extension of the braid group developed here may be fruitful in further work on such an idea.

\appendix*

\section{Knotted graphs and their associated manifold}

\begin{figure}[hbt]
\begin{tikzpicture}
\matrix[column sep = 1.5 cm, row sep = 0.5 cm]{


	\draw (-1, 0) circle (0.15);
	\draw (1, 0) circle (0.15);
	
	\draw [rounded corners] (-0.15, 0) -- (-0.85, 0) (-1.15, 0) -- (-1.5, 0) -- (-1.5, 1) -- node [above, pos = 0.6] {$-1$} (1, 1) -- (1, -0.5)
		-- (0, -0.5) -- (0, 0.5) -- node [above] {$+2$} (-1, 0.5) -- (-1, -1) -- node [below, pos = 0.4] {$-1$} (1.5, -1) -- (1.5, 0) -- (1.15, 0)
		(0.85, 0) -- node [above] {$+2$} (0.15, 0);
	
&


	\draw [fill = gray!30] (-1.5, 1) node [above] {$0$} -- (-0.5, 0) node [right] {$1$} --
		(-1.5, -1) node [below] {$2$} -- (-2.5, 0) node [left] {$3$} -- cycle
		(-2.5, 0) -- (-0.5, 0);
	\draw [dashed] (-1.5, 1) -- (-1.5, -1);
	
	\node at (-1.5, -2) {Tet 0};

	\draw [fill = gray!30] (1.5, 1) node [above] {$0$} -- (2.5, 0) node [right] {$1$} --
		(1.5, -1) node [below] {$2$} -- (0.5, 0) node [left] {$3$} -- cycle
		(0.5, 0) -- (2.5, 0);
	\draw [dashed] (1.5, 1) -- (1.5, -1);
	
	\node at (1.5, -2) {Tet 1};
	
\\
};
\end{tikzpicture}
\caption{\label{3-sph-detail}Graph diagram and dual triangulation for the sequence $3^\ell 5^\ell 1^+$ with the same choice of edge twists given in Figure \ref{3-sph-graphs}.}
\end{figure}
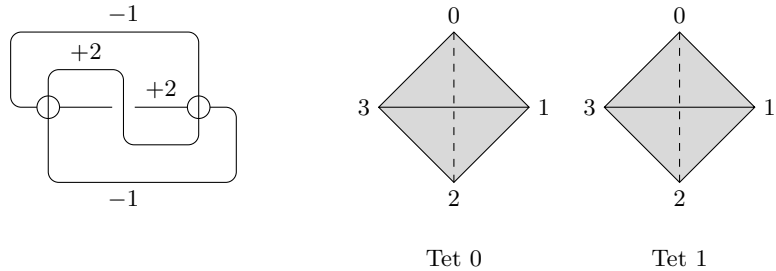

Here, the details of how to go from a graph diagram of a DT sequence to a dual triangulation are spelled out. This is done using the notation of the software package Regina~\cite{Regina}, where $t(abc)$ gives the face of the tetrahedron with label $t = 0, 1$, and the face vertices labelled as $a, b, c \in \{0, 1, 2, 3\}$. Recall each twist is a rotation of one dual face relative to another when they are identified; this is a permutation of the face vertices. In Section \ref{invariant}, the two sequences $3^\ell 5^\ell 1^+$ and $3^\ell 5^\ell 1^-$ which are dual to triangulations of the 3-sphere are studied. A graph diagram for $3^\ell 5^\ell 1^+$, and its dual tetrahedra, are displayed in Figure \ref{3-sph-detail}. The sequence $3^\ell 5^\ell 1^-$ has the same dual triangulation as $3^\ell 5^\ell 1^+$, with only the graph embedding differing between the two, so we do not need to treat the sequence $3^\ell 5^\ell 1^-$ separately. We now see what the edge twist information says about how the faces and edges of the dual tetrahedra are associated with each other. The straight center edge of the graph diagram indicates a gluing between the faces $0(012)$ and $1(023)$, while the $+2$ twist shows we must rotate one face by $2\pi/3$ radians as we glue it to the other. If we imagine moving one face towards the other, the positive twist means the ``moving'' face is rotated counterclockwise based on an axis pointing towards the other face. Thus, the 0 vertex of tetrahedron 0 is identified with vertex 3 of tetrahedron 1; the other two gluings are vertices 1 and 2 of tetrahedron 0 to vertices 2 and 0 of tetrahedron 1, respectively. Notice this automatically gives a gluing between the edges 01, 02, and 12 of tetrahedron 0, and their counterparts 32, 30, and 20 of tetrahedron 1 -- the order of the vertices for each edge gives an orientation of the edge. The other framed edge with a $+2$ twist identifies the faces $0(013)$ and $1(123)$. These faces are not drawn opposite each other in Figure \ref{3-sph-detail}, unlike the earlier case of the faces $0(012)$ and $1(023)$. For ease of analysis, it may help to move one face -- say $0(013)$ -- by translation only so it is drawn opposite the other, before applying the required twist. After this rotation, we have the identification of face $0(013)$ with $1(321)$, where each vertex glued to its partner appearing at the same location in the face description. The other two edges have twists of $-1$, so one face must be rotated $\pi/3$ radians clockwise, relative to the other face. In these cases, if we move one face to be opposite the other, the normal vectors to the faces being glued together must point in opposite directions; this is why a simple translation was sufficient for $0(013)$ and $1(123)$. Doing so gives the face identifications $0(023)$ and $1(301)$, as well as $0(123)$ and $1(201)$. These gluings are collected together in Table \ref{sph-glue}.

\begin{table}[hbt]
\begin{tabular}{c|cccc}
Tetrahedron	& Face 012	& Face 013	& Face 023	& Face 123	\\
\hline
0			& 1(320)		& 1(321)		& 1(301)		& 1(201)	\\
1			& 0(231)		& 0(230)		& 0(210)		& 0(310)	\\
\end{tabular}
\caption{\label{sph-glue}The gluings of the two tetrahedra giving the 3-sphere, for the sequences $3^\ell 5^\ell 1^+$ and $3^\ell 5^\ell 1^-$. The tetrahedra are labeled the same as those shown in Figure \ref{3-sph-detail}, and the notation is that used in the software package Regina.}
\end{table}


\end{document}